\RequirePackage{lineno}
\documentclass[aps,prd,onecolumn,showpacs,superscriptaddress,groupedaddress, nofootinbib,11pt]{revtex4}  
\usepackage{graphicx}
\usepackage{epstopdf}
\usepackage{amsmath}
\usepackage{amsfonts}
\usepackage{amssymb}
\usepackage{appendix}
\usepackage{mathtools}
\usepackage{comment}
\usepackage{bbold}
\usepackage{color}
\usepackage{slashed}
\usepackage{subfigure}
\usepackage{setspace}
\usepackage{footnote}
\usepackage[T1]{fontenc}
\usepackage{multirow}
\usepackage{makecell}
\usepackage{verbatim}
\usepackage{graphicx}

\begin{document}

\singlespacing


\title{Monte Carlo Tuning for $e^+e^-\rightarrow\mbox{Hadrons}$ and Comparison with Unfolded LEP Data }

\author{Jennifer Kile}
\address{Centro de F\'isica Te\'orica de Part\'iculas--CFTP\\Instituto Superior T\'ecnico--IST, Universidade de Lisboa, Av. Rovisco Pais,\\P-1049-001 Lisboa, Portugal\\{\normalfont email:}  jenkile@protonmail.com}

\author{Julian von Wimmersperg-Toeller} 
\address{{\normalfont email:}  jvonwimm@gmail.com}

\date{\today}

\begin{abstract}
We perform two tunes of the SHERPA Monte Carlo generator for the generation of $e^+e^-\rightarrow\mbox{hadrons}$ using the publicly-available LEP analyses in Rivet.  In each of these tunes, we generate events at $\sqrt{s}=91.25\mbox{ GeV}$ using matrix elements for final states containing up to six partons.  In the first, ``LO'' tune, matrix elements for all final states are generated at leading order; in the second, ``NLO'' tune, matrix elements for final states with up to four partons are generated at next-to-leading order using BlackHat, while those for states with five and six partons are generated at leading order.  The tunes are accomplished using Professor, and comparisons with unfolded LEP1 and LEP2 data are produced with Rivet.  We also compare the data with events generated with KK2f interfaced to PYTHIA using the standard ALEPH tune.  We find that both SHERPA samples show improvement relative to KK2f for observables related to four-jet final states, while all three samples produce comparable results for event-shape variables.  Overall, the agreement with data is best for the LO tune.  We provide our tuning parameters and many data-Monte Carlo comparisons.
\end{abstract}


\maketitle


\section{Introduction}
\label{intro}

The LEP experiments were instrumental in studies of QCD properties \cite{Heister:2003aj,Abreu:1996na,Achard:2004sv,Pfeifenschneider:1999rz} such as confirming its non-Abelian structure and measuring $\alpha_s(M_Z)$.   In addition to giving insight into the nature of QCD, these studies provided input for the simulation of QCD events which constituted backgrounds to other LEP analyses.  In particular, four-jet QCD events were an important background in many analyses, such as Higgs searches and studies of $W$ pair production.

During the time of the LEP operation, the Standard Model (SM) QCD simulation was typically generated using the $q\bar{q}$ matrix element (ME) interfaced to a parton shower (PS) generated with PYTHIA \cite{Sjostrand:2000wi}, HERWIG \cite{Corcella:2000bw,Corcella:2002jc}, or ARIADNE \cite{Lonnblad:1992tz}; good description of three-jet states was accomplished by matching to the LO $q\bar{q}g$ ME.  PYTHIA and HERWIG additionally were able to produce events using the four-parton ME, but without simultaneously including the two- and three-parton contributions.  Toward the end of LEP running, APACIC++ \cite{Kuhn:2000dk} was developed, which could generate events with up to five partons, matched to a parton shower; for an experimental application, see Ref. \cite{Abdallah:2004uq}.  For a description of the state of QCD generators at the close of the LEP era, see Ref. \cite{Ballestrero:2000ur}.

In the fifteen years since the LEP shutdown, there have been significant developments in the simulation of SM QCD processes.  For a review of several general-purpose MC generators, see Ref. \cite{Buckley:2011ms}.  Among these is the SHERPA generator \cite{Hoeche:2011fd,Hoeche:2009rj,Gleisberg:2008ta,Schonherr:2008av,Gleisberg:2008fv,Berger:2008sj,Gleisberg:2007md,Schumann:2007mg,Krauss:2001iv}, which has been compared to LEP $e^+e^-\rightarrow\mbox{hadrons}$ data unfolded from detector effects \cite{Hoeche:2009rj,Hoche:2010kg,Gehrmann:2012yg,Frederix:2010ne}.  Particularly important features for our purposes here are the MEPS \cite{Catani:2001cc,Lonnblad:2001iq,Mangano:2001xp,Krauss:2002up,Hoeche:2009rj,Hamilton:2009ne,Giele:2011cb,Lonnblad:2011xx} method, which combines LO MEs of different multiplicities with parton showers to produce inclusive hadronic samples, and MC@NLO\cite{Hoeche:2011fd,Frixione:2002ik,Nason:2004rx,Frederix:2011ig}, which combines fixed-multiplicity NLO MEs to a PS.  SHERPA combines the advantages of MEPS and those of MC@NLO and can produce inclusive samples of $e^+e^-\rightarrow\mbox{hadrons}$ with full merging and matching, with some final states calculated at NLO, if desired.  

In an accompanying paper \cite{paper3}, we discuss a four-jet excess observed in hadronic events in the archived ALEPH data from LEP2. Hadronic events are forced into four jets. When the four jets in these events are paired into dijets which minimize the difference between the two dijet masses, the mean dijet mass shows a cluster around $M_1+M_2\sim 110$ GeV, where $M_1$ and $M_2$ are the masses of the dijet systems, and $M_1$ is the mass of the dijet which contains the most energetic jet in the event.  The excess shows a particular concentration of events where $M_1\sim 80$ GeV and  $M_2\sim 25$ GeV; the local significance of the excess in this region is between $4.7\sigma$ and $5.5\sigma$, depending on hadronization uncertainty assumptions.  The rest of the events are located in a broad excess near $M_1\sim M_2\sim 55\mbox{ GeV}$, with a local significance of $4.1-4.5\sigma$.  For details, we refer the reader to Ref. \cite{paper3}.

The SM background in the region where this excess is observed is dominated by QCD processes.  In order to study this excess, we require the state-of-the-art in simulation of the QCD background and, in particular, an accurate description of four-jet states.  To attain this, it is especially desirable that the MC includes the four-parton ME. This is the first of two papers in which we describe the procedure to obtain this improved description of the SM QCD background using the SHERPA generator.  The output of QCD MC generators, however, is not uniquely determined a priori.  Instead, QCD MC generators have parameters which can be dialed to adjust, for example, the details of showering and fragmentation.  Tuning the values of these parameters requires input from experiment.  

In this paper, we tune the SHERPA QCD MC parameters using LEP1 data and compare the resulting QCD MC to data at both LEP1 and LEP2 energies.  All tuning and data-MC comparisons are accomplished using the publicly-available data in the Rivet v. 2.0.0 \cite{Buckley:2010ar,Cacciari:2011ma} package, which provides distributions of predefined quantities unfolded from experimental effects such as detector effects, selection cuts, and initial-state radiation (ISR).  These quantities include event-shape variables such as thrust, hemisphere masses and jet rates, as well as variables related to the flavors and energies of specific particles in the event.  Tuning the MC consists of finding values for the generator parameters such that the distributions for these variables in the MC closely reproduce those in the data.  

 In Ref. \cite{paper2}, we take the resulting SHERPA MC and put it through the full ALEPH detector simulation and compare it with the archived ALEPH LEP1 and LEP2 datasets.  This latter comparison is necessary in order to evaluate the MC for use in experimental analyses, and it allows us to look at arbitrary observables, including those more directly related to the observed excess.  Full details of the detector-level data-MC comparison are available in Ref. \cite{paper2}; our discussion here will be limited to results obtained with Rivet.

Our tuning of the SHERPA MC proceeds roughly as follows.  A large number (approximately $10^2-10^3$) of sets of values for the generator parameters are chosen.  For each set of values, a sample of MC is generated using SHERPA.  For each generated sample, Rivet produces distributions for its built-in observables, which are then compared to the same distributions for data.  Using the Professor package v. 1.3.3 \cite{Buckley:2009bj}, a fit is performed to determine values for the SHERPA generator parameters which produce good agreement with data.

We perform two tunes of the SHERPA MC, which we will refer to as the ``LO'' and ``NLO'' tunes.  In the case of the LO tune, we generate $e^+e^-\rightarrow\mbox{hadrons}$ events at $\sqrt{s}=91.25\mbox{ GeV}$  using the LO MEs for final states with up to six partons. For the NLO tune, we generate events using MEs for final states with up to six partons, but MEs for states with two, three or four final-state partons are generated at NLO, while those for final states with five or six partons are generated at LO.  For the LO tune we used SHERPA v. 2.2.0, while for the NLO tune we used a slightly modified version of SHERPA v. 2.0.beta.\footnote{We found that only with SHERPA version 2.0.beta were we able to reproduce the results in \cite{Gehrmann:2012yg}, in particular relating to the stability of the NLO results under renormalization scale variations.  However, using v. 2.0.beta results in a loss of the (default) option in CSSHOWER \cite{Schumann:2007mg} of SHERPA versions $\ge$2.0.0 to use modified transverse momentum ordering (CSS\_EVOLUTION\_SCHEME=1). Our version of 2.0.beta thus uses modified transverse momentum ordering as implemented in v. 2.0.0.}  We utilized BlackHat v. 0.9.9 \cite{Berger:2008sj} for higher-order corrections.  In both the LO and NLO tunes, we use PYTHIA v. 6.4.18 \cite{Sjostrand:2006za} for fragmentation. 

The resulting MC samples are then compared to ALEPH LEP1 data using Rivet distributions based on the results in \cite{Barate:1996fi,Heister:2003aj}; these analyses include distributions of numerous event-shape and flavor variables.  We provide details of our tuning procedure and report the resulting values of the tune parameters.  We then generate samples using these parameter values and use Rivet to compare both the LO and NLO MC with LEP1 and LEP2 data.  We also compare the LO and NLO MC with MC generated with KK2f v. 4.19 \cite{Jadach:2000ir,Jadach:1999vf,Jadach:1998jb}, generated with PYTHIA 6.156 \cite{Sjostrand:2000wi} using the standard ALEPH tune.\footnote{For simplicity, we will just refer to this generation as being done with KK2f below.}    We provide plots of many such distributions in the text.

We find a large improvement in our LO tuned SHERPA MC over KK2f for distributions closely related to four-jet final states, while the LO MC and KK2f are comparable for more general event-shape variables.  While the NLO sample also showed improvement over KK2f for four-jet variables, we were unable to obtain NLO results which were competitive with the LO generation.  Differences between SHERPA v. 2.2.0 and SHERPA v. 2.0.beta, the inclusion of the $b$ quark mass in the ME and PS for the LO tune but not for the NLO tune, and differences in the choice of merging scale may complicate comparisons between our LO and NLO tunes, however.  In our subsequent analysis of the four-jet excess, we will use the MC resulting from our LO tune, and will use the NLO tune MC and KK2f for systematic studies.  We may follow up in the future with studies of NLO tuning utilizing later versions of SHERPA.

Our main purpose in tuning and generating this MC is to produce a reliable sample of the QCD background with which to analyse the excess seen in Ref. \cite{paper3}, with a particular emphasis on the correct modelling of four-jet events.  Thus, the tune we present here is a necessary step in producing a MC sample which can be compared directly to data at detector level.  However, our results here are useful for other purposes as well.  In particular, it demonstrates the utility of MC generators with multi-parton MEs and proper merging and matching for other LEP analyses and analyses at future lepton colliders; this is of particular interest for the numerous analyses which consider four-jet final states.  To facilitate future use by the LEP collaborations, we provide our tuning parameters.  We do not necessarily suggest that other experiments use our tuning parameters ``off-the-shelf'' in their MC generation; rather, we hope that our results will provide a baseline for what behavior is expected from the LO and NLO MC at full simulation.  We also hope that our results may provide some useful feedback to the SHERPA authors or to people doing future LEP QCD studies.  Lastly, we hope our work highlights the value in applying new simulation technologies, not available at the time of data taking, to pre-existing datasets, including, but not limited to, those from LEP.

The remaining sections of this paper are organized as follows.  In Section \ref{procedure}, we describe the tuning procedure and give the results of the LO and NLO tunes.  We compare data and MC using Rivet at both LEP1 and LEP2 energies and compare the data-MC agreement of the LO and NLO generations to that of KK2f in Section \ref{sec:rivet}.  Scale variations are discussed in Section \ref{syst}, and we conclude in Section \ref{conc}.  Our Professor weight files are provided in the Appendix.

\section{Tuning Procedure}
\label{procedure}

In this section, we describe how we obtain two sets of tune parameters for generation of the process $e^+e^-\rightarrow \mbox{hadrons}$.  The first of these, which we will call the ``LO'' tune, is for the case where the process $e^+e^-\rightarrow \mbox{hadrons}$ is generated using the LO MEs for final states of up to six partons.  The second, ``NLO'' tune produces parameters for generation of MC using MEs for up to six final-state partons, with MEs for two-, three-, and four-parton final-states generated at NLO, and MEs for five- and six-parton final states at LO.  For both tunes, we generate samples of weighted events with SHERPA using PYTHIA for fragmentation.  

We tune a total of 19 generator parameters.  As this is a large number of parameters to tune simultaneously, the tunes are accomplished via a two-step procedure.  In the first step, which we refer to as the ``flavor'' tune, we obtain values for the parameters which are directly related to the flavors of individual hadrons in the jets; these parameters are all inputs to PYTHIA.  Experimental distributions in Rivet which are sensitive to these parameters, such as multiplicities for specific hadrons, are included in this stage of the tune, along with a subset of the event-shape variables discussed below.  The values of the PYTHIA flavor parameters obtained in this stage are then held fixed for the remainder of the tuning procedure.  The second stage of the tune is what we call the ``event-shape'' tune.  The Rivet variables used in this phase of the tune include observables such as thrust, jet rates, and hemisphere masses as well as inclusive charged particle distributions; these variables are more directly relevant for our purposes than the flavor variables above.  We do separate event-shape tunes for the cases of LO and NLO MC generation.  The SHERPA parameters tuned in this procedure are the strong coupling constant $\alpha_s$ and parameters relevant to the parton shower and Lund string fragmentation.  For all phases of the tune, the relevant SHERPA parameters will be specified below, and the Rivet variables and their respective weights are given in the appendices.

All tuning parameters are defined in SHERPA run.dat files.  Additionally, there are five SHERPA parameters, shown in Table \ref{tab:fragparam}, set to (non-default) values kept constant throughout the tuning.  CSS\_EW\_MODE=1 allows quark splittings to photons in the parton shower, and PDF\_SET=None turns off ISR.   The remaining three parameters relate to fragmentation and are similar to those used in previous tunes by ALEPH \cite{Barate:1996fi}.  The $b$ quark mass and QCUT, which defines the SHERPA merging scale (defined in \cite{Gehrmann:2012yg}), are set to values which differ for the LO and NLO tunes; see below.
\begin{table}
\begin{tabular}{|c|c|c|}
\hline
Parameter & Parameter Description & Set Value\\
\hline
CSS\_EW\_MODE & photon splitting & 1\\
\hline
PDF\_SET &  PDF & None\\
\hline
MSTJ(11) & longitudinal fragmentation function & 3\\
\hline
PARJ(54) & $-1\times$ Peterson fragmentation parameter $\epsilon_c$ & -0.04\\
\hline
PARJ(55) & $-1\times$ Peterson fragmentation parameter $\epsilon_b$ & -0.0025\\
\hline
\end{tabular}
\caption{Fixed SHERPA parameters and their values used for all tunes.}
\label{tab:fragparam}
\end{table}

\subsection{Tune of Flavor Parameters}

The tune of flavor parameters was accomplished via an iterative process.  We take an initial set of flavor parameter values and do a tune on the event-shape parameters.  Then, setting the event-shape parameters to this set of approximate values, we then did a tune varying the flavor parameters.  We retain the resulting values for the flavor parameters for the rest of the tuning procedure, but the intermediate values of the event-shape parameters are abandoned.  Each of these steps was done at LO; we will limit our discussion here to the final stage of the flavor tune. 

We tune 13 flavor parameters.  All of these parameters are inputs to PYTHIA for the fragmentation step.  Values for the tune parameters were sampled randomly; approximately 1000 MC samples of 300,000 events each were generated using SHERPA v. 2.0.0.  These MC samples were generated at LO for up to five partons in the final state ME.  The $b$ quark mass was set to the default value $4.8$ GeV, and the QCUT which determines the merging scale was set to $\mbox{QCUT}^2=s\times 10^{-2.25}$.  

These samples were then compared to data using Rivet.  Most of the distributions used were from the Rivet analysis ALEPH\_1996\_S3486095, which contains data from LEP1 \cite{Barate:1996fi}.  Additionally, four variables were used from the analysis ALEPH\_2004\_S5765862; this analysis \cite{Heister:2003aj} contains both LEP1 and LEP2 distributions, but those used were only from LEP1.  The weights for this analysis are given in Appendix \ref{flavtundet}.  The tune ranges and final values of the flavor parameters output from Professor are shown in Table \ref{tab:flavtun}.  We set the flavor parameters to these output values for both tunes described below, which are over the event-shape parameters only.

\begin{table}
\begin{tabular}{|c| c| c| c|}
\hline
Parameter & Parameter Description & Tune Range & Tune Value \\
\hline
PARJ(1) & diquark suppression & 0.01-0.21 & 0.1290\\
PARJ(2) & strange quark suppression & 0.01-0.61 & 0.2220\\
PARJ(3) & strange diquark suppression & 0.01-1.01 & 0.8231\\
PARJ(4) & spin-1 diquark suppression & 0.01-0.21 & 0.0746\\
PARJ(11) & light meson $s=1$ probability & 0.01-1.01 & 0.4243\\
PARJ(12) & strange meson $s=1$ probability & 0.01-1.01 & 0.3755\\
PARJ(13) & charm meson $s=1$ probability & 0.01-1.01 & 0.3497\\
PARJ(14) & $l=1$ probability for $s=0$ & 0.0-1.0 & 0.1501\\
PARJ(15) & $l=1,j=0$ probability for $s=1$ & 0.0-1.0 & 0.\\
PARJ(16) & $l=1,j=1$ probability for $s=1$ & 0.0-1.0 & 0.1983\\
PARJ(17) & $l=1,j=2$ probability for $s=1$ & 0.0-0.4 & 0.\\
PARJ(19) & baryon suppression factor & 0.01-1.01 & 0.6869\\
PARJ(26) & $\eta'$ suppression factor & 0.0-0.5 & 0.1806\\
\hline
\end{tabular}
\caption{Flavor parameter results from a LO tune keeping event-shape variables fixed.  The resulting values for these parameters were then used as input for both the LO and NLO tunes.}
\label{tab:flavtun}
\end{table}

\subsection{LO tune}
\label{sec:lotune}

The event-shape tunes determine the values of six parameters in the SHERPA generator; these parameters, their ranges, and their final values output from Professor for the LO tune are shown in Table \ref{tab:lotun}.  $\alpha_s(M_Z)$ is the strong coupling constant evaluated at the $Z$ peak.  The parameter CSS\_FS\_PT2MIN controls the cutoff scale between the parton shower and fragmentation, while CSS\_FS\_AS\_FAC is a scale factor for the evaluation of the strong coupling constant in the parton shower.  PARJ(21), PARJ(41), and PARJ(42) are PYTHIA Lund string fragmentation parameters.  Values for five of these parameters were sampled randomly within their respective ranges.  The values for $\alpha_s(M_Z)$, however, were constrained to discrete points as different values of $\alpha_s(M_Z)$ require separate, time-consuming integrations.\footnote{For both the LO and NLO tunes, the value of $\alpha_s(M_Z)$ output by the tune was close to one of the discrete input values.  In each case, subsequent tunings were then done where the value of $\alpha_s(M_Z)$ alternatively was fixed to that discrete value or allowed to float.  In both cases, we concluded that fixing $\alpha_s(M_Z)$ to that discrete value was sufficient.}  

Approximately 500 MC samples, each containing one million weighted events, were generated with SHERPA v. 2.2.0\footnote{It may be noted that this is a different version of SHERPA than that used for the flavor tune.  Performing a tune is a very time-consuming process; during the time this work was being done, newer versions of SHERPA were released.  We do not expect the difference in version to significantly affect our results.} using LO MEs for final states with up to six partons.  The $b$ quark mass was set to the default of $4.8$ GeV.  As we want the region explored in Ref. \cite{paper3} to be simulated using the ME, the QCUT was set to $\mbox{QCUT}^2=s\times 10^{-3}$.

LEP1 data contained in Rivet analyses ALEPH\_1996\_S3486095 and ALEPH\_2004\_S5765862 were used for the tuning; a weight file is given in Appendix \ref{lotundet}.  Arriving at these weights was a somewhat iterative processs; particular attention was paid to correctly reproducing the jet rates.  We used Professor's quartic interpolation for the tuning procedure.

\begin{table}
\begin{tabular}{|c| c| c| c|}
\hline
Parameter & Parameter Description & Tune Range & Tune Value \\
\hline
$\alpha_s(M_Z)$ & strong coupling constant at $M_Z$ & \makecell{0.1100, 0.1150,\\ 0.1180, 0.1200,\\ 0.1220, 0.1250, 0.1300} &  0.1180\\
\hline
CSS\_FS\_PT2MIN & lower shower evolution cutoff/$\mbox{GeV}^2$  & 0.1-2.5  & 1.849653\\
\hline
CSS\_FS\_AS\_FAC & scale factor for evaluation of $\alpha_s$ & 0.1-2.5 & 0.6367039\\
\hline
PARJ(21)& \makecell{primary hadron transverse\\momentum width $\sigma$/GeV} & 0.01-1.01 & 0.3840116\\
\hline
PARJ(41)& Lund fragmentation parameter $a$ & 0.01-2.01 & 1.312967\\
\hline
PARJ(42)& Lund fragmentation parameter $b/\mbox{GeV}^{-2}$  & 0.01-2.01 & 1.584376 \\
\hline
\end{tabular}
\caption{Event-shape tune parameter results for LO tune.}
\label{tab:lotun}
\end{table}


\subsection{NLO tune}

Here, we tune the same six parameters as we did for the LO tune.  The parameter ranges and final tune values are given in Table \ref{tab:nlotun}.  As in the LO tune, we sampled $\alpha_s(M_Z)$ at discrete points; for the NLO tune, the spacing between these points was $0.001$.  All other parameters were sampled randomly within their specified ranges.  We note that the parameter ranges shown for the NLO tune in Table \ref{tab:nlotun} are somewhat tighter than those for the LO tune; in initial NLO tune explorations, we found that the output SHERPA parameters typically stayed within these reduced ranges, and we thus restricted the range for the tune parameters accordingly.  

We used a modified version of SHERPA v. 2.0.beta to generate 91 MC samples, each containing $1.5\times 10^5$ partially-unweighted events.  The events were generated using NLO MEs for final states with up to four partons using BlackHat and with LO MEs for final states of five or six partons.  The QCUT was set to a constant $6$ GeV to accommodate the region of interest in Ref. \cite{paper3}.  The $b$ quark was left massless in the matrix element and parton shower, as BlackHat v. 0.9.9 is restricted to the case of massless quarks.

As for the LO tune, Rivet analyses ALEPH\_1996\_S3486095 and ALEPH\_2004\_S5765862 were used to compare the generated MC with LEP1 data.  Due to greater statistical uncertainty in comparison with the LO samples\footnote{The NLO samples, unlike the LO samples, have events with negative weights.}, quadratic interpolation was used in Professor.  A weight file can be found in Appendix \ref{nlotundet}.  In performing this tune, special attention was given to the thrust variable, which we found to be correlated with the mass sum variable important in Ref. \cite{paper3}.  

We note that the variables $p_{\perp in}$, $p_{\perp out}$ and oblateness are included in the LO weight file, but absent from the NLO weight file.  We initially included these variables in our NLO tune, but had difficulty achieving data-MC agreement for these variables without degrading the thrust distribution.  Additionally, we saw indications that agreement or disagreement between data and MC was correlated between these three variables.   As we prioritized being able to correctly reproduce the data thrust distribution, and as there is a long history \cite{Barate:1996fi,Heister:2003aj} of disagreement between data and MC for $p_{tout}$ at LEP, we decided to omit these variables from the NLO tune.  Such difficulties were smaller in the LO tune.  


\begin{table}
\begin{tabular}{|c| c| c| c|}
\hline
Parameter & Parameter Description & Tune Range & Tune Value \\
\hline
$\alpha_s(M_Z)$ & strong coupling constant at $M_Z$ & $0.115-0.121$  & $0.120$ \\
\hline
CSS\_FS\_PT2MIN & lower shower evolution cutoff/$\mbox{GeV}^2$  & $0.4-1.6$ & $0.9500309$\\
\hline
CSS\_FS\_AS\_FAC & scale factor for evaluation of $\alpha_s$ & $1.0-2.0$ & $1.814162$\\
\hline
PARJ(21)& primary hadron transverse momentum width $\sigma$/GeV & $0.34-0.42$ & $0.3891697$\\
\hline
PARJ(41)& Lund fragmentation parameter $a$ & $0.4-1.6$ & $1.239345$\\
\hline
PARJ(42)& Lund fragmentation parameter $b/\mbox{GeV}^{-2}$  & $0.4-1.6$ & $1.211087$\\
\hline
\end{tabular}
\caption{Ranges and final values of the parameters for the NLO tune.  All parameters except $\alpha_s(M_Z)$ are randomly sampled within their specified ranges.  $\alpha_s(M_Z)$ is sampled in its range at intervals of $0.001$.}
\label{tab:nlotun}
\end{table}


\section{Rivet Data-MC comparisons}
\label{sec:rivet}

We now compare our tuned MC to LEP data using Rivet.  Here, we concentrate on event-shape and four-jet variables, as these are more important than flavor observables for studying the excess observed in Ref. \cite{paper3}.\footnote{We find the overall behavior of the LO and NLO tunes with respect the flavor variables to be similar to that from KK2f.}  We generate MC events using the SHERPA LO and NLO tunes as well as with KK2f.  We compare them with unfolded data from the analyses used in the tuning, ALEPH\_1996\_S3486095 and ALEPH\_2004\_S5765862, as well as analyses from DELPHI and OPAL, DELPHI\_1996\_S3430090  \cite{Abreu:1996na} and OPAL\_2001\_S4553896 \cite{Abbiendi:2001qn} and OPAL\_2004\_S6132243 \cite{Abbiendi:2004qz}.  Distributions from LEP1 and from LEP2 will be studied separately below.

\subsection{LEP1 Event-Shape Observables}
\label{evtshp}

Here, we compare data and MC for event-shape variables commonly used in QCD analyses.  Definitions of many of these variables can be found in Refs. \cite{Abreu:1996na,Heister:2003aj}.  For each plot in this section, the upper pane gives the MC expectation for KK2f and the LO and NLO SHERPA samples, along with the Rivet data.  The bottom pane shows the MC/data, along with the yellow error band resulting from experimental and unfolding uncertainties. For these plots, we generated $2\times10^6$ unweighted events with SHERPA using the LO tune, $3\times10^6$ partially unweighted events with the NLO tune, and $10^6$ events using KK2f.  The three MC samples have been normalized to the number of events in data.

\begin{figure}[h]
\begin{center}
\subfigure[]{\includegraphics[width=3.0in,bb=0 0 330 330]{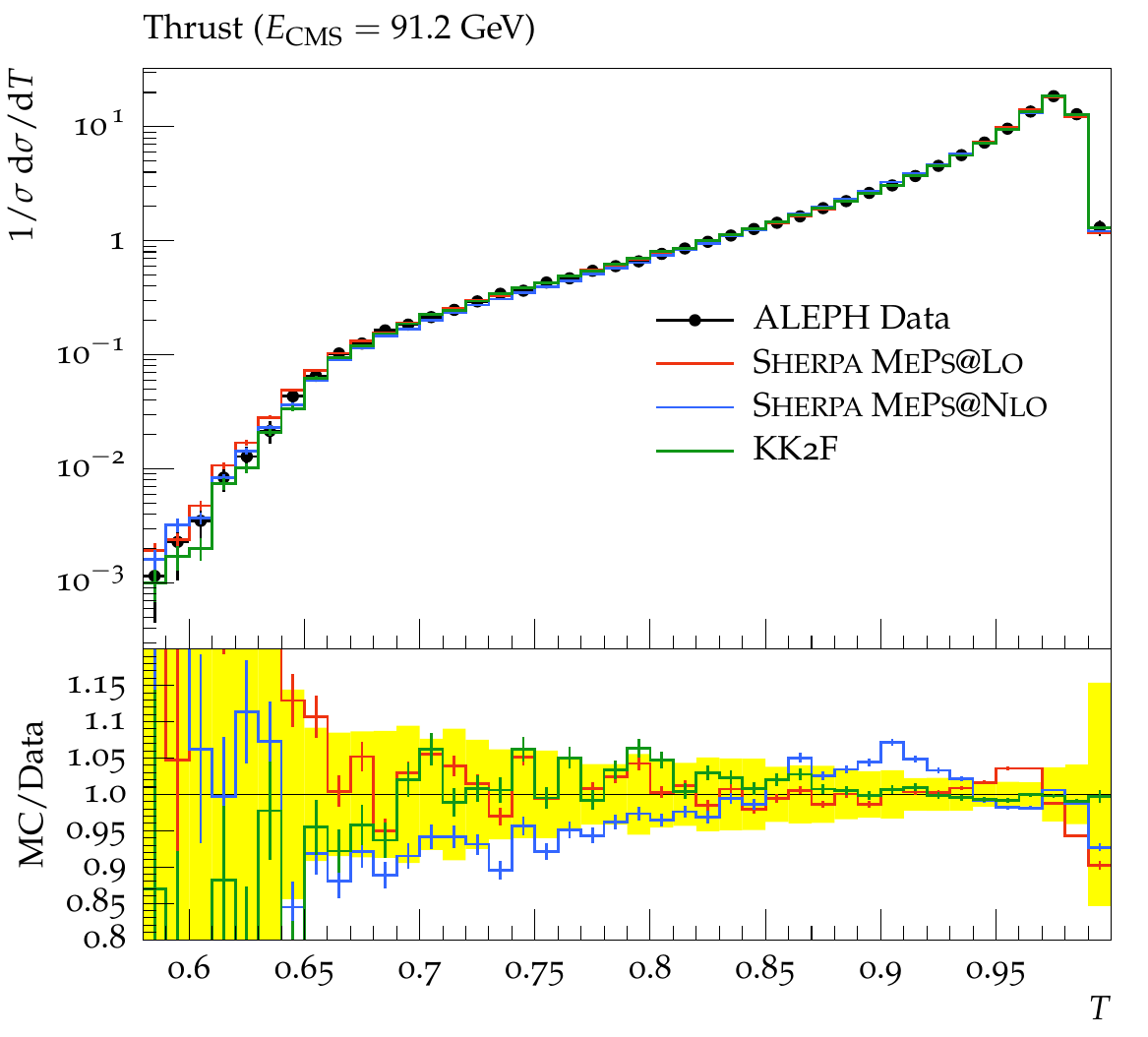}}\hspace{.4in}
\subfigure[]{\includegraphics[width=3.0in,bb=0 0 330 330]{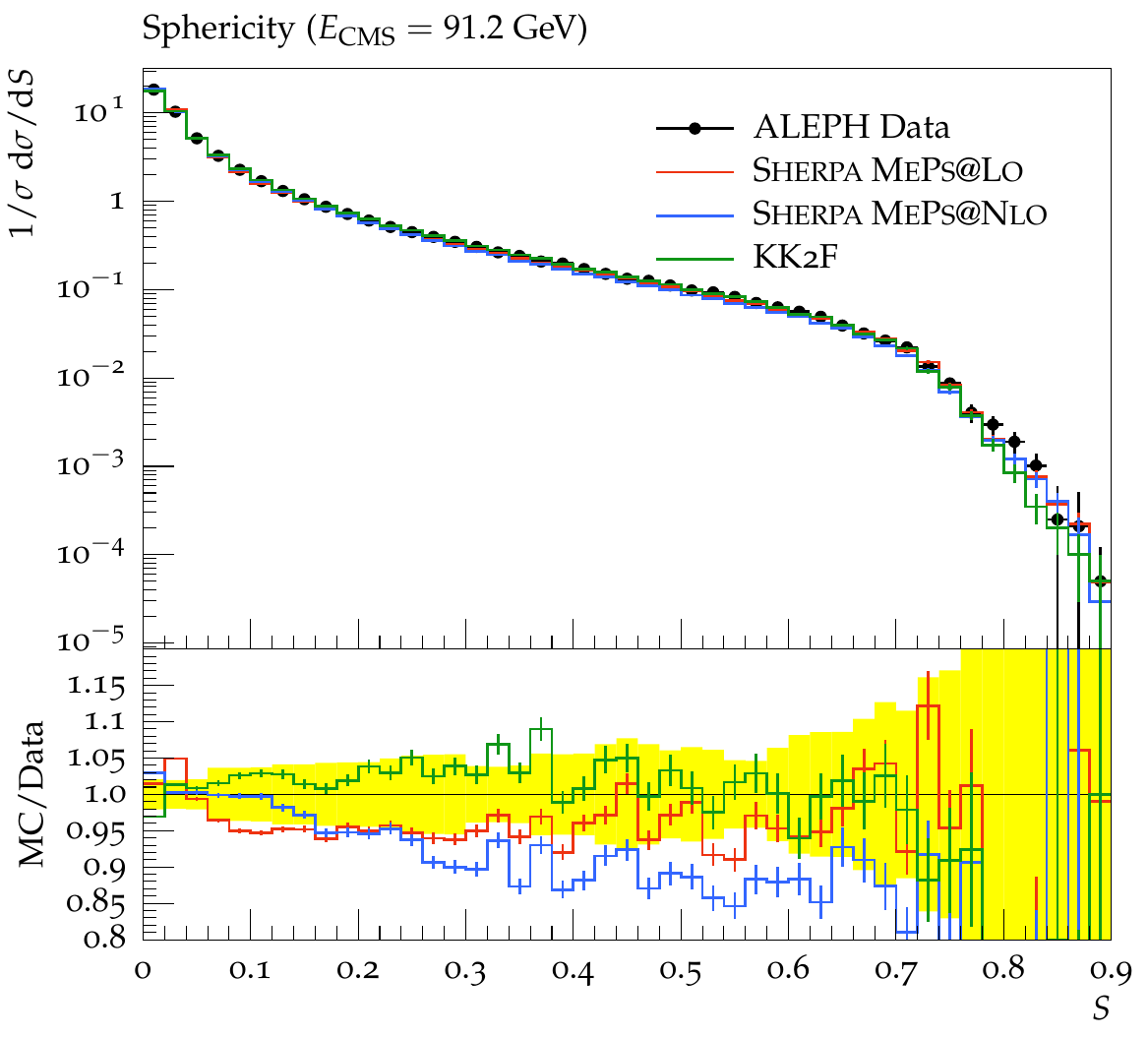}}\\
\subfigure[]{\includegraphics[width=3.0in,bb=0 0 330 330]{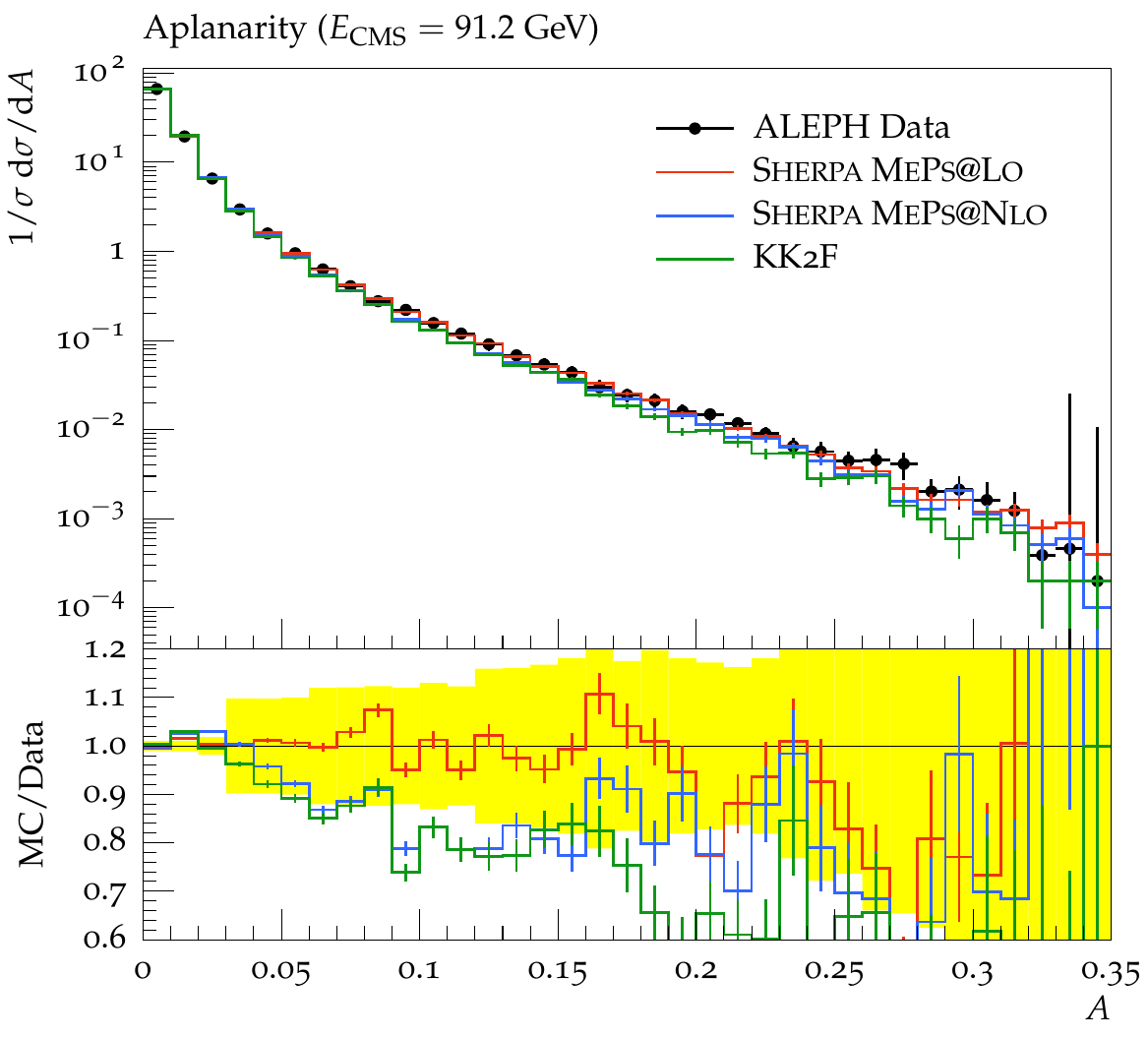}}\hspace{.4in}
\subfigure[]{\includegraphics[width=3.0in,bb=0 0 330 330]{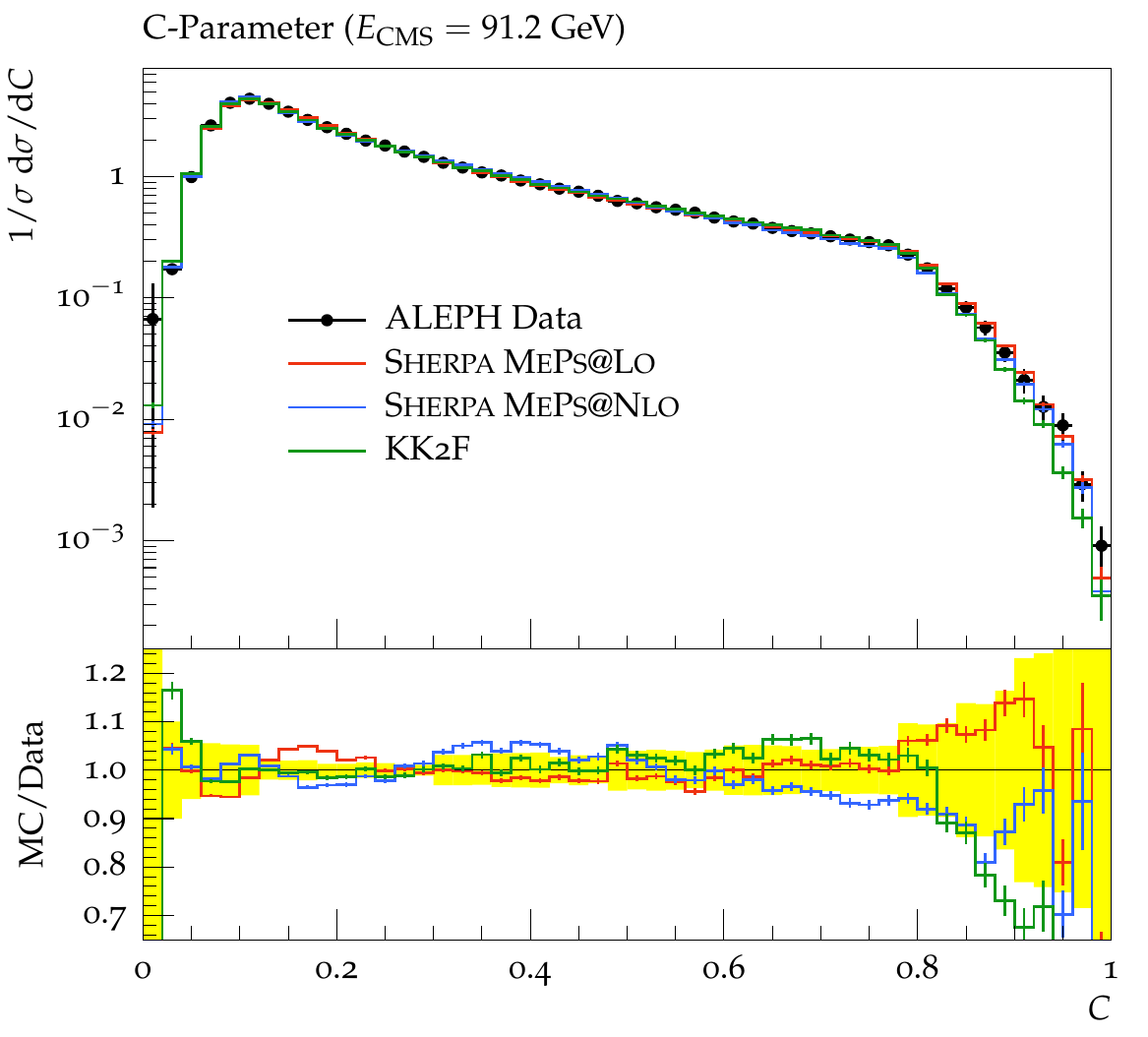}}\\
\end{center}
\caption{Plots of event-shape variables thrust $T$, sphericity $S$, aplanarity $A$, and the C-parameter at LEP1 from ALEPH.}
\label{fig:eventshp1}
\end{figure}

We plot four of these variables from ALEPH in Fig. \ref{fig:eventshp1}.  In the case of the thrust $T$ in Fig. \ref{fig:eventshp1} (a), we find that all three samples are largely within the error band.  In Fig. \ref{fig:eventshp1} (b), we find that the tail of the sphericity distribution is modelled best by the KK2f and LO SHERPA samples; the NLO SHERPA sample falls somewhat outside the yellow error band, but performs well in the peak region where most of the events are located.  The aplanarity, shown in Fig. \ref{fig:eventshp1} (c), is best modelled by the LO SHERPA MC, with the KK2f falling signficantly outside the error band away from the peak region; all three samples perform well near the peak.  The three samples perform comparably and lie mostly within the error band in the case of the C-parameter in Fig. \ref{fig:eventshp1} (d), although some deviation is seen in the LO SHERPA sample in the peak region.


\begin{figure}[h]
\begin{center}
\subfigure[]{\includegraphics[width=3.0in,bb=0 0 330 330]{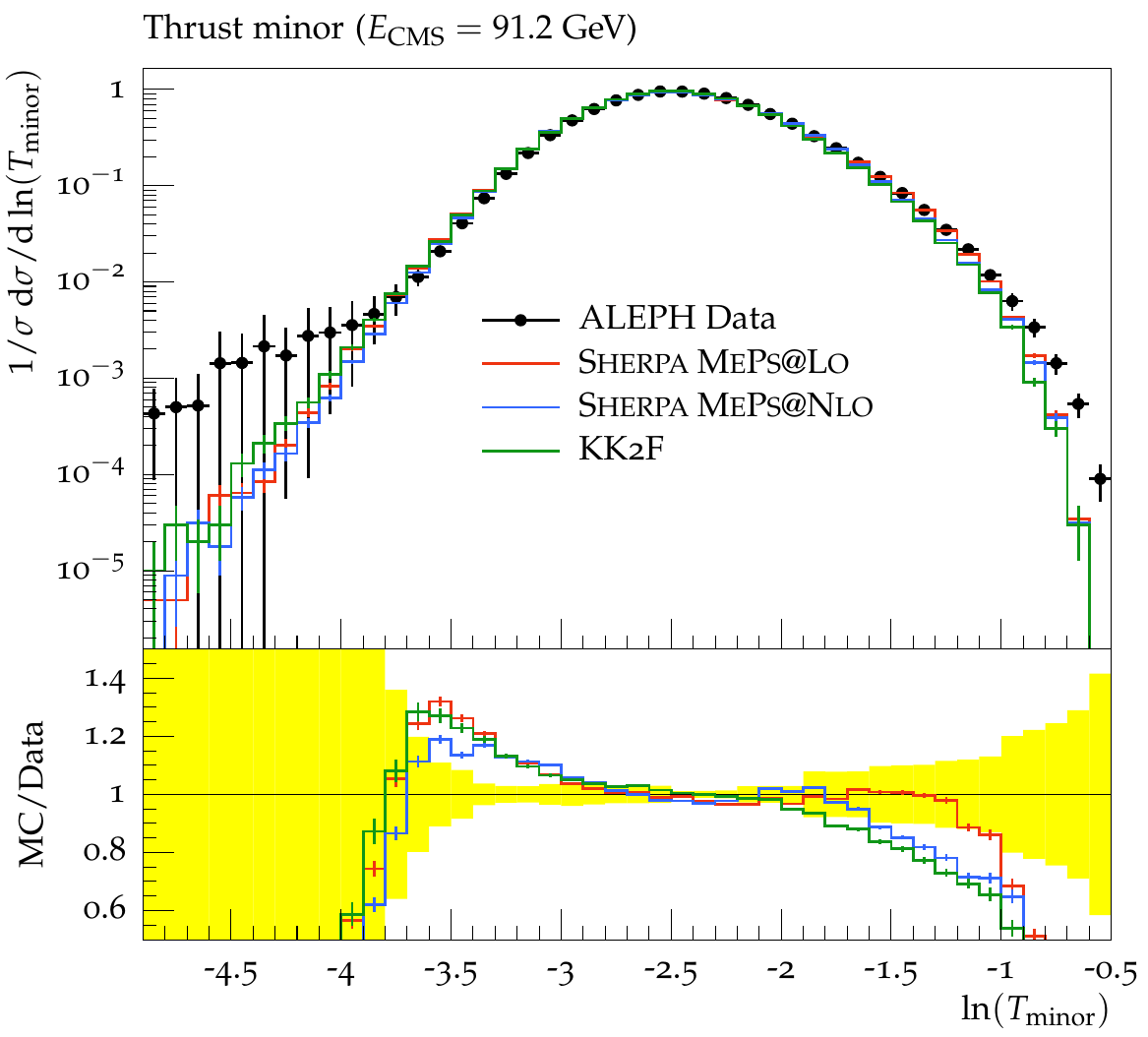}}\hspace{.4in}
\subfigure[]{\includegraphics[width=3.0in,bb=0 0 330 330]{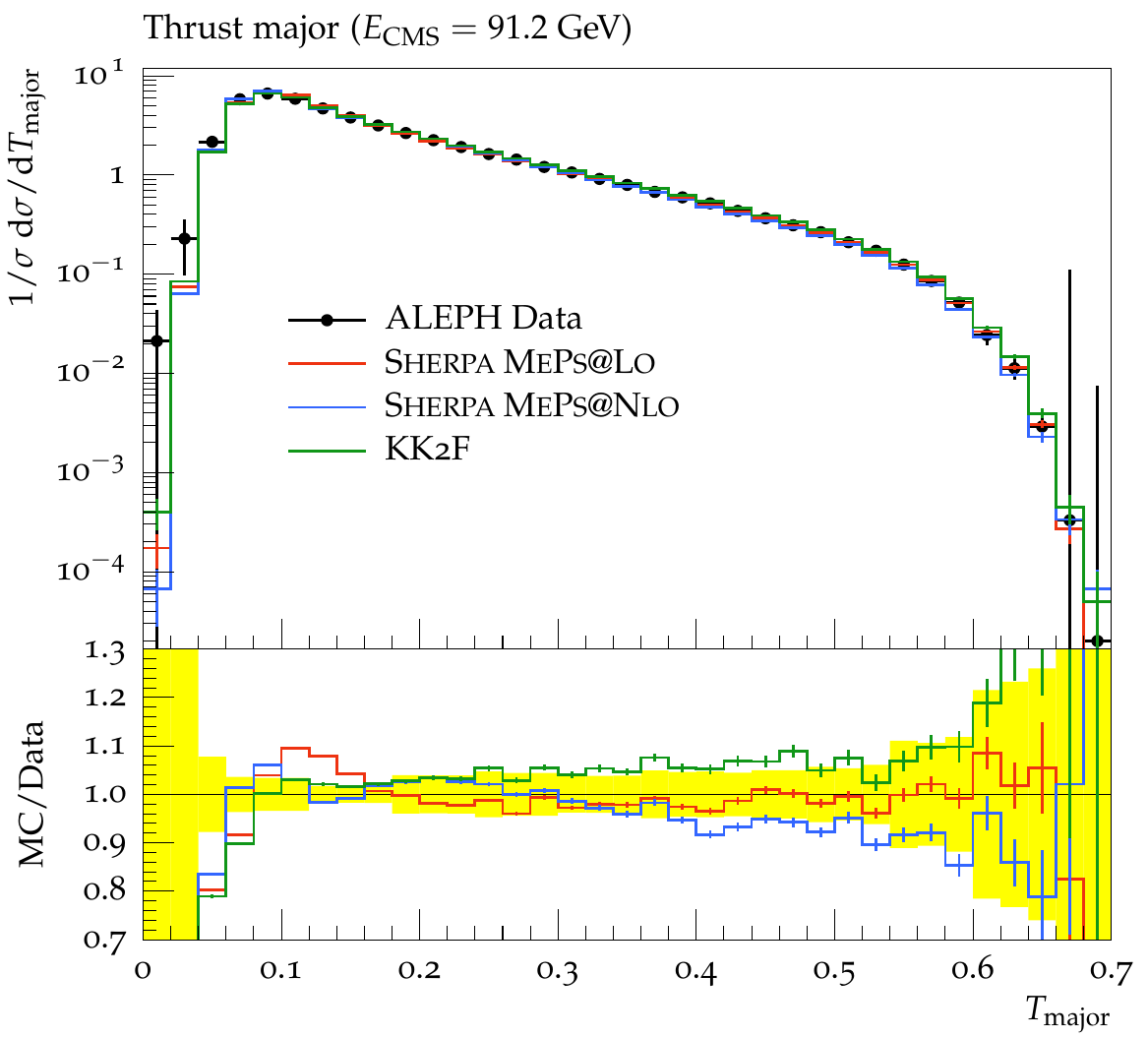}}\\
\subfigure[]{\includegraphics[width=3.0in,bb=0 0 330 330]{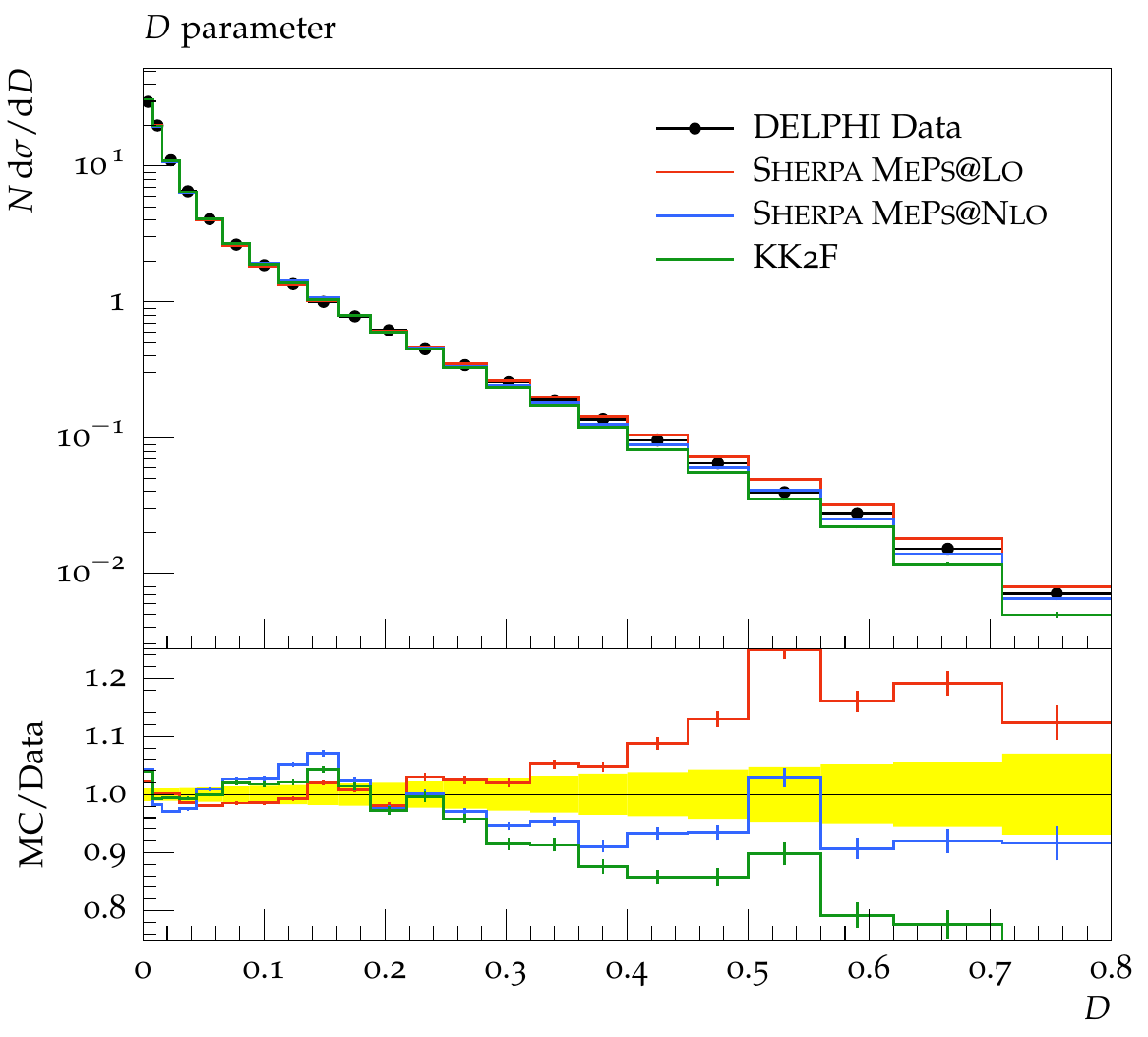}}\hspace{.4in}
\subfigure[]{\includegraphics[width=3.0in,bb=0 0 330 330]{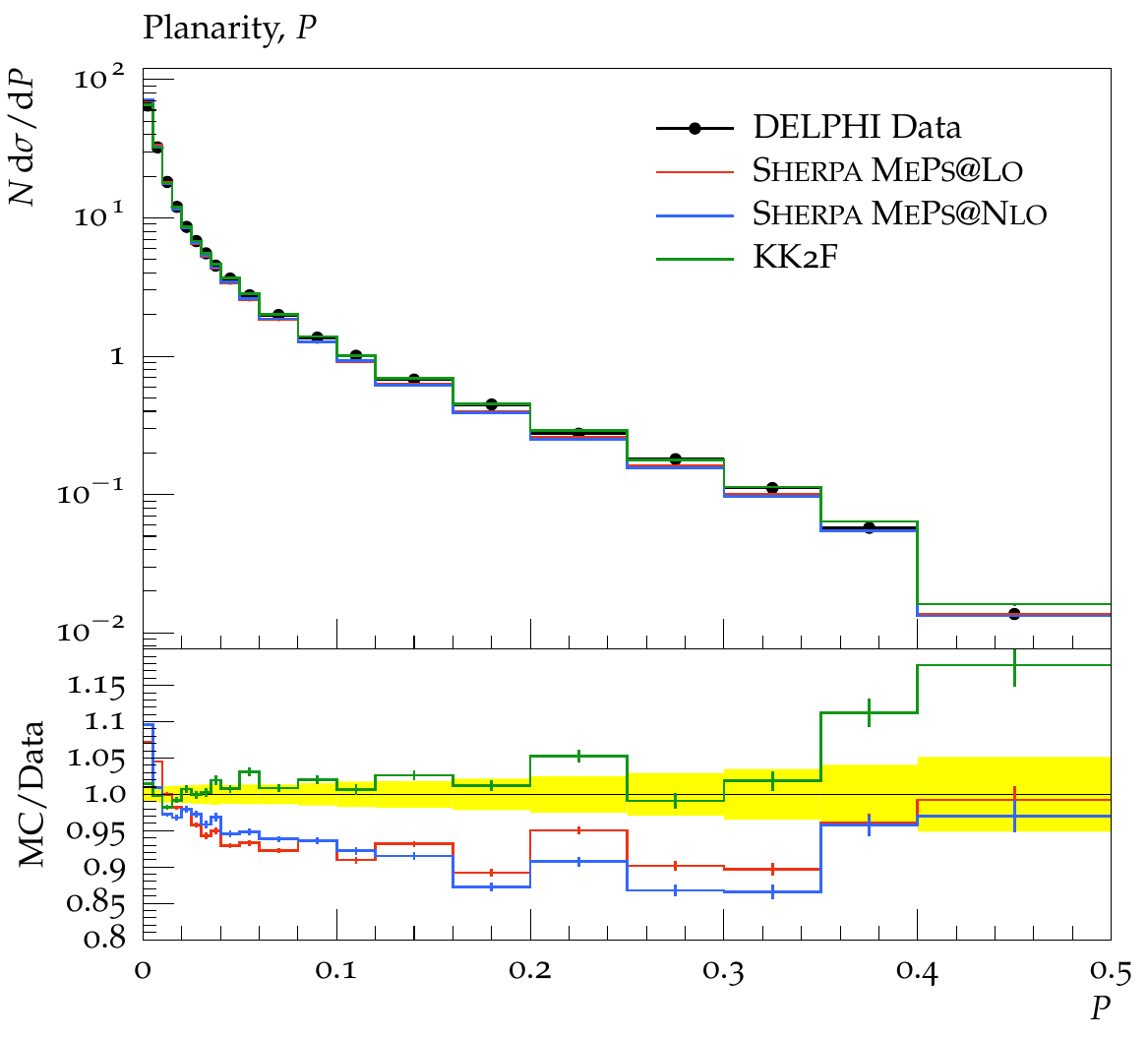}}
\end{center}
\caption{Plots of event-shape variables at LEP1.  (a) $\ln{(T_{minor})}$ and (b) $T_{major}$ from ALEPH; (c) $D$-parameter and (d) planarity $P$ from DELPHI.}
\label{fig:eventshp2}
\end{figure}

Fig. \ref{fig:eventshp2} displays two additional event-shape variables from ALEPH as well as two from DELPHI.  The $T_{minor}$ distribution in Fig. \ref{fig:eventshp2} (a) shows improvement in data-MC agreement in moving from KK2f to the two SHERPA samples.  For $T_{major}$ in Fig. \ref{fig:eventshp2} (b), the three samples are again comparable.  The peak region of the $D$ parameter in Fig. \ref{fig:eventshp2} (c) is described reasonably well by all three MCs, but only the SHERPA NLO MC does well in the tail of the distribution.  The planarity distribution in Fig. \ref{fig:eventshp2} (d) is described well only by the KK2f MC.

\begin{figure}[h]
\begin{center}
\subfigure[]{\includegraphics[width=3.0in,bb=0 0 330 330]{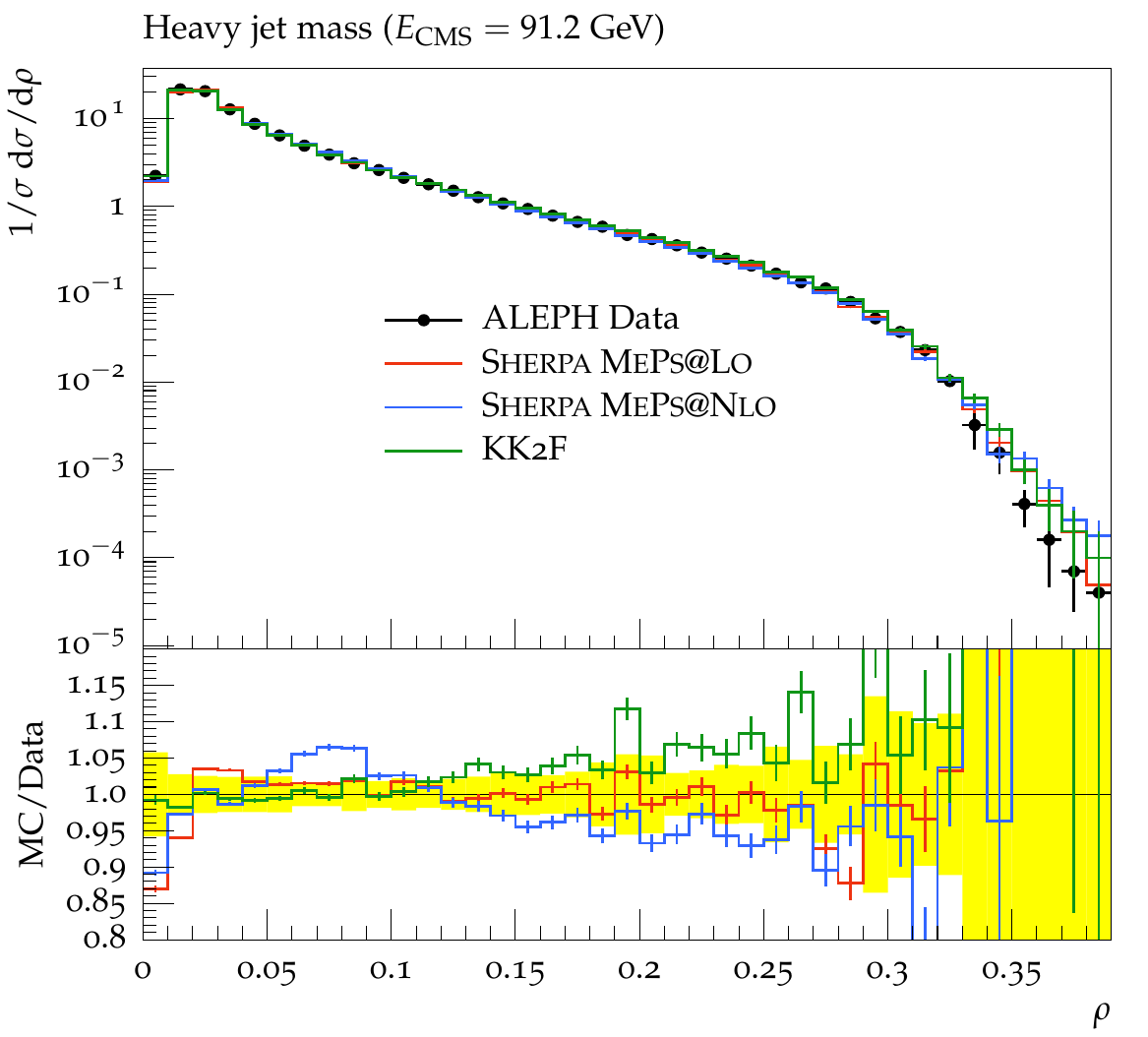}}\hspace{.4in}
\subfigure[]{\includegraphics[width=3.0in,bb=0 0 330 330]{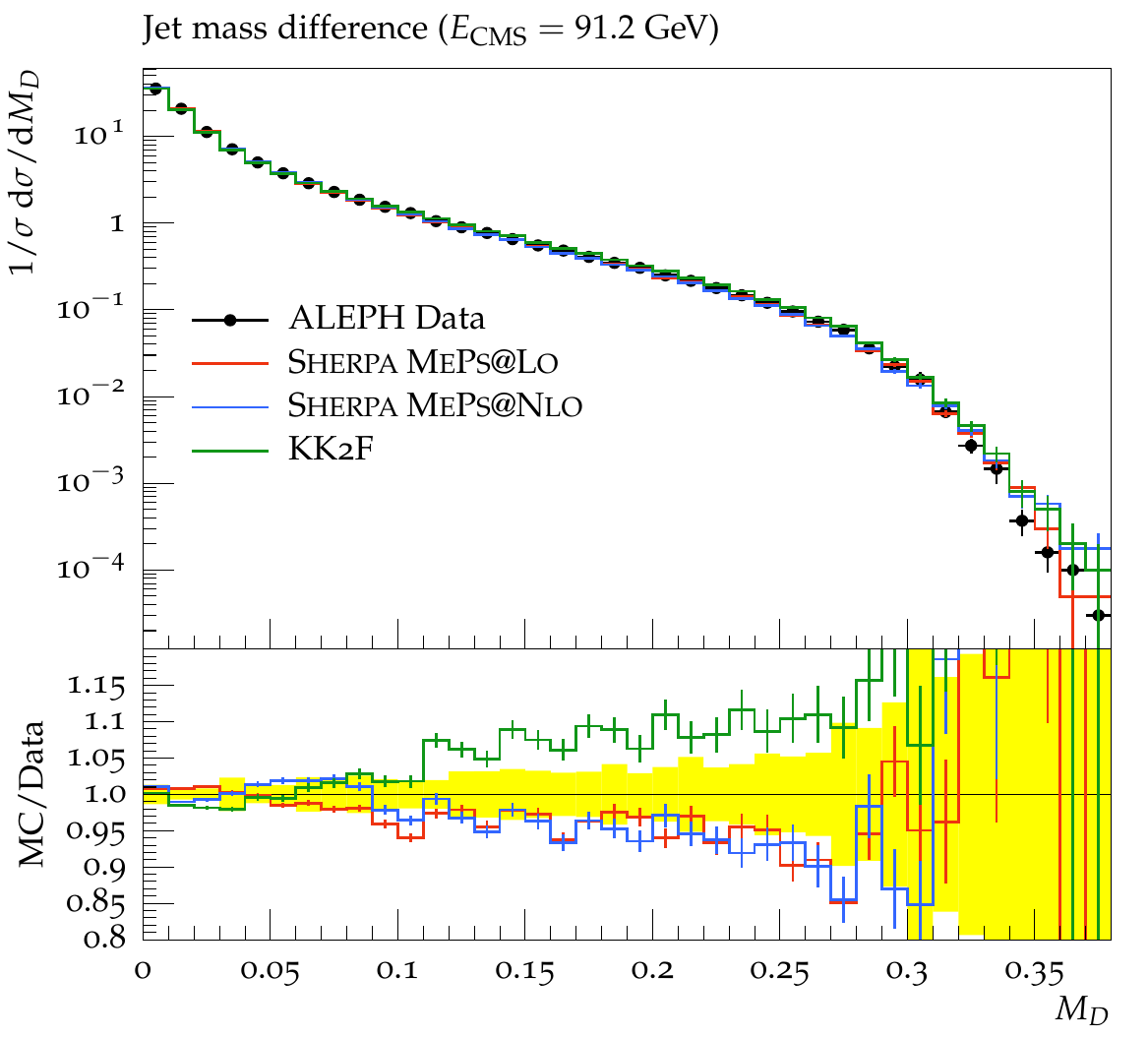}}\\
\subfigure[]{\includegraphics[width=3.0in,bb=0 0 330 330]{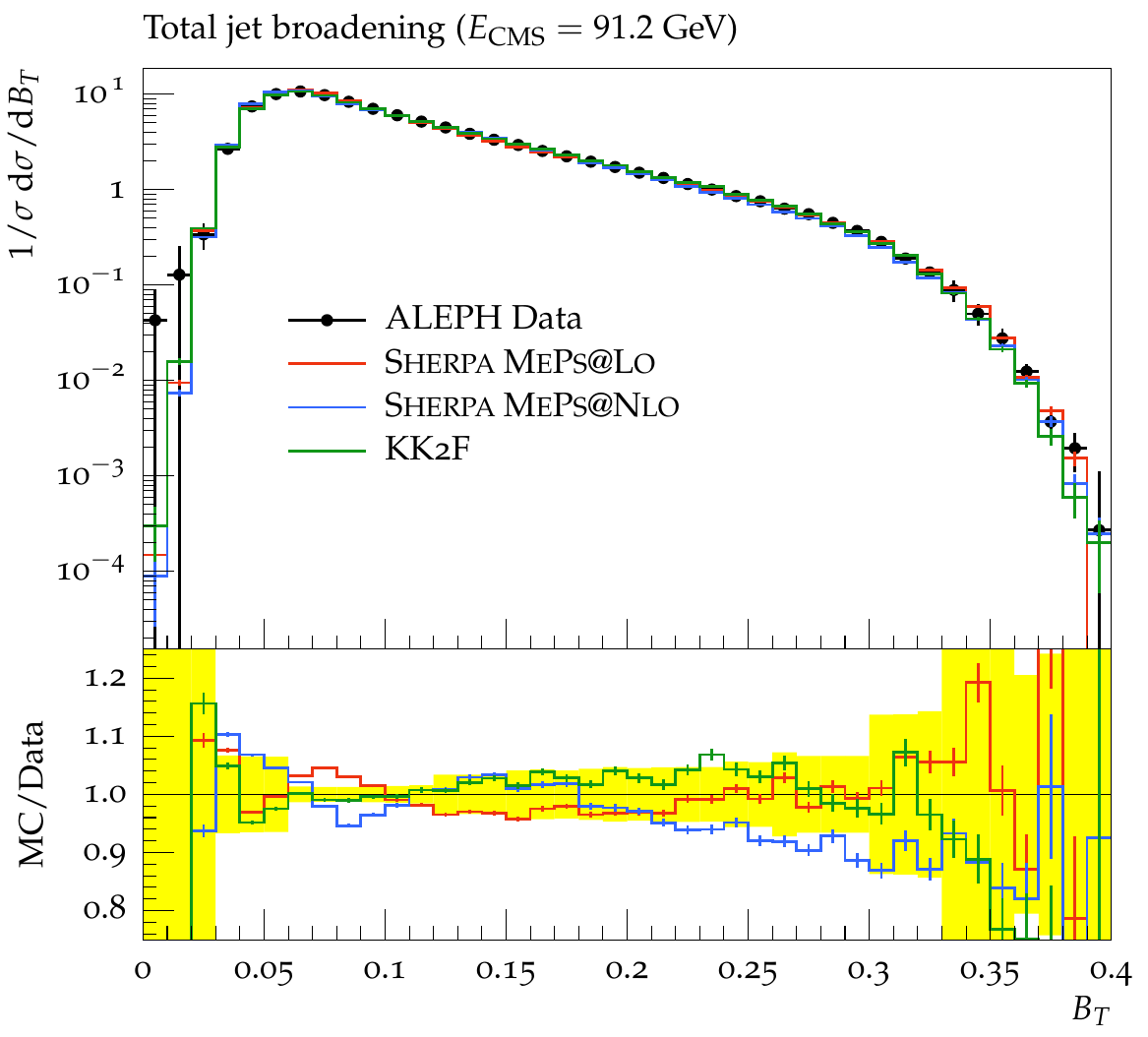}}\hspace{.4in}
\subfigure[]{\includegraphics[width=3.0in,bb=0 0 330 330]{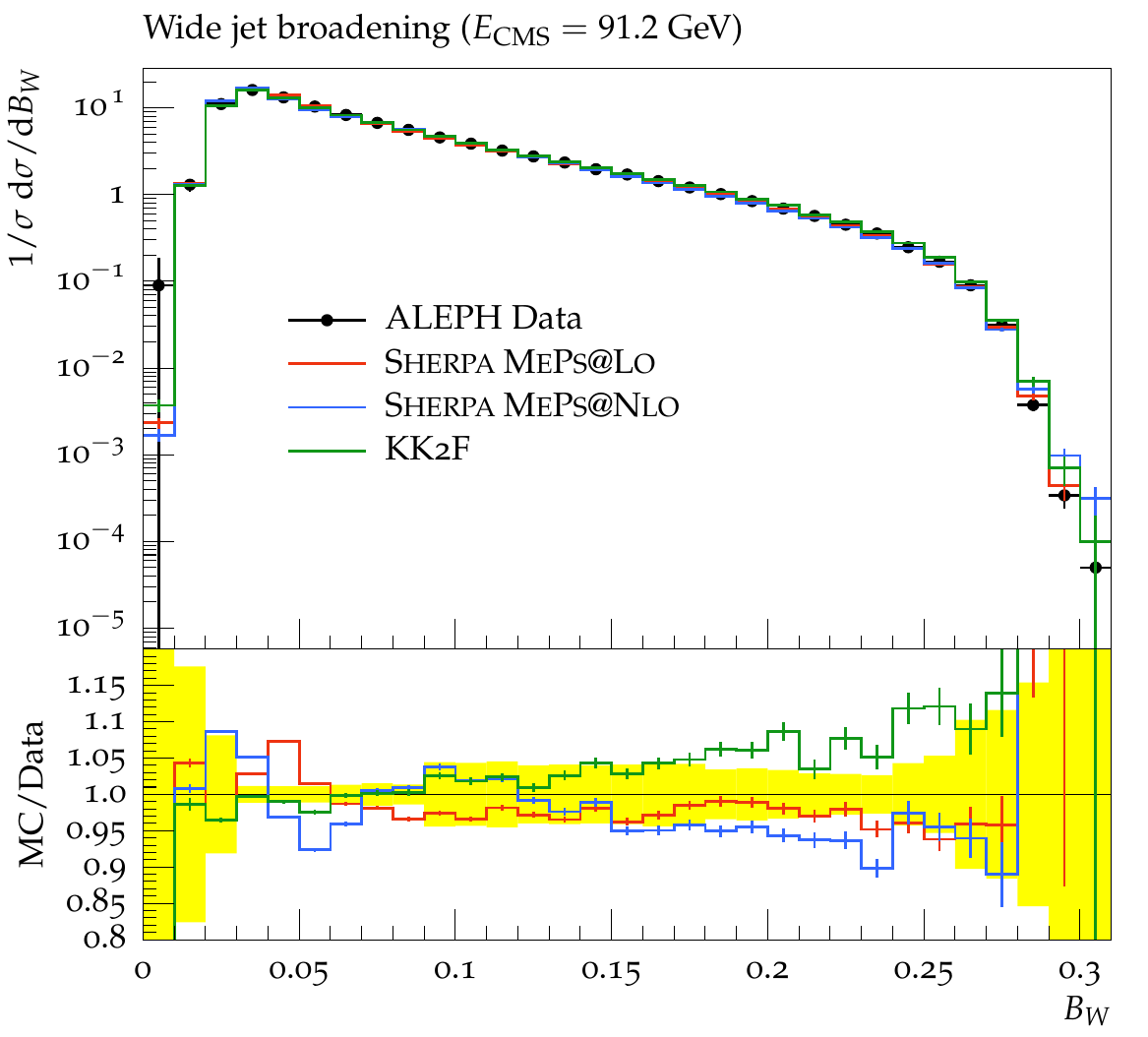}}\\
\end{center}
\caption{Plots of jet mass and broadening variables at LEP1 from ALEPH.}
\label{fig:eventshp3}
\end{figure}

\begin{figure}[h]
\begin{center}
\subfigure[]{\includegraphics[width=3.0in,bb=0 0 330 330]{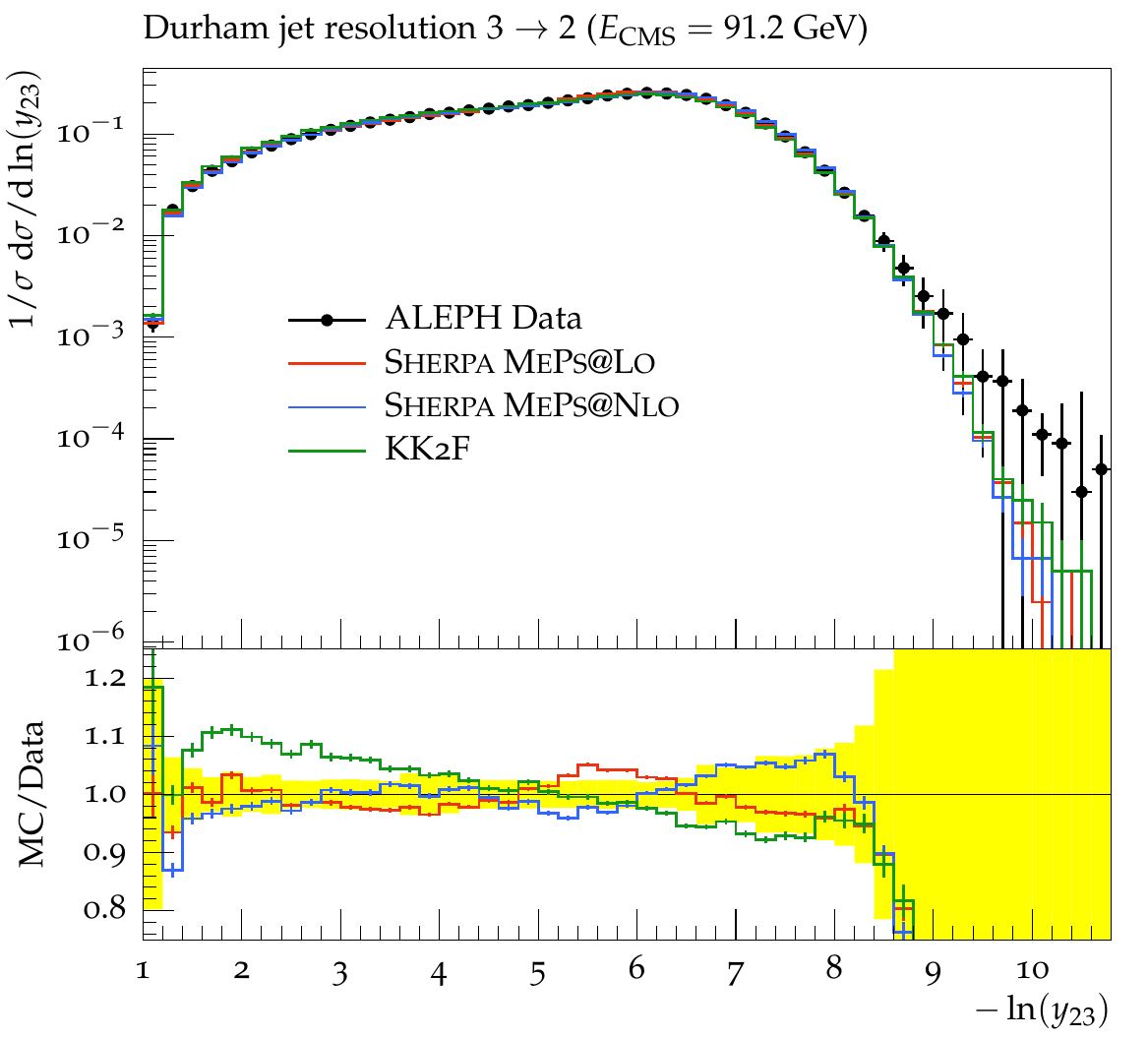}}\hspace{.4in}
\subfigure[]{\includegraphics[width=3.0in,bb=0 0 330 330]{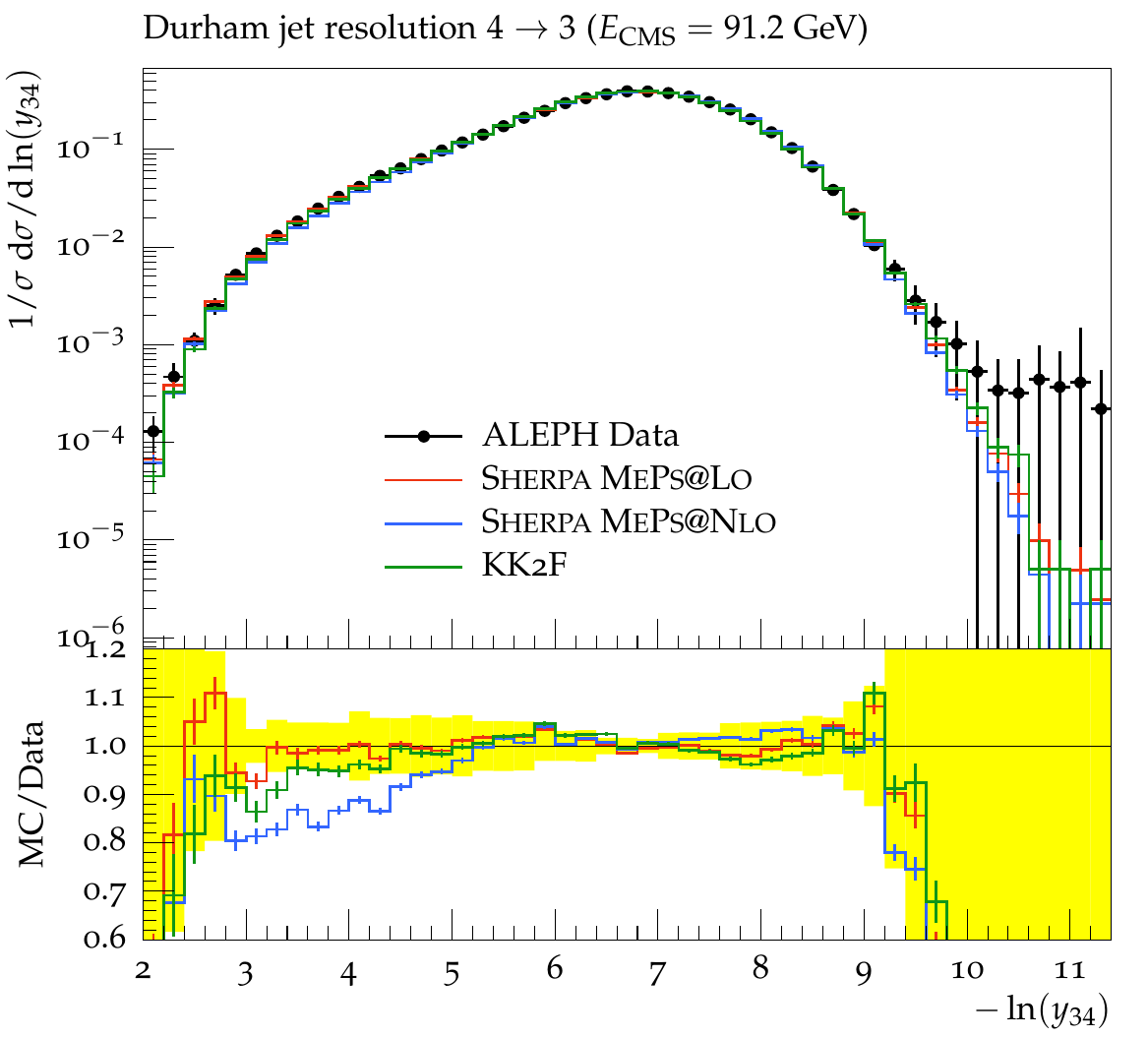}}\\
\subfigure[]{\includegraphics[width=3.0in,bb=0 0 330 330]{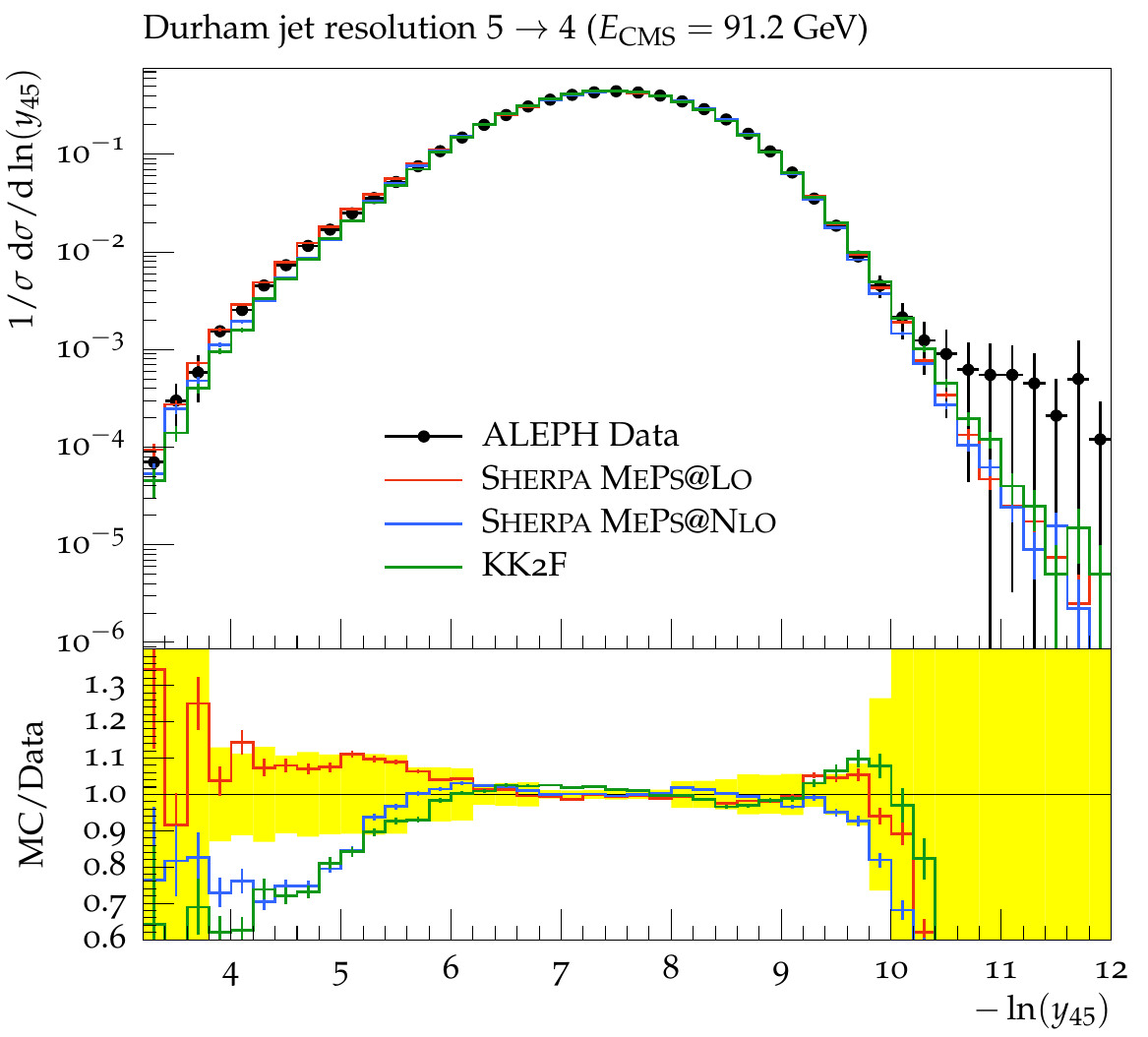}}\hspace{.4in}
\subfigure[]{\includegraphics[width=3.0in,bb=0 0 330 330]{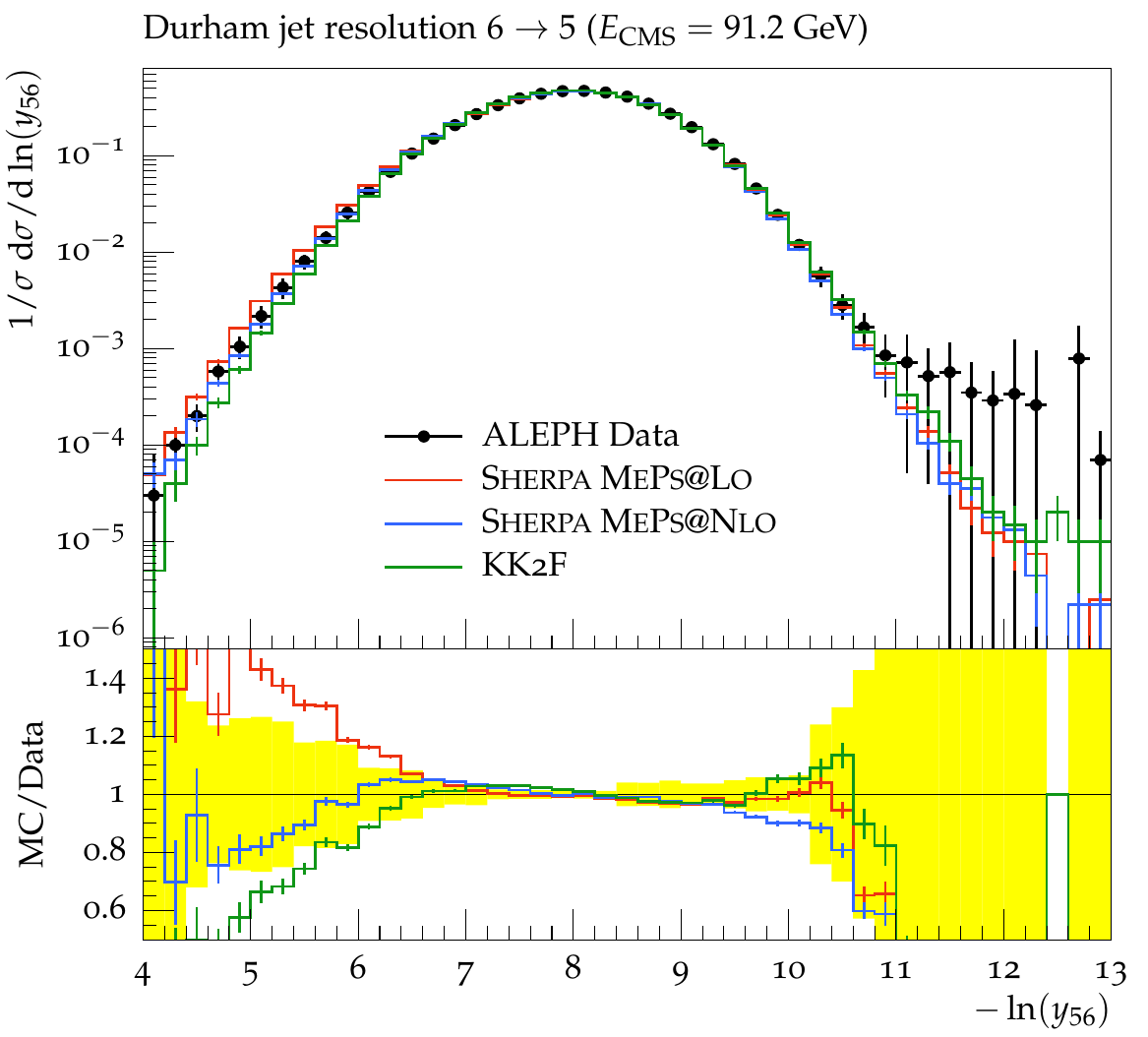}}\\
\end{center}
\caption{Plots of Durham jet resolution parameters $y_{ij}$ from ALEPH.}
\label{fig:ycuts}
\end{figure}

Plots of jet mass and broadening variables from ALEPH are shown in Fig. \ref{fig:eventshp3}.  These distributions are produced by dividing the event into two jets along the thrust axis.  In all four plots, all three MCs reproduce the data reasonably well.  In Fig. \ref{fig:ycuts}, we plot the Durham jet resolution parameters $y_{ij}$ from ALEPH.  We see that the data distribution for $y_{23}$ is described similarly well by the two SHERPA samples and somewhat less well by KK2f.  The $y_{34}$ and $y_{45}$ distributions are best described by the LO SHERPA MC, while  $y_{56}$ is best described by the NLO SHERPA MC.  While the NLO SHERPA generation does not describe  $y_{34}$ as well as the KK2f MC, in most cases we see that the SHERPA MC is comparable to or better than KK2f in describing the data.

We note that, for many of these event-shape variables, the NLO and LO SHERPA MC and the KK2f MC perform comparably.  Even within a given distribution, the relative performances of the three generations can depend upon the value of the observable in question.  Overall, however, we see no particularly concerning behavior in the new SHERPA samples in comparison to the KK2f generation with respect to the event-shape variables.

\subsection{LEP1 Four-Jet Observables}
\label{4jetdists}

We now turn to four-jet variables.  For all of these observables, the event is forced into four jets.  These variables are of particular interest because they probe the four-jet structure of the event and give an important comparison between SHERPA, which contains the four-jet matrix element, and KK2f, which uses only the parton shower.  

We compare the MC generation to data from OPAL via the Rivet analysis OPAL\_2001\_S4553896 \cite{Abbiendi:2001qn}.  Unlike the previous Rivet analyses, this analysis has hadronization effects unfolded and is compared at parton level; these plots thus require a separate MC generation.  We generate $4\times10^6$ unweighted events with the LO SHERPA tune, and approximately $4\times10^6$ partially unweighted events using the NLO tune.  We also generate $2\times10^6$ unweighted events with KK2f.

\begin{figure}[h]
\begin{center}
\subfigure[]{\includegraphics[width=3.0in,bb=0 0 330 330]{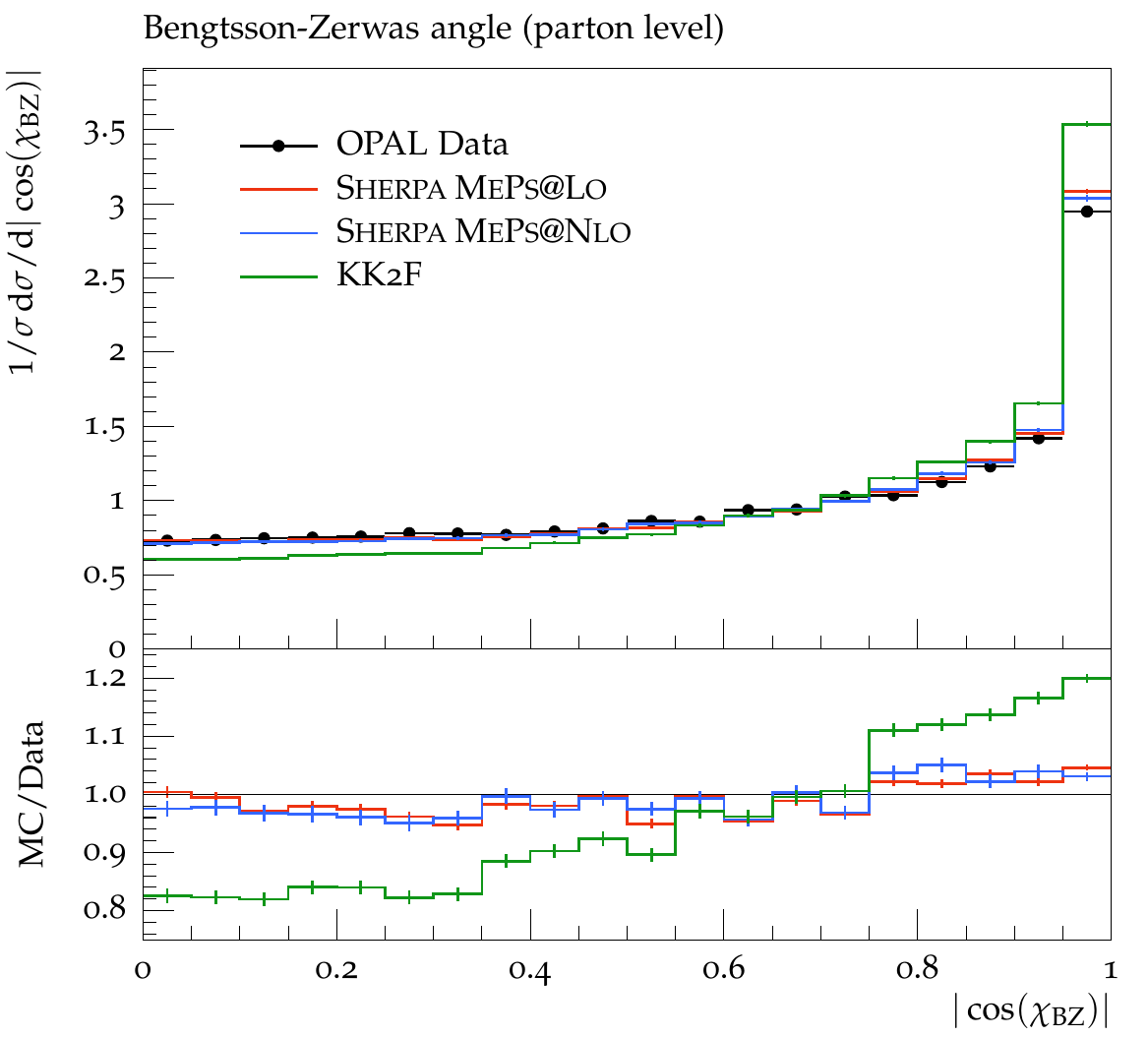}}\hspace{.4in}
\subfigure[]{\includegraphics[width=3.0in,bb=0 0 330 330]{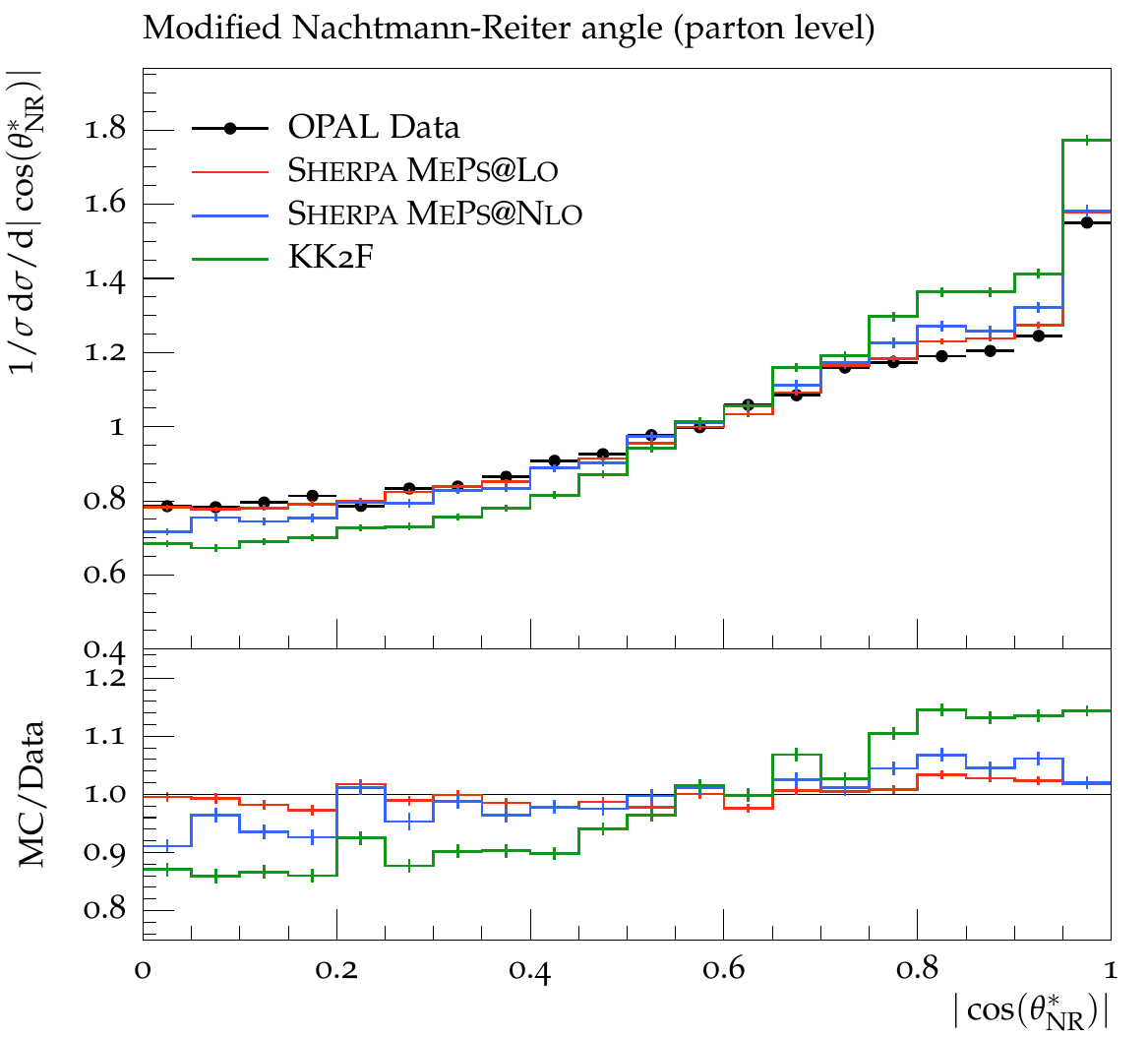}}\\
\subfigure[]{\includegraphics[width=3.0in,bb=0 0 330 330]{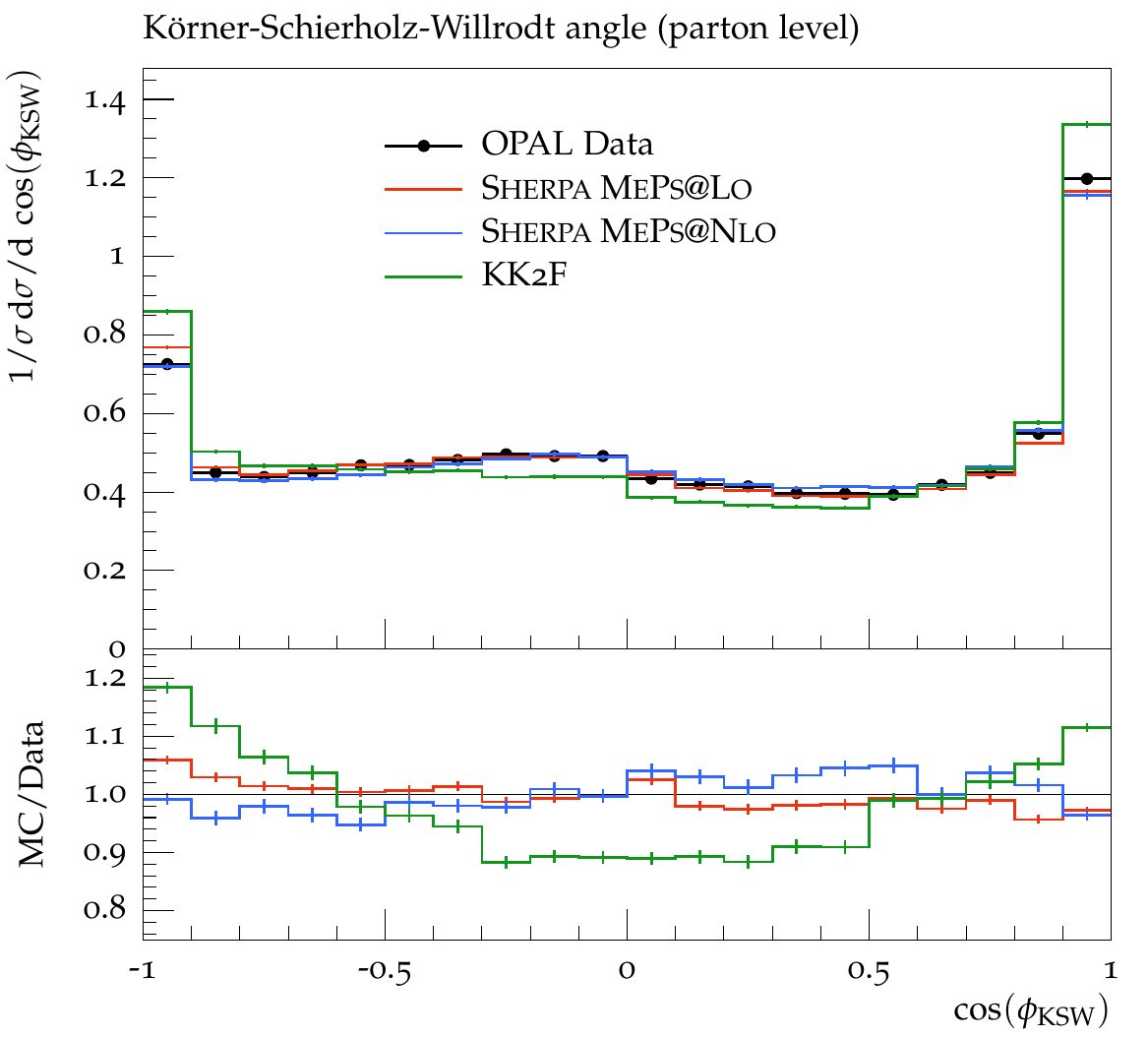}}\hspace{.4in}
\subfigure[]{\includegraphics[width=3.0in,bb=0 0 330 330]{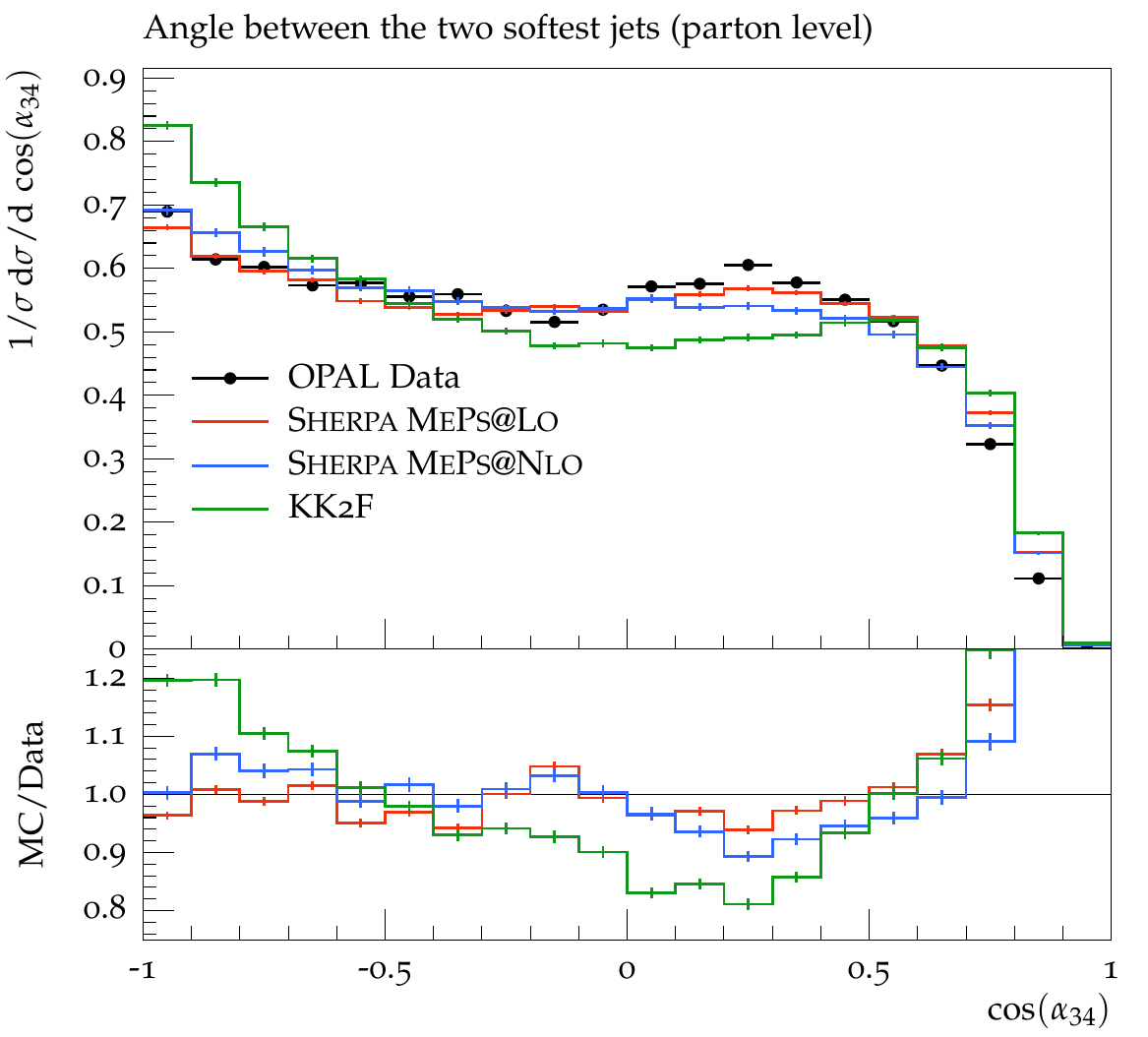}}\\
\end{center}
\caption{Plots of four-jet observables at LEP1 from OPAL.}
\label{fig:fourjet}
\end{figure}

Plots of the Bengtsson-Zerwas angle \cite{Bengtsson:1988qg}, the modified Nachtman-Reiter angle \cite{Nachtmann:1982xr}, the K\"{o}rner-Schierholz-Willrodt angle \cite{Korner:1980pv}, and the angle between the two softest jets are shown in Fig. \ref{fig:fourjet}.\footnote{We note that the analysis OPAL\_2001\_S4553896 provides no systematic errors for the unfolded data for these plots.  \cite{Abbiendi:2001qn} gives statistical errors only, which are small compared to the difference between KK2f and the SHERPA lines shown in Fig. \ref{fig:fourjet}.  Their experimental and hadronization correction factors are typically less than $10\%$, so it is reasonable to expect that their systematic errors are smaller yet.  We will study these variables, along with other jet angular variables, at full simulation in Ref. \cite{paper2}.}  We see that in all cases, both the LO and NLO SHERPA generation perform similarly and out-perform the KK2f generation.

\subsection{Other LEP1 distributions}
\label{otherdists}

\begin{figure}[h]
\begin{center}
\subfigure[]{\includegraphics[width=3.0in,bb=0 0 330 330]{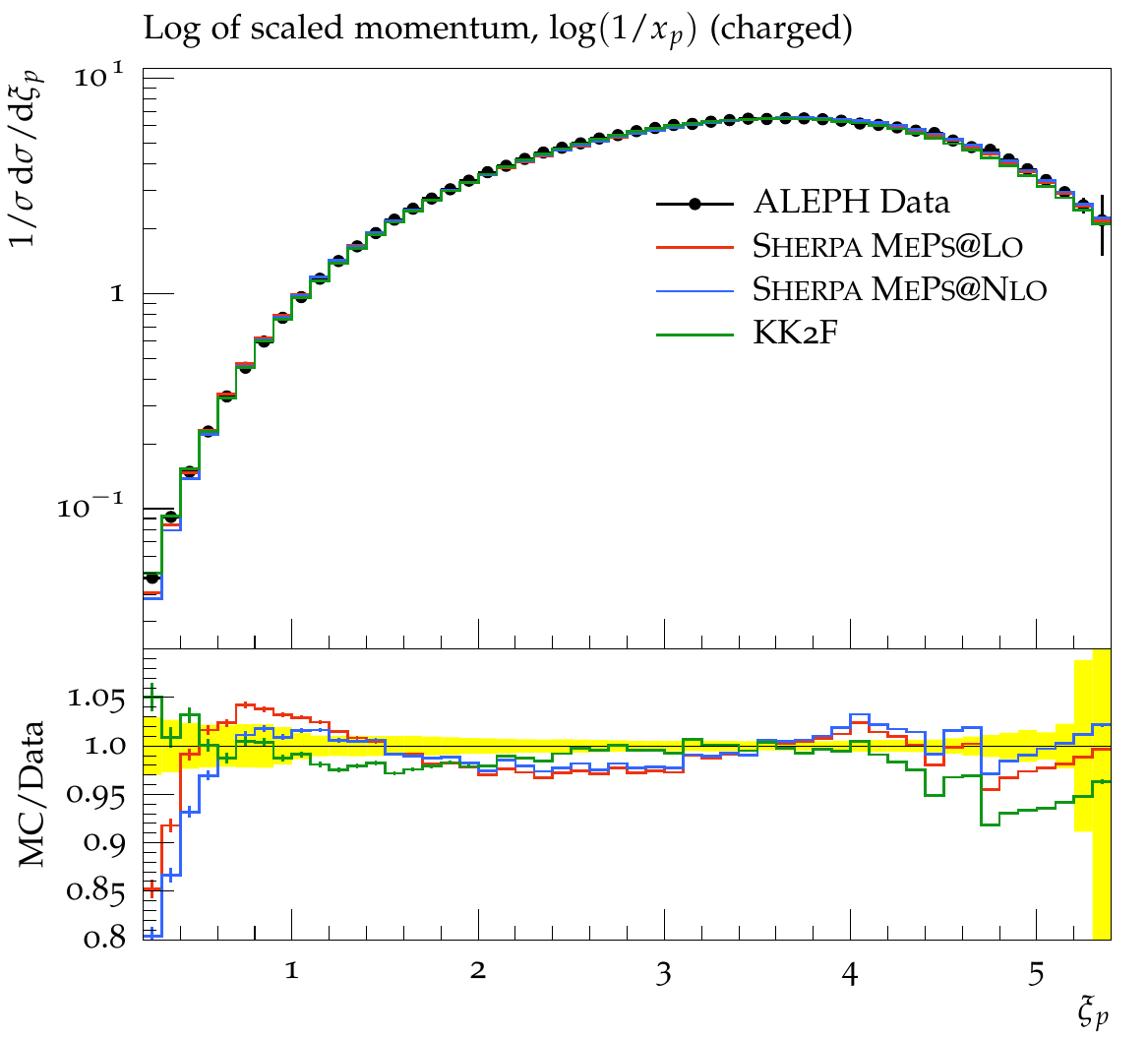}}\hspace{.4in}
\subfigure[]{\includegraphics[width=3.0in,bb=0 0 330 330]{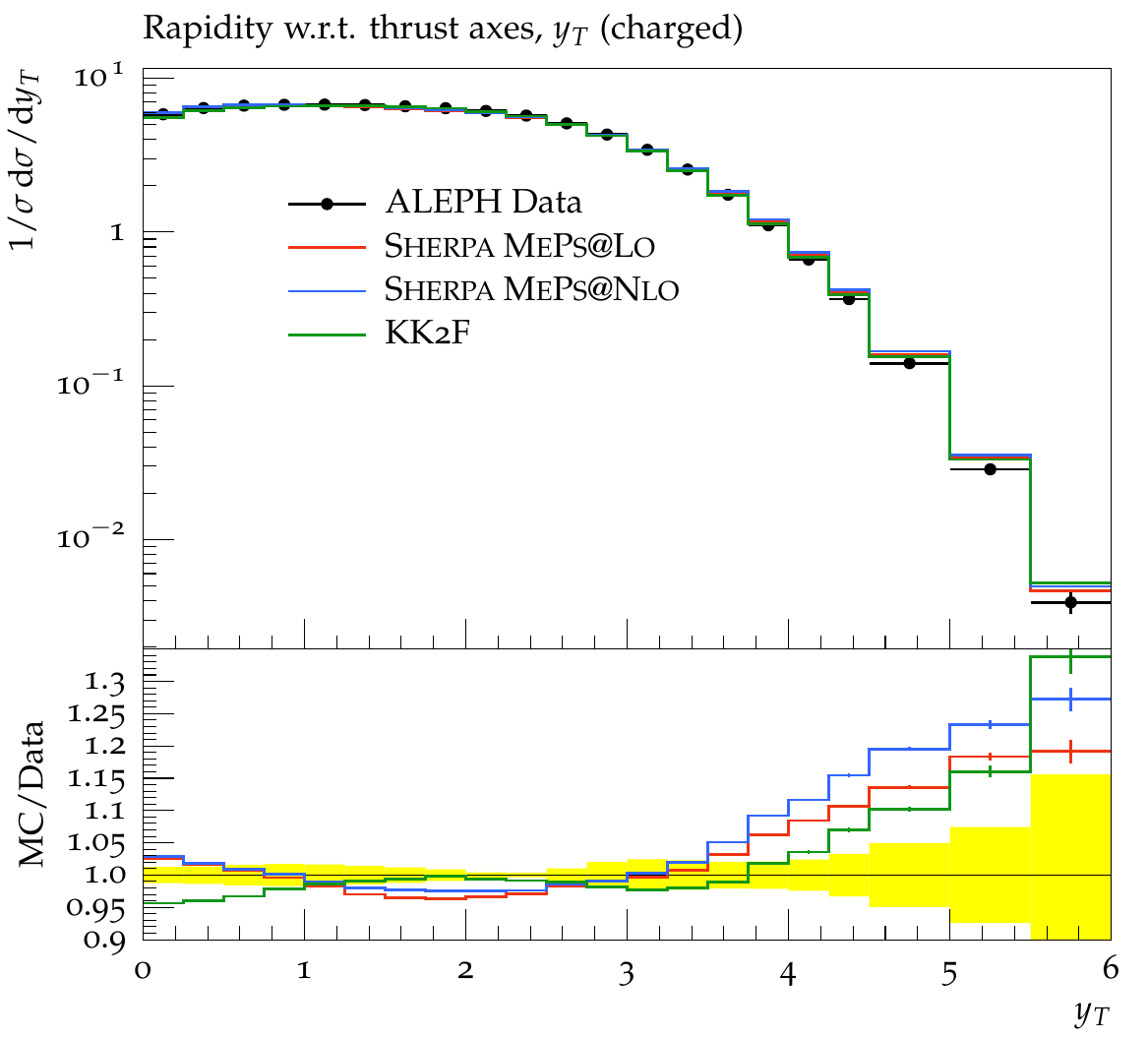}}
\end{center}
\caption{Plots of (a) $\xi_p=-\log{x_p}$ and (b) rapidity with respect to thrust axes, $y_T$ at LEP1 from ALEPH.}
\label{fig:logxprap}
\end{figure}

Lastly, we consider a few more LEP1 distributions.  The numbers of generated events are as for the event-shape variables above.  In Fig. \ref{fig:logxprap} (a), we compare the MC to data from ALEPH for the distribution of the log of the scaled momentum; we see that the three generations are comparable for this variable.  The rapidity with respect to the thrust axis in Fig. \ref{fig:logxprap} (b) behaves similarly for the three MC samples, with good agreement for the region containing the bulk of the events but with deviations in the tail.  

\begin{figure}[h]
\begin{center}
\end{center}
\subfigure[]{\includegraphics[width=3.in,bb=0 0 330 330]{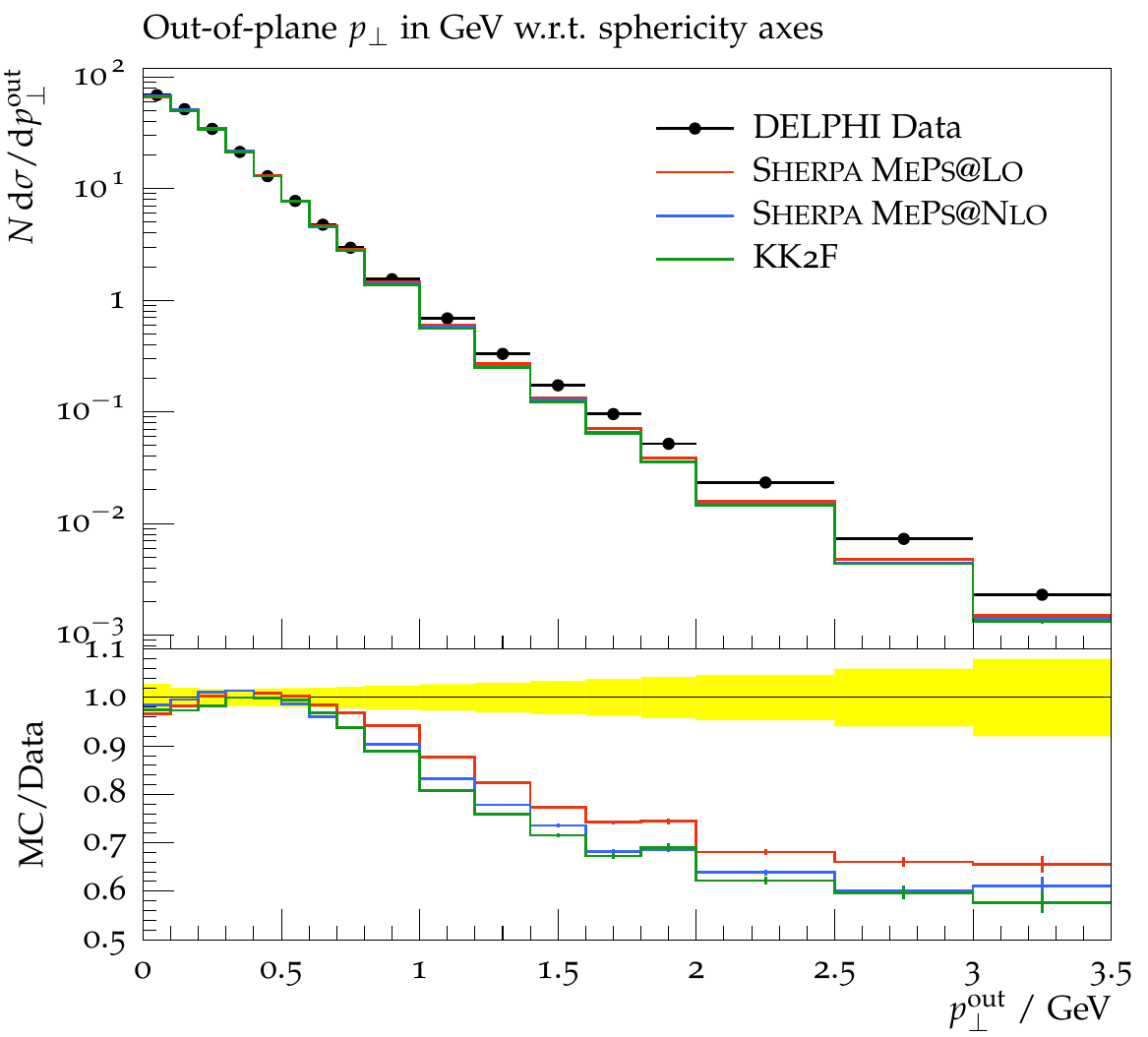}}\hspace{.4in}
\subfigure[]{\includegraphics[width=3.in,bb=0 0 330 330]{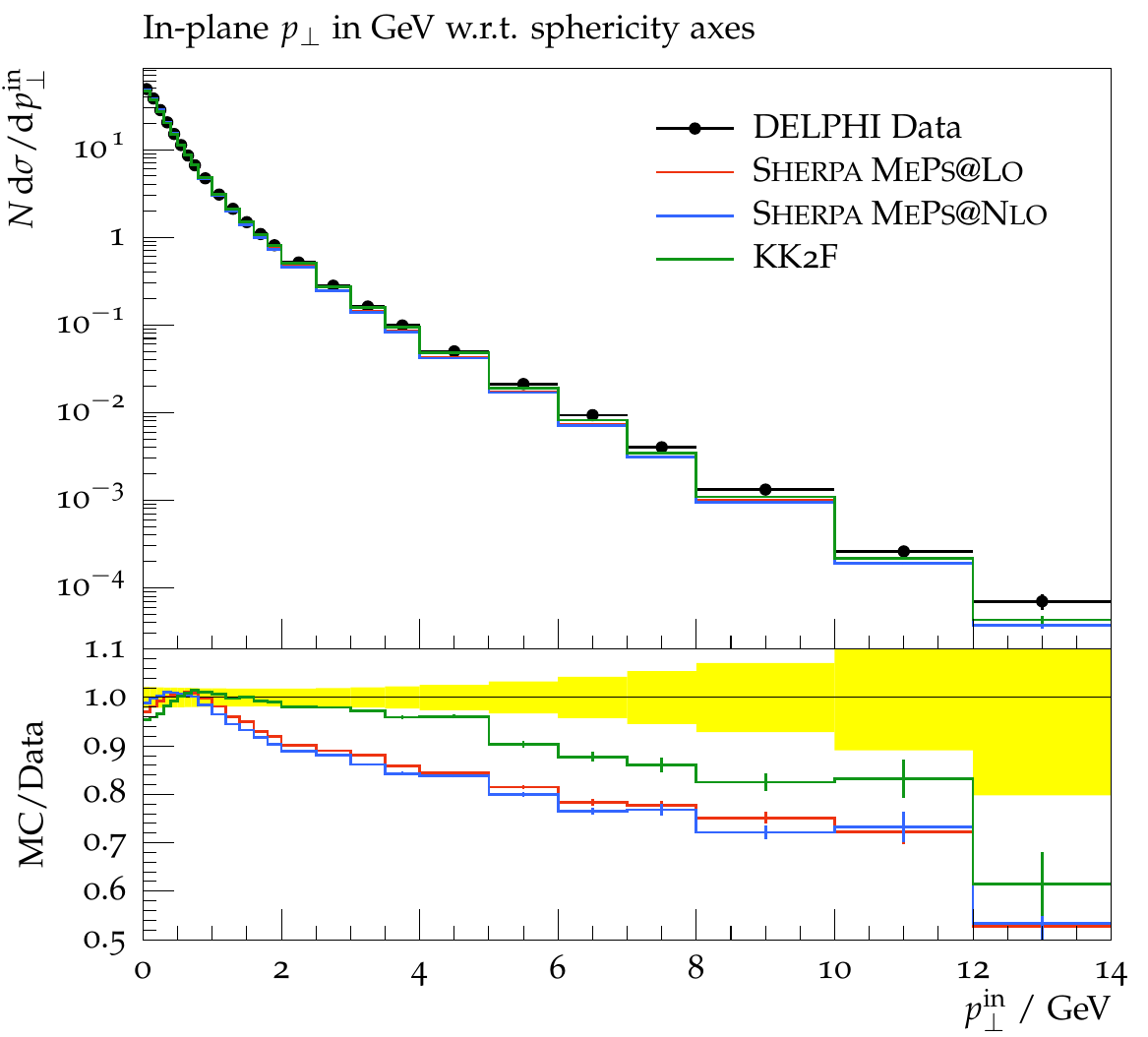}}
\caption{Plots of $p_{\perp out}$ and $p_{\perp in}$ at LEP1 from DELPHI.}
\label{fig:ptinout}
\end{figure}

In Fig. \ref{fig:ptinout}, we compare MC distributions for $p_{\perp out}$ and $p_{\perp in}$, defined relative to the sphericity axes, to data from DELPHI.  For $p_{\perp out}$, we see good agreement between the three MC generators, but significant disagreement between MC and data.  $p_{\perp in}$ also shows significant disagreement between data and simulation.  We note that this disagreement has been observed previously by the LEP experiments \cite{Barate:1996fi}. 

\subsection{Selected LEP2 distributions}
\label{lepiidists}

\begin{figure}[h]
\begin{center}
\end{center}
\subfigure[]{\includegraphics[width=3.in,bb=0 0 330 330]{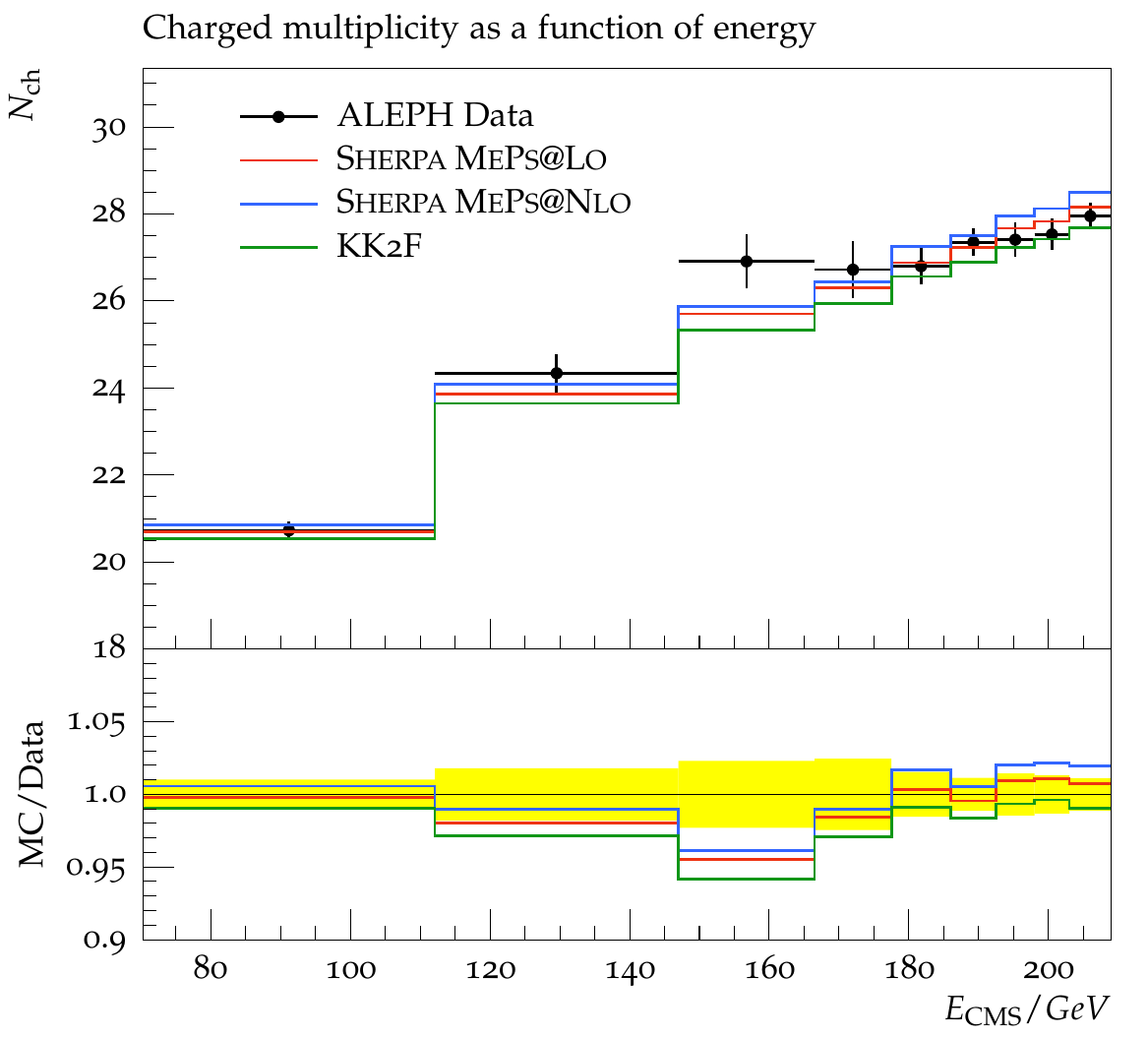}}
\caption{Plot of mean charged multiplicity as a function of $\sqrt{s}$ from ALEPH.}
\label{fig:charmult}
\end{figure}

\begin{figure}[h]
\begin{center}
\end{center}
\subfigure[]{\includegraphics[width=3.in,bb=0 0 330 330]{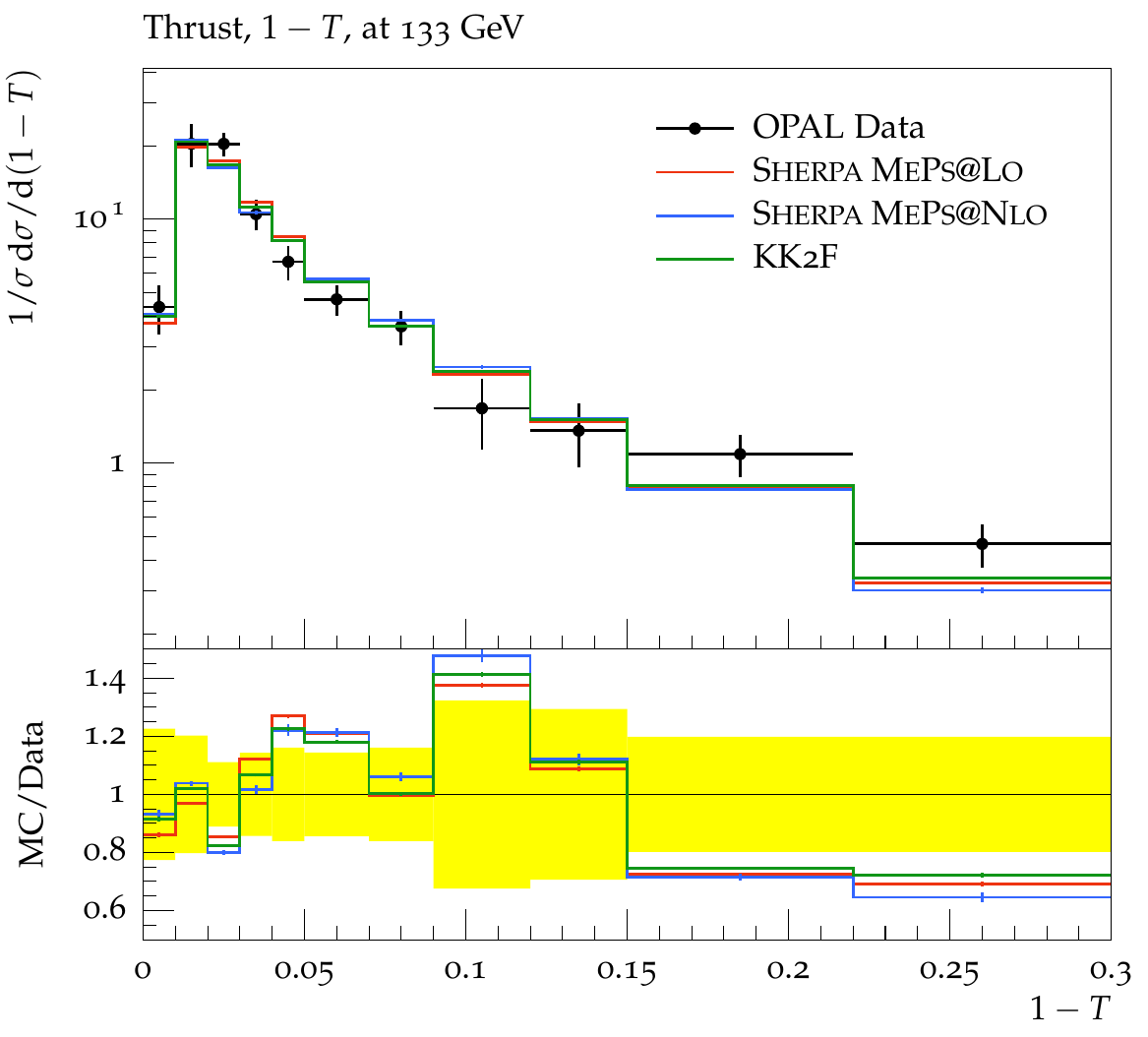}}\hspace{.4in}
\subfigure[]{\includegraphics[width=3.in,bb=0 0 330 330]{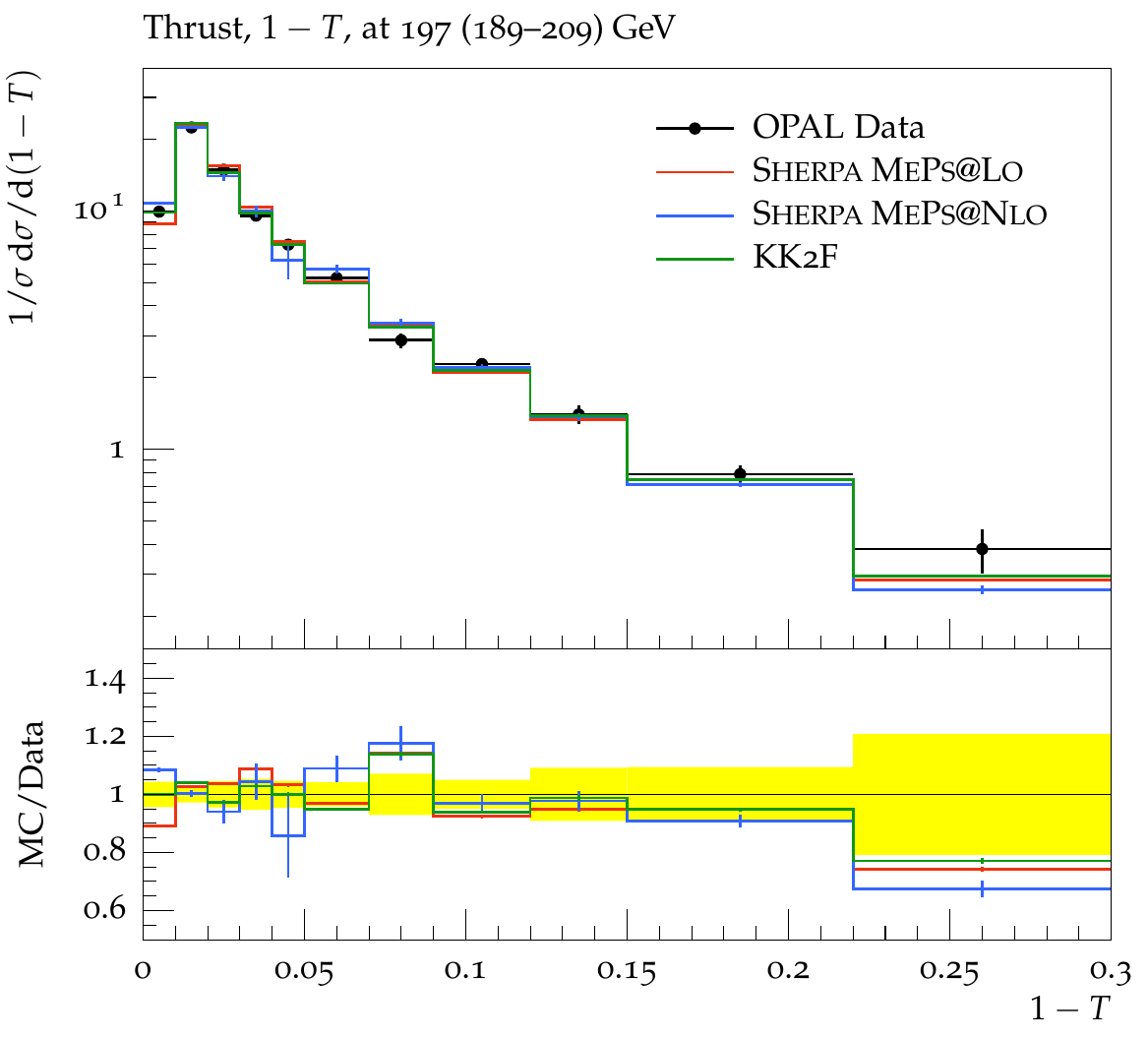}}
\caption{Plots of $1-T$ at LEP2 from OPAL for ranges of LEP2 energies.  Plots on the left are for $<\sqrt{s}>=133$ GeV, and those on the right are for $<\sqrt{s}>=197$ GeV.}
\label{fig:l2thrust}
\end{figure}

\begin{figure}[h]
\begin{center}
\end{center}
\subfigure[]{\includegraphics[width=3.in,bb=0 0 330 330]{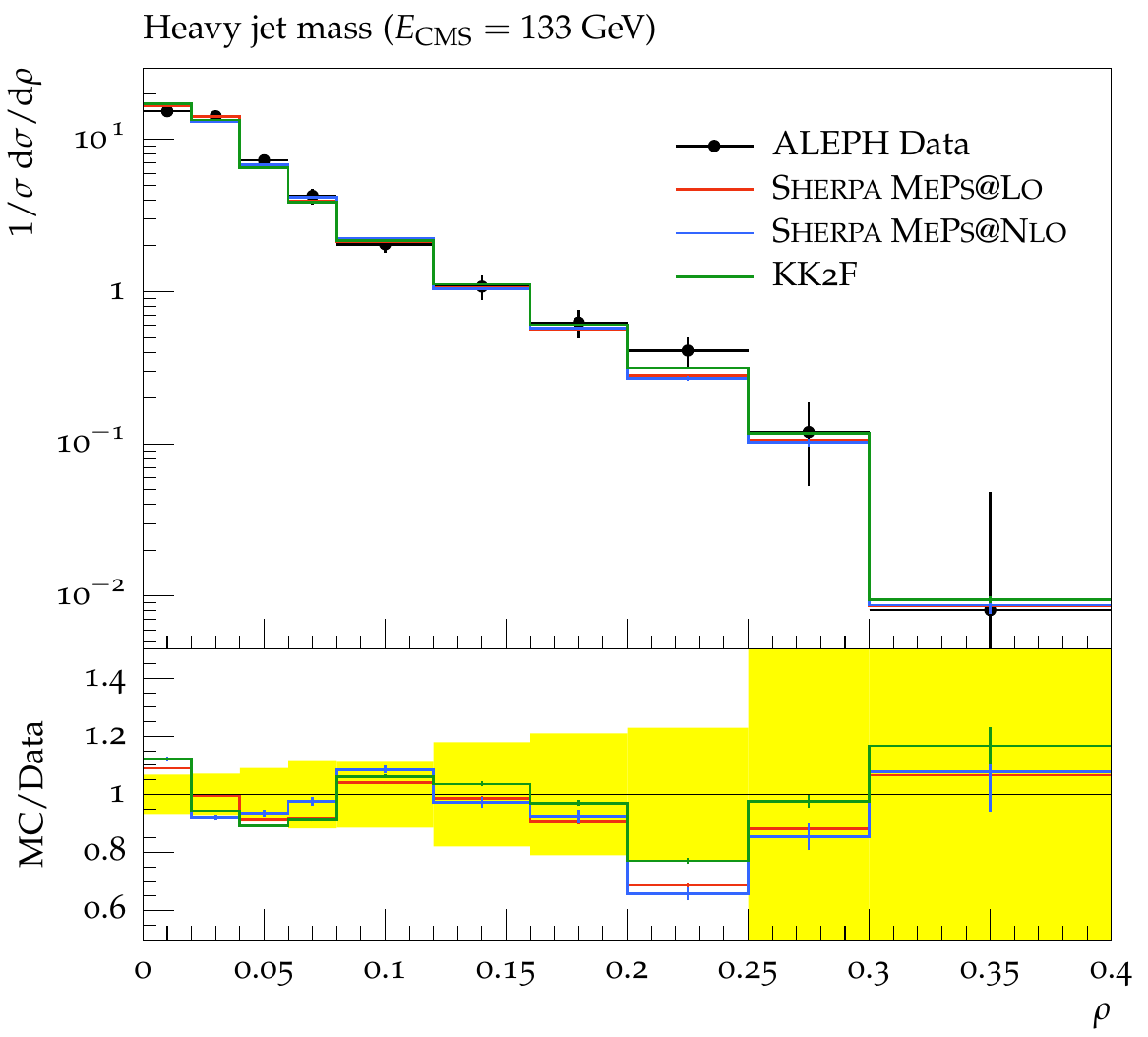}}\hspace{.4in}
\subfigure[]{\includegraphics[width=3.in,bb=0 0 330 330]{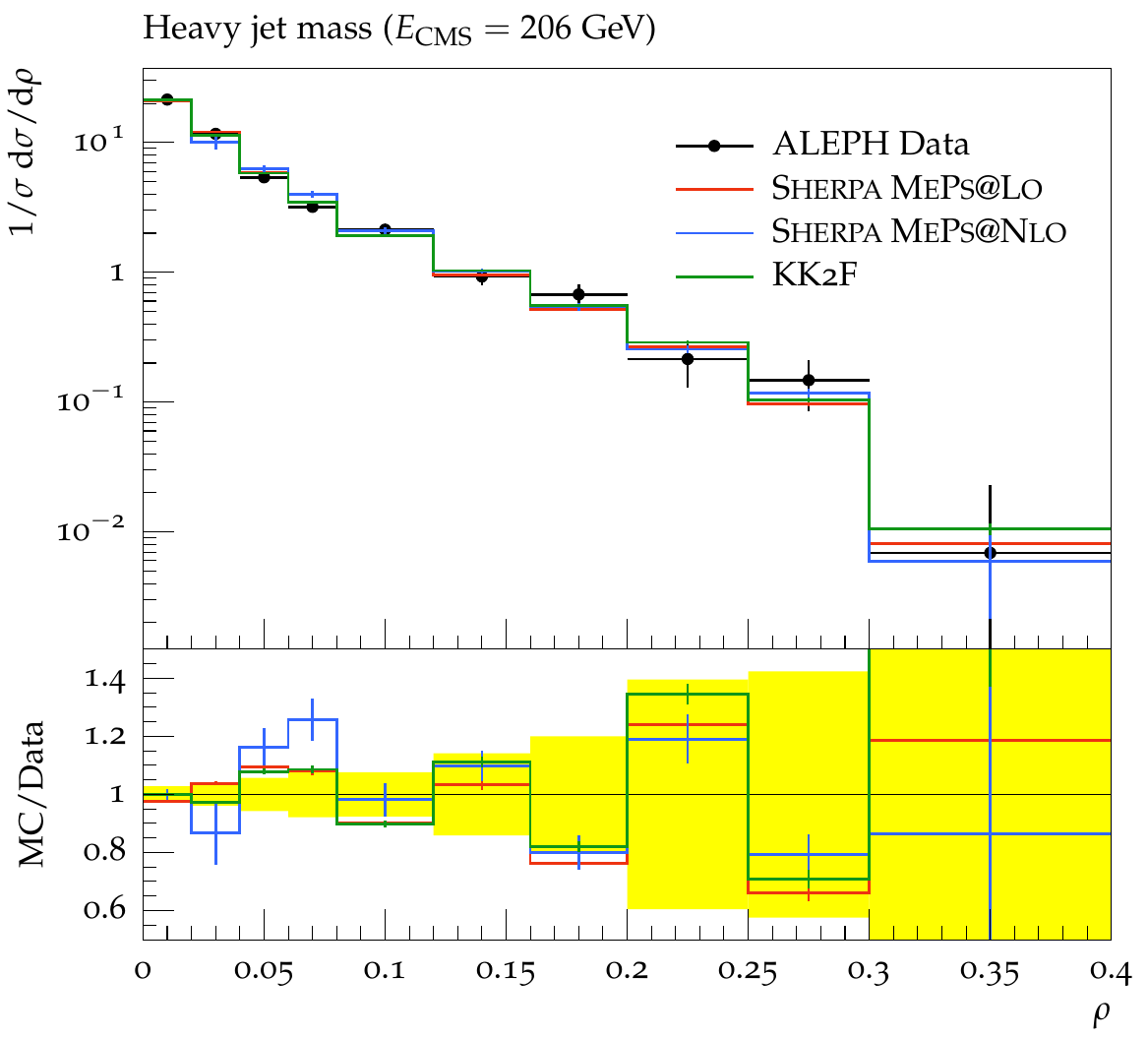}}\\
\subfigure[]{\includegraphics[width=3.in,bb=0 0 330 330]{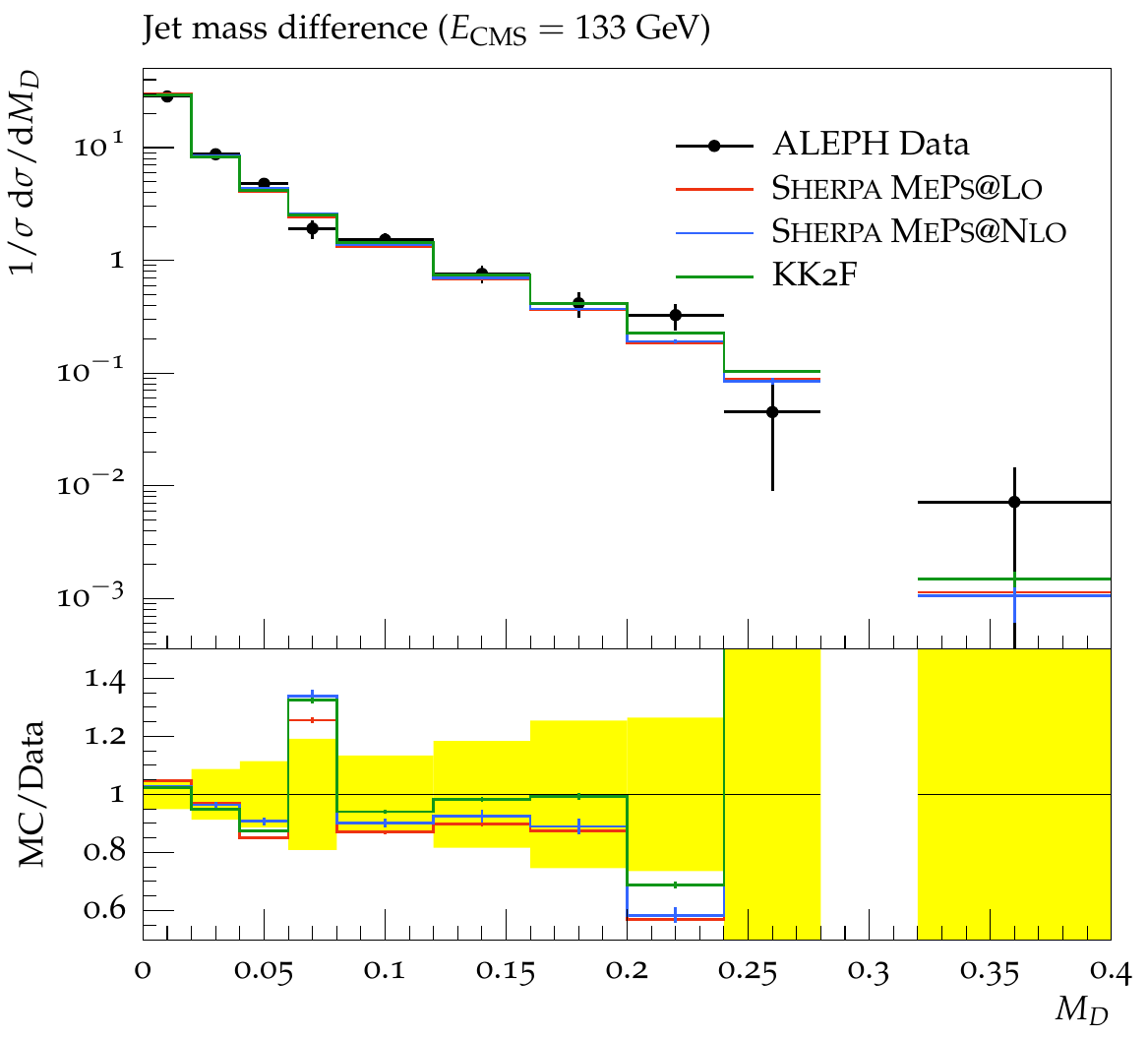}}\hspace{.4in}
\subfigure[]{\includegraphics[width=3.in,bb=0 0 330 330]{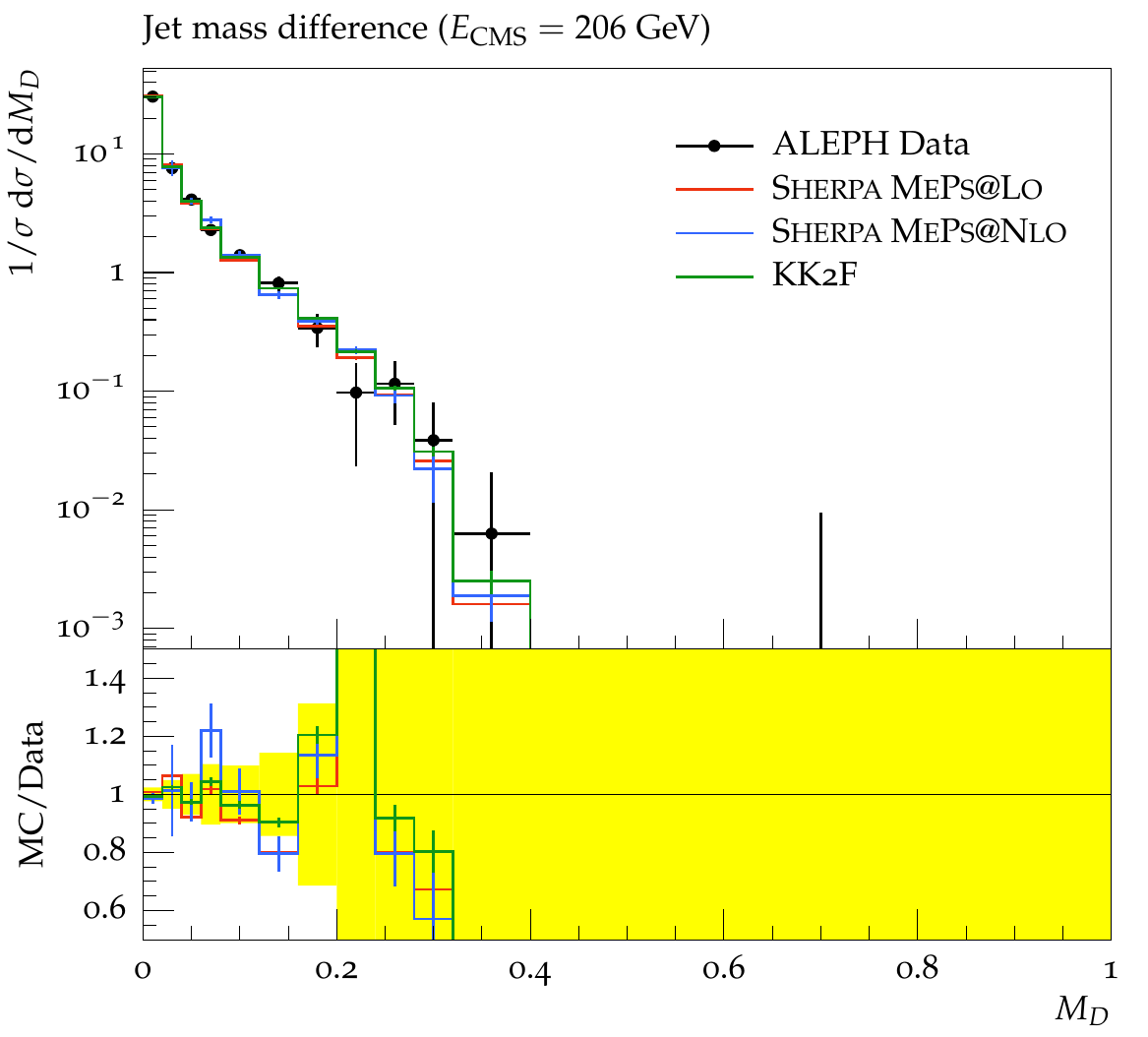}}
\caption{Plots of the heavy jet mass $\rho$ and the jet mass difference at LEP2 from ALEPH.  Plots on the left are for $<\sqrt{s}>=133$ GeV, and those on the right are for $<\sqrt{s}>=206$ GeV.}
\label{fig:mhmassdiff}
\end{figure}

\begin{figure}[h]
\begin{center}
\end{center}
\subfigure[]{\includegraphics[width=3.in,bb=0 0 330 330]{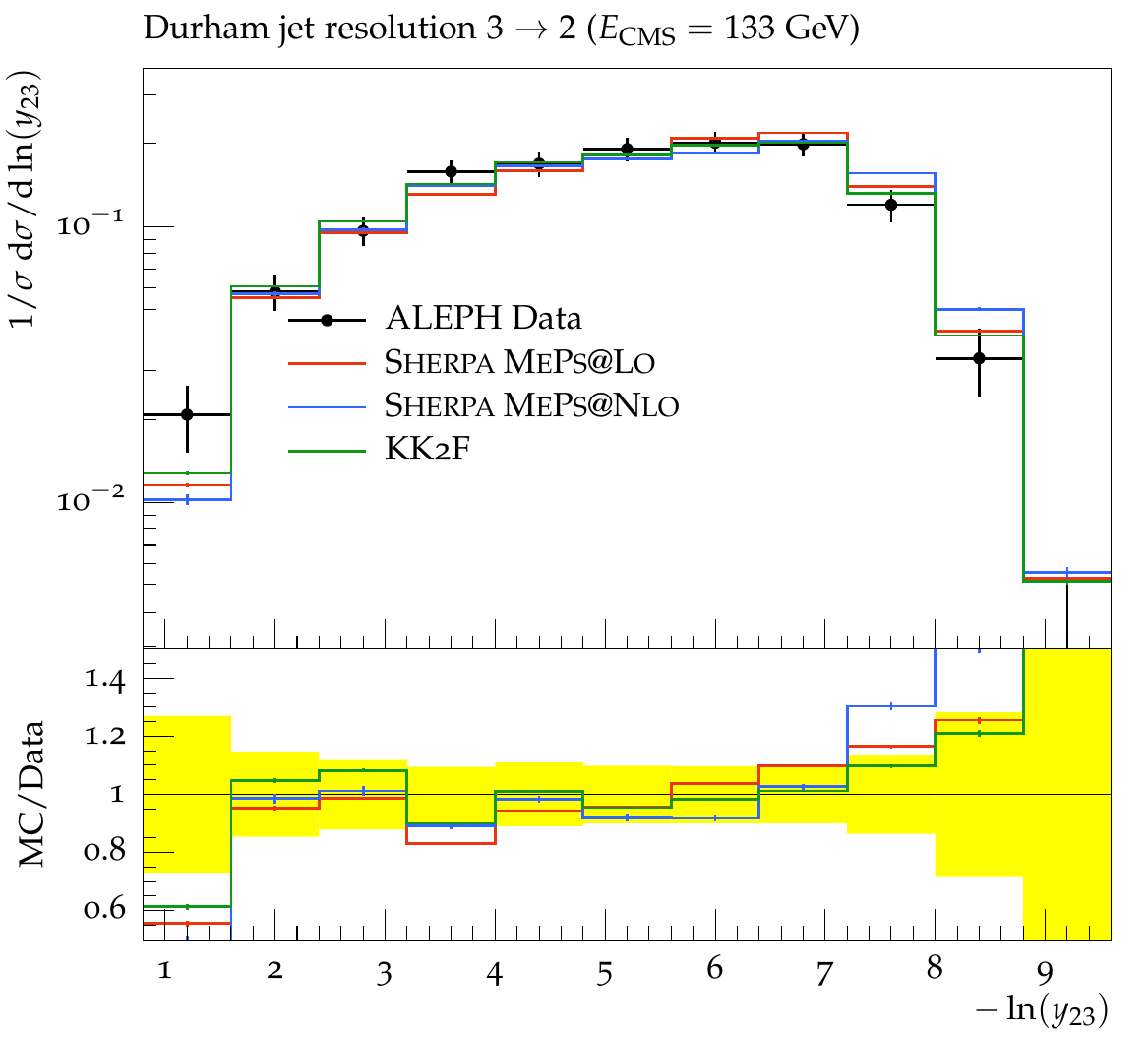}}\hspace{.4in}
\subfigure[]{\includegraphics[width=3.in,bb=0 0 330 330]{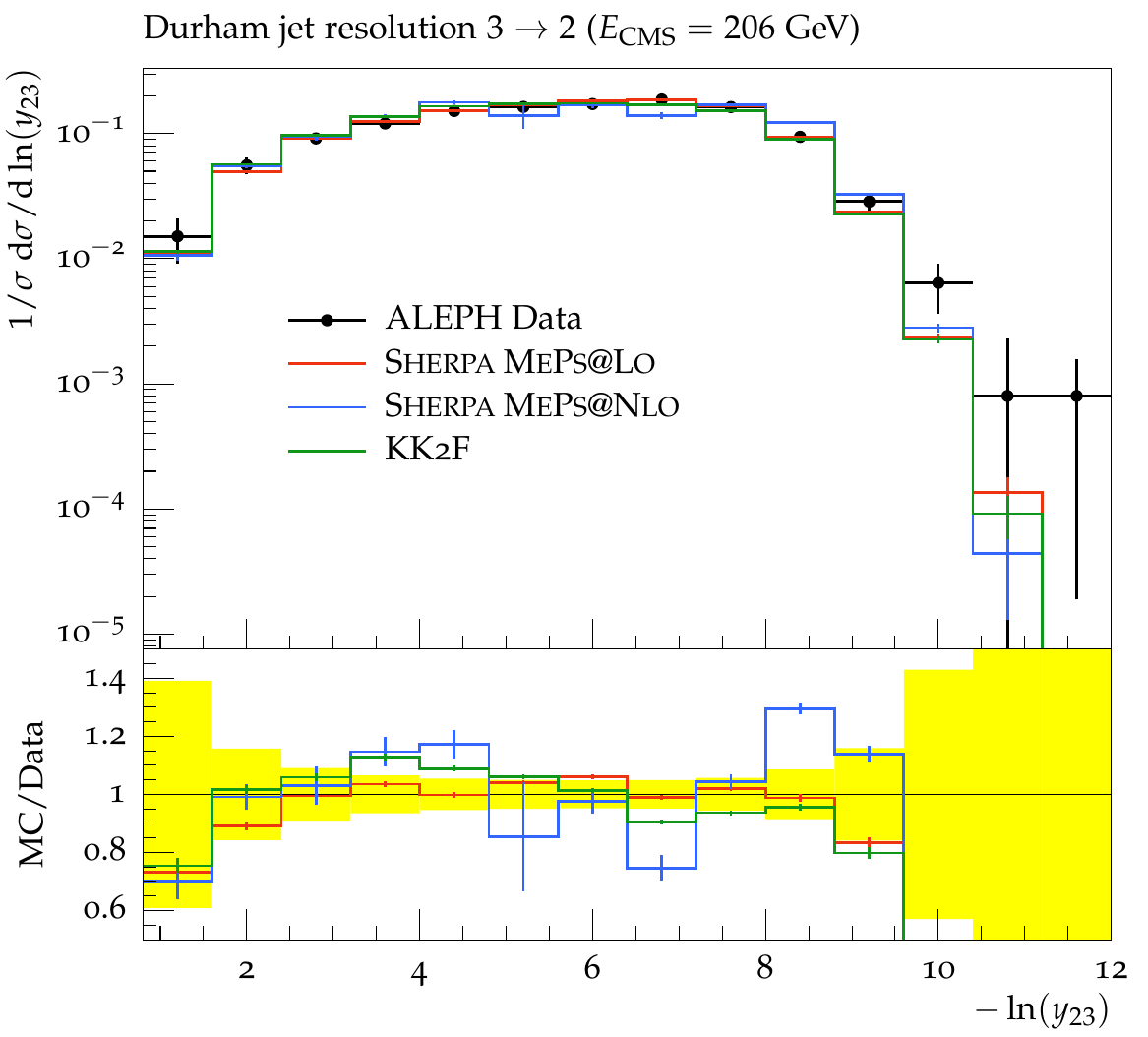}}\\
\subfigure[]{\includegraphics[width=3.in,bb=0 0 330 330]{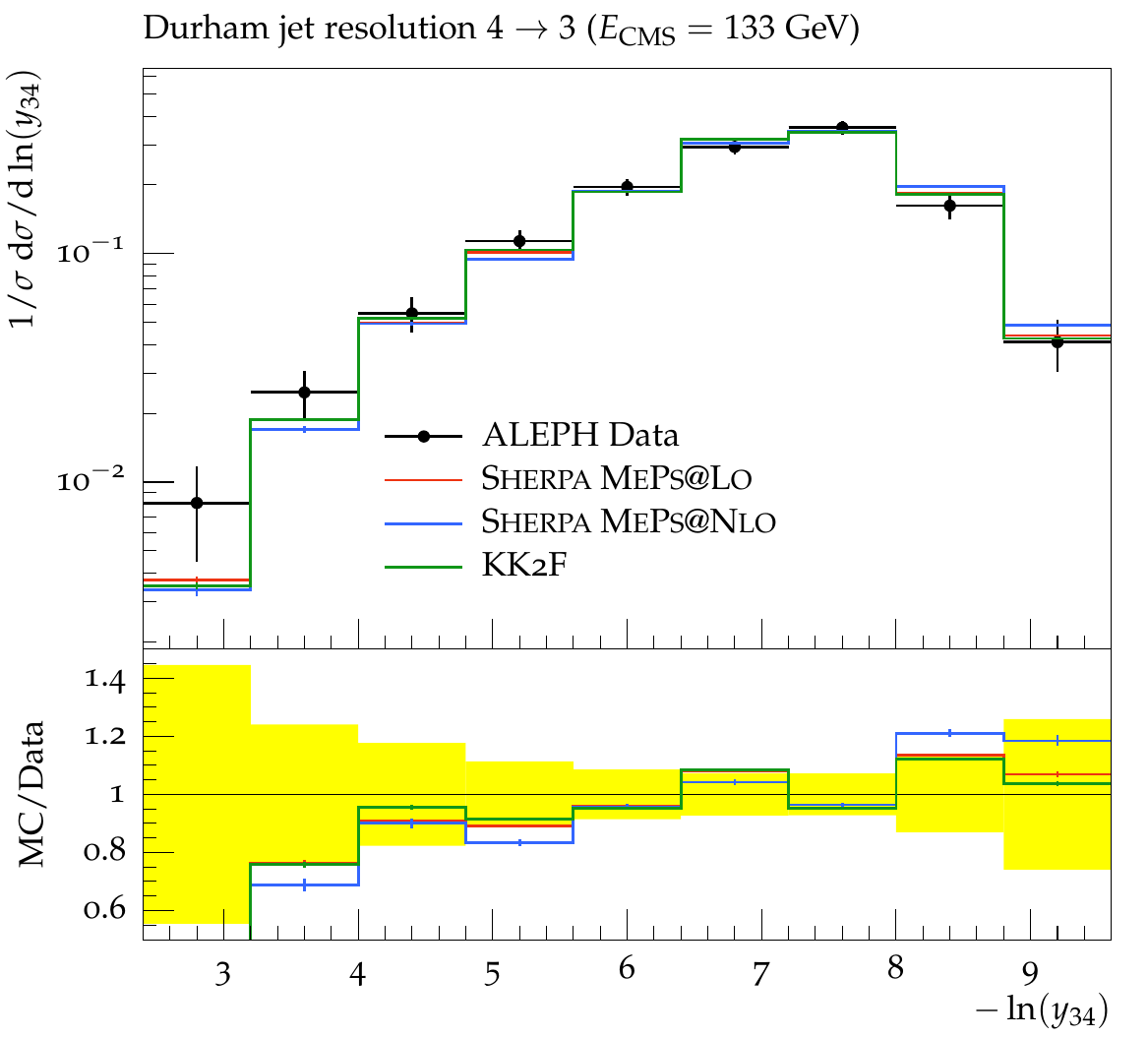}}\hspace{.4in}
\subfigure[]{\includegraphics[width=3.in,bb=0 0 330 330]{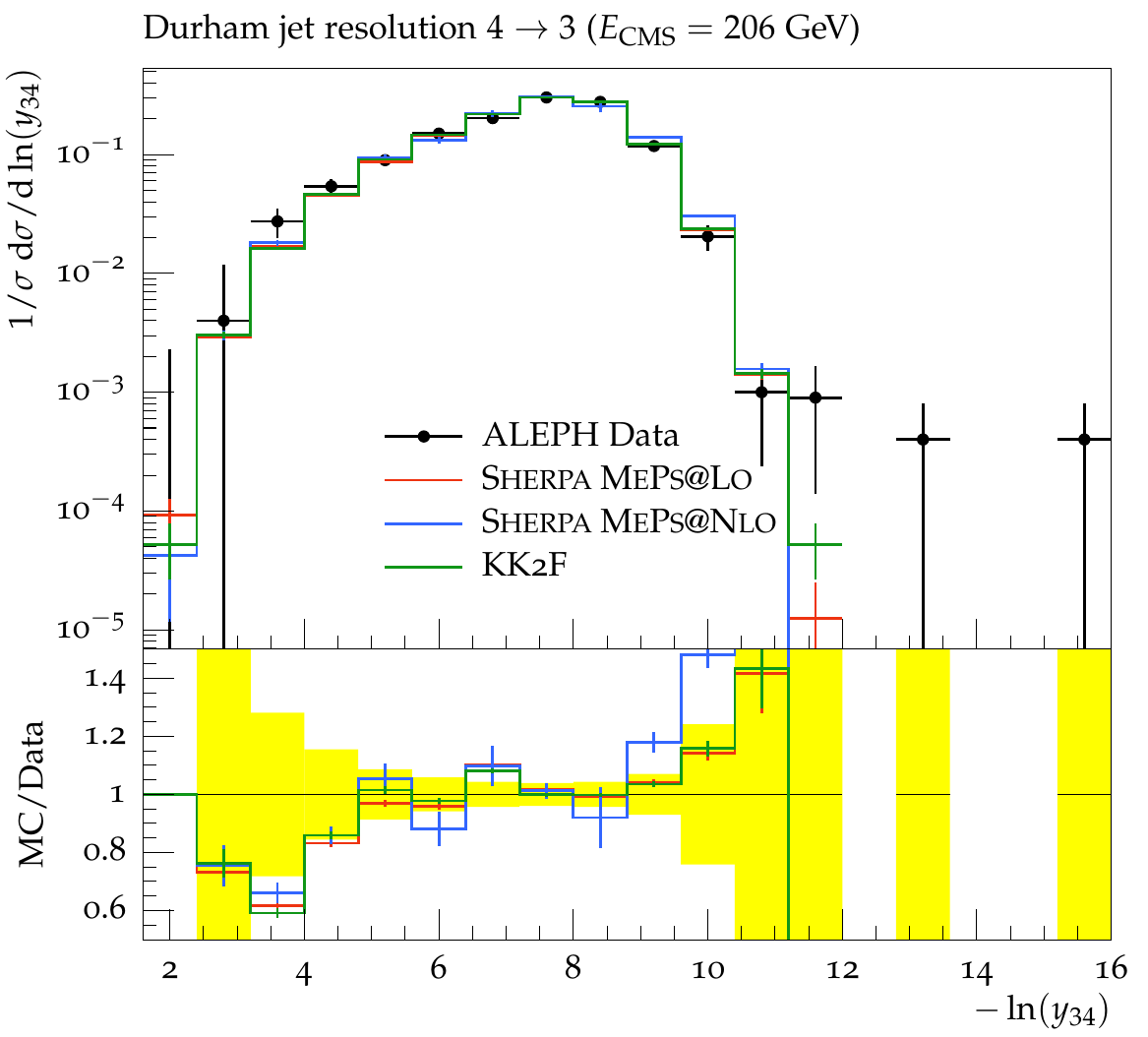}}
\caption{Plots of $y_{23}$ and $y_{34}$ at LEP2 from ALEPH.  Plots on the left are for $<\sqrt{s}>=133$ GeV, and those on the right are for $<\sqrt{s}>=206$ GeV.}
\label{fig:l2y23y34}
\end{figure}

\begin{figure}[h]
\begin{center}
\end{center}
\subfigure[]{\includegraphics[width=3.in,bb=0 0 330 330]{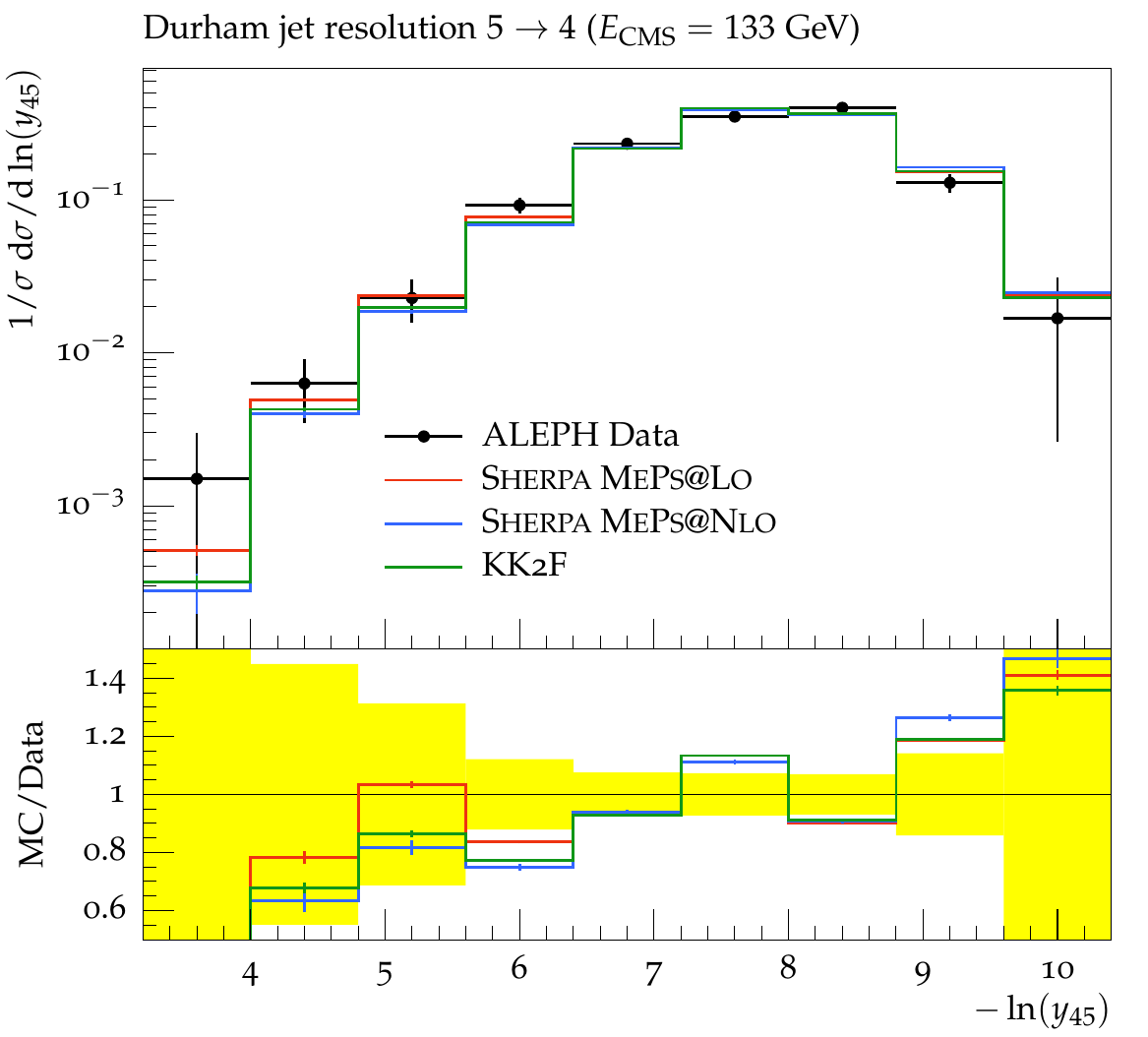}}\hspace{.4in}
\subfigure[]{\includegraphics[width=3.in,bb=0 0 330 330]{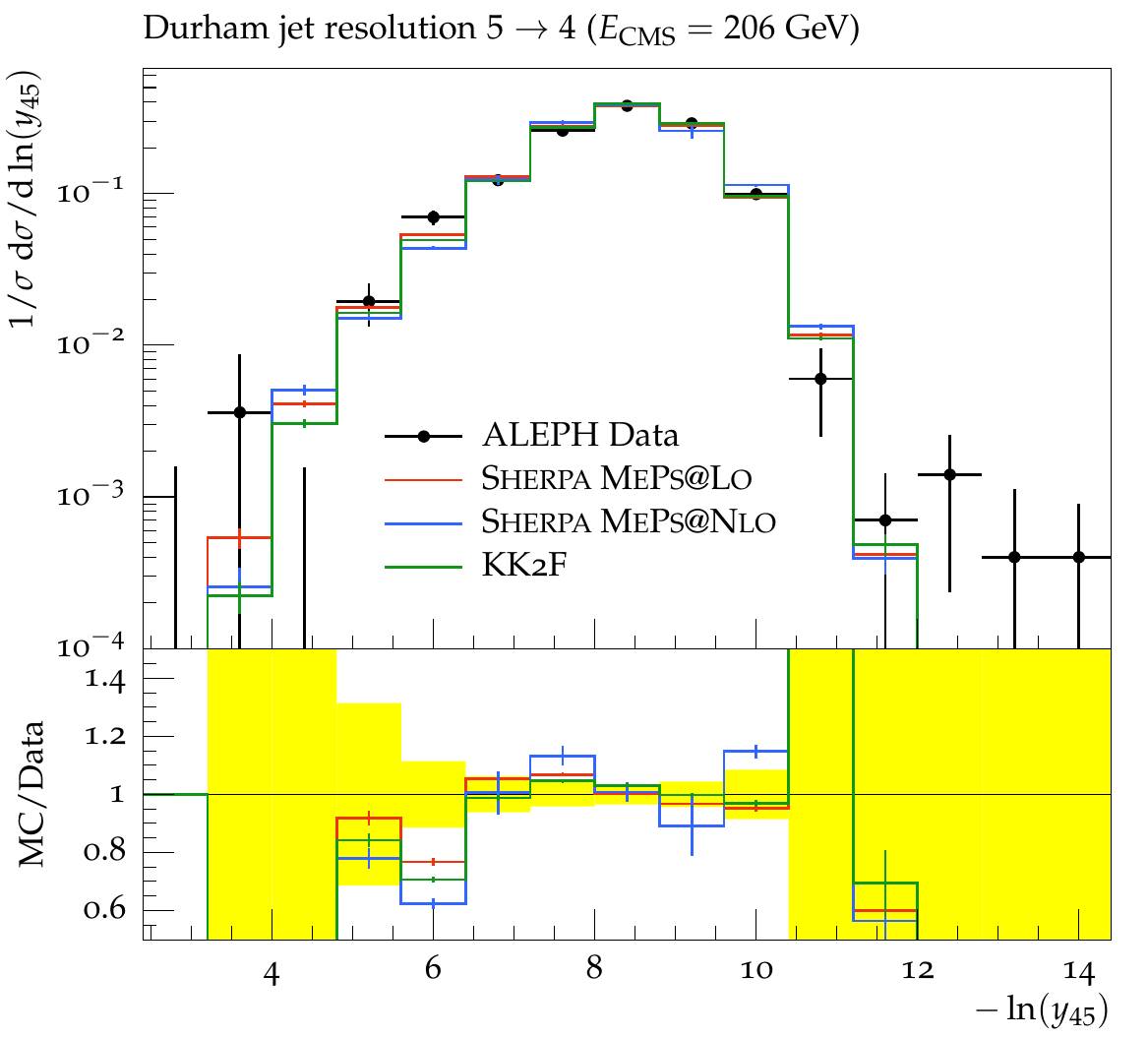}}\\
\subfigure[]{\includegraphics[width=3.in,bb=0 0 330 330]{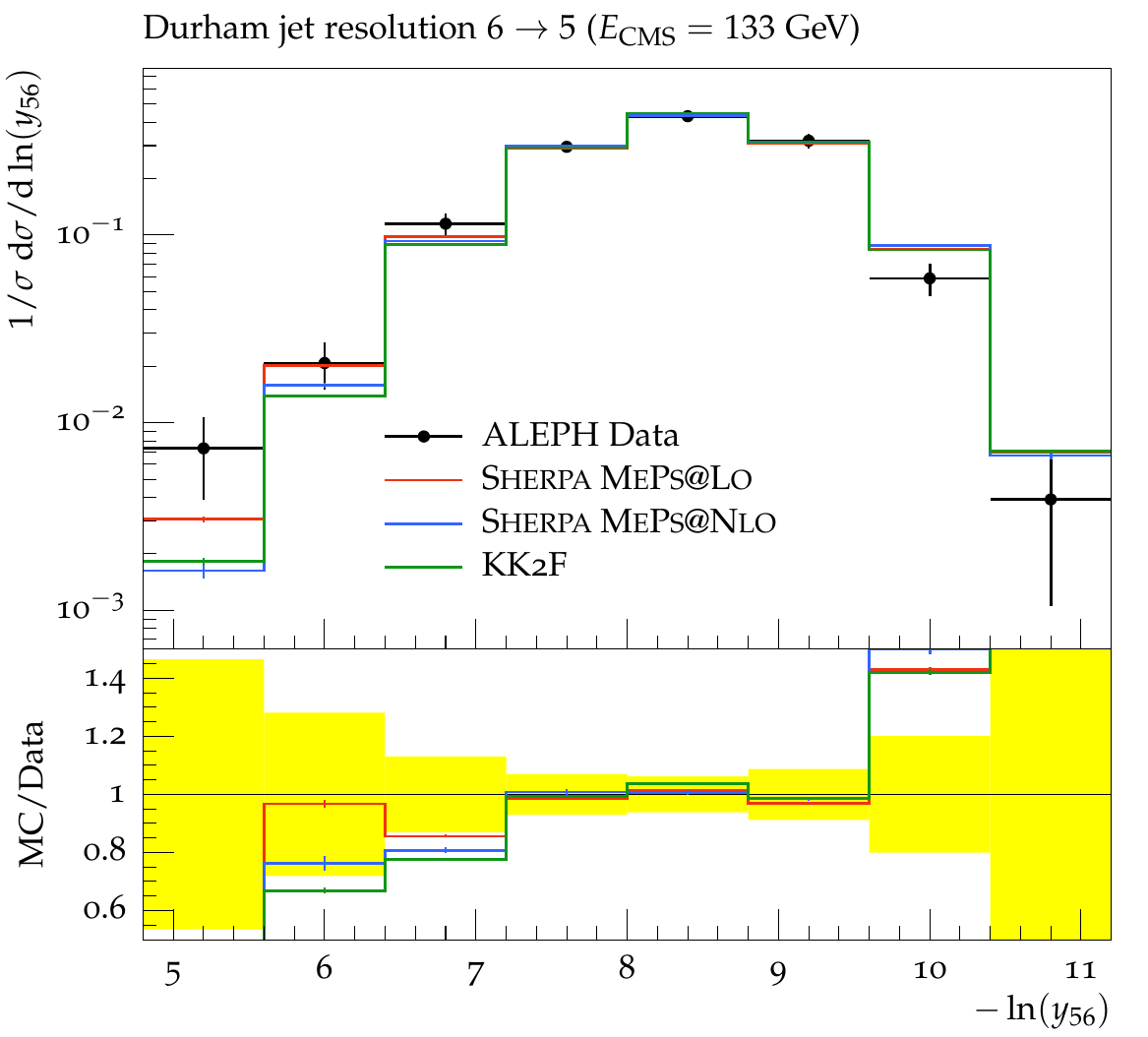}}\hspace{.4in}
\subfigure[]{\includegraphics[width=3.in,bb=0 0 330 330]{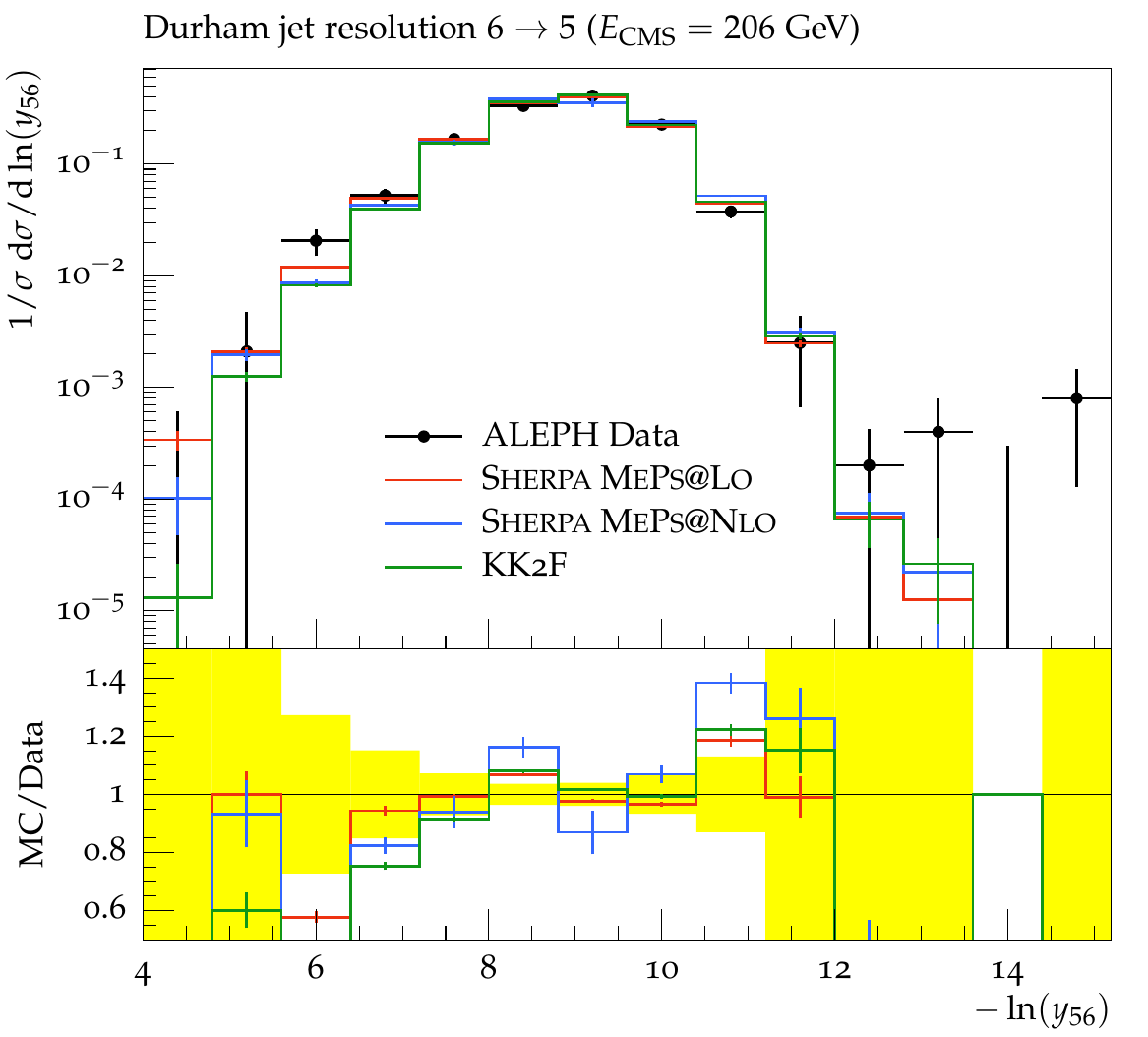}}
\caption{Plots of $y_{45}$ and $y_{56}$ at LEP2 from ALEPH.  Plots on the left are for $<\sqrt{s}>=133$ GeV, and those on the right are for $<\sqrt{s}>=206$ GeV.}
\label{fig:l2y45y56}
\end{figure}

We also consider some selected variables at LEP2 energies.  Plots of several quantities are given in Figs. \ref{fig:charmult}-\ref{fig:l2y45y56}.  Events are generated at center-of-mass energies from $\sqrt{s}=133$ GeV to $\sqrt{s}=208$ GeV.  The NLO tune SHERPA events are partially unweighted, and the LO tune and KK2f events are unweighted.  Effective luminosities for the MC generation are given in Table \ref{tab:l2mclum}.  As before, the MC samples are all normalized to the data.  Due to the decreased statistics of hadronic events at LEP2, the yellow error bands on the unfolded data are considerably larger than those for LEP1.  We see broad agreement between the data and the three MC generations, but will make a few comments on specific distributions.      

\begin{table}
\begin{tabular}{|c| c| c| c|}
\hline
$\sqrt{s}$ & KK2f & LO SHERPA  Tune & NLO SHERPA tune\\
\hline
$133$ GeV & $363\times$ data  & $363\times$ data & $181\times$ data \\
\hline
$161.3-182.6$ GeV & $138\times$ data  & $138\times$ data & $69\times$ data \\
\hline
$188.6-208.0$ GeV & $25\times$ data  & $25\times$ data & $25\times$ data \\
\hline
\end{tabular}
\caption{Effective luminosities for MC generation for the LEP2 data-MC comparisons.  }
\label{tab:l2mclum}
\end{table}

A plot of the mean charged multiplicity as a function of $\sqrt{s}$ from ALEPH is given in Fig. \ref{fig:charmult}.  We see that the LO SHERPA, NLO SHERPA, and KK2f samples all reproduce the unfolded data reasonably well.  Next, we give some event-shape variables.  Plots of the thrust from OPAL at $<\sqrt{s}>=133$ GeV and $<\sqrt{s}>=197$ GeV are given in Fig. \ref{fig:l2thrust}.  Jet mass variables from ALEPH are similarly displayed in Fig. \ref{fig:mhmassdiff}.  In all cases, we see reasonable agreement between the unfolded data and the KK2f and SHERPA MC samples.  In Figs. \ref{fig:l2y23y34}-\ref{fig:l2y45y56}, we show the Durham jet resolutions from ALEPH at $<\sqrt{s}>=133$ and $206$ GeV.  Broad agreement between data and all three MC samples is observed.

\section{Renormalization Scale Variations}
\label{syst}

Here we discuss changes in the LO and NLO SHERPA MC distributions under variations in renormalization scale.  We do this by multiplying the square of the renormalization scale $\mu_R^2$ by a factor $SCF=0.5, 2.0$.    As before, the MC samples are normalized; only variations in the shapes are studied here.  All MC lines shown here are compared to data at LEP1.

\begin{figure}[h]
\begin{center}
\includegraphics[width=3.0in,bb=0 0 330 330]{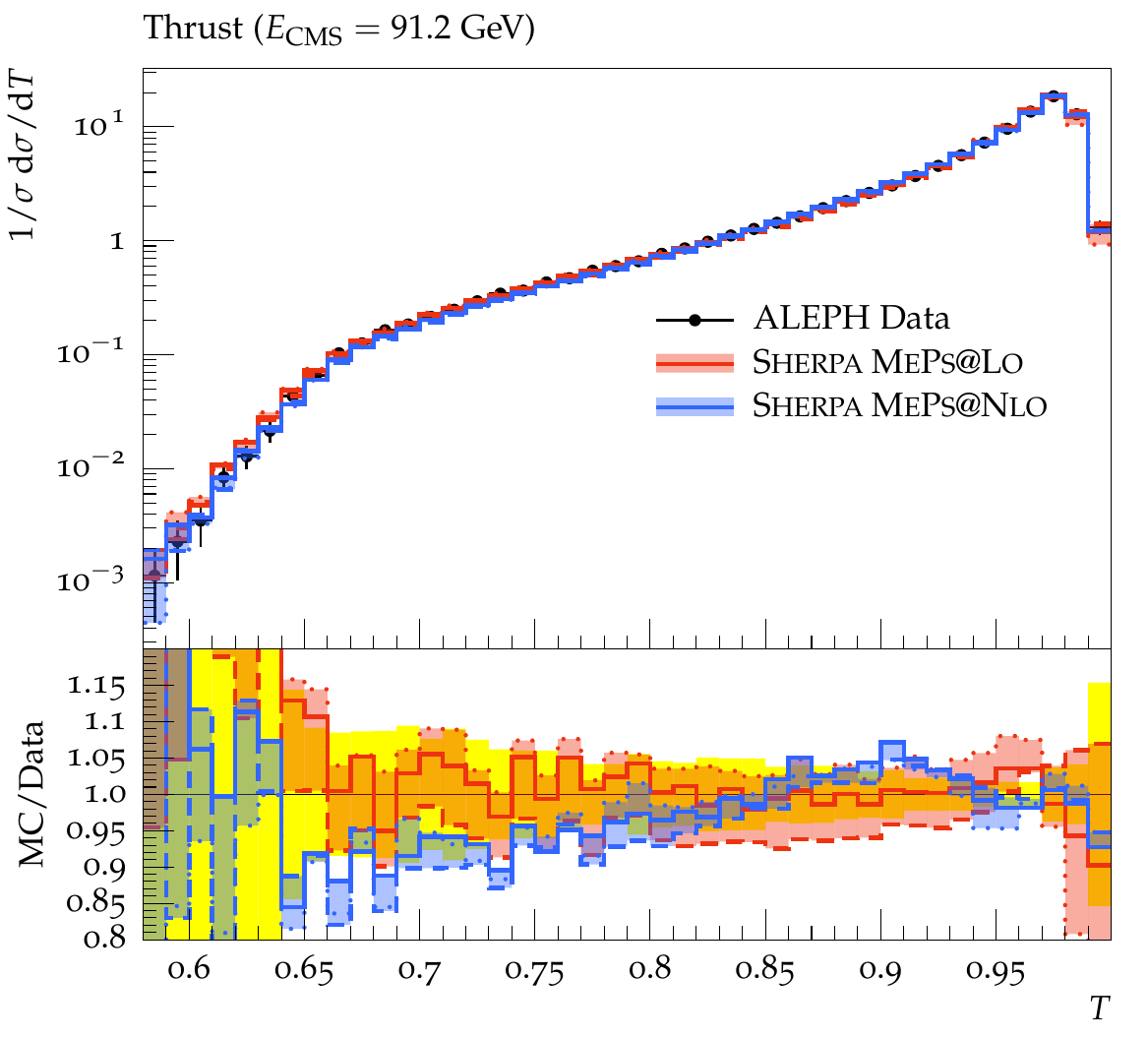}
\end{center}
\caption{Variation of thrust at LEP1 from ALEPH under scale variations.  SCF is varied from $0.5$ (dotted lines) to $2.0$ (dashed lines) for both the LO and NLO curves.}
\label{fig:scal1}
\end{figure}

\begin{figure}[h]
\begin{center}
\subfigure[]{\includegraphics[width=3.0in,bb=0 0 330 330]{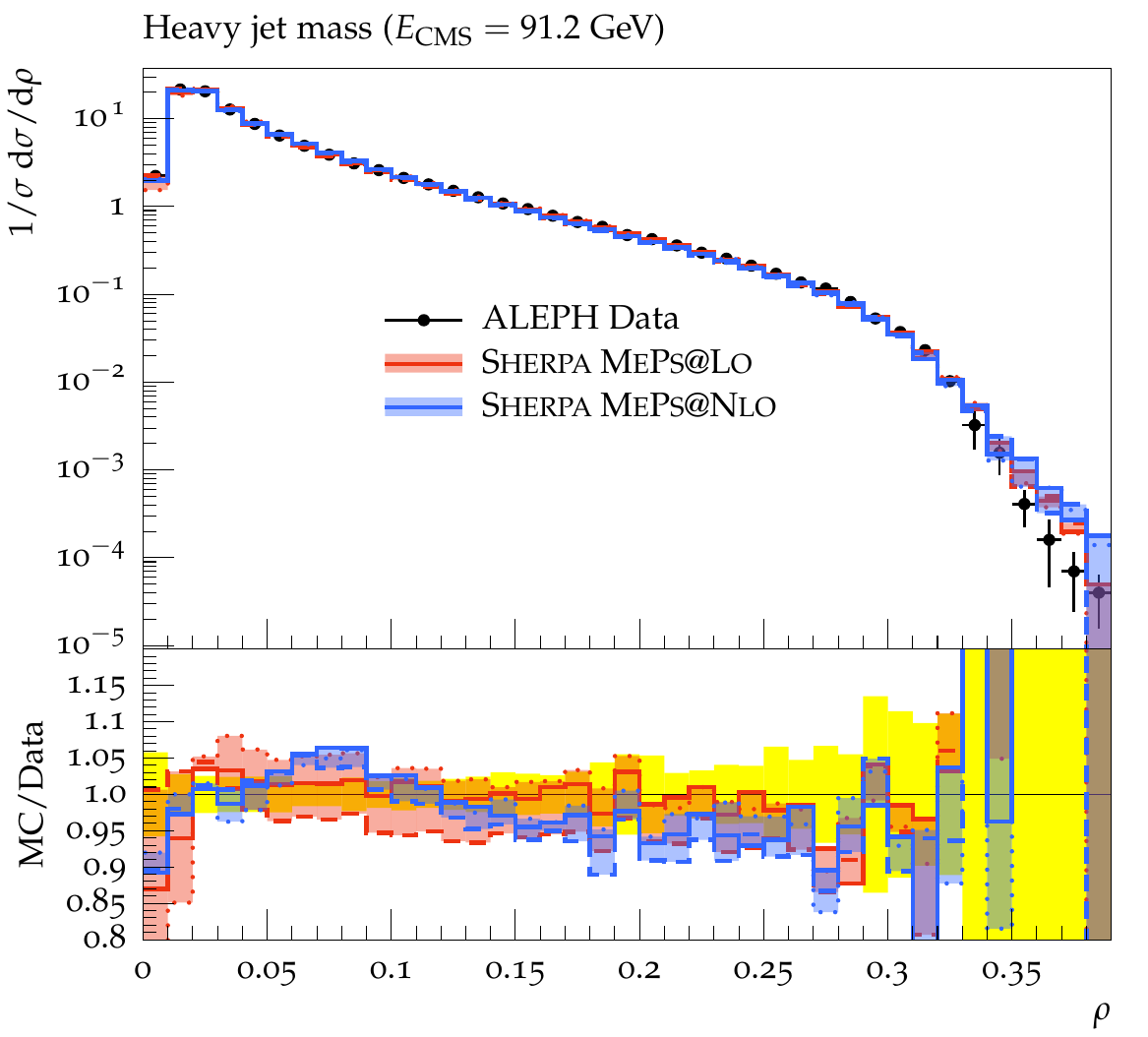}}\hspace{.4in}
\subfigure[]{\includegraphics[width=3.0in,bb=0 0 330 330]{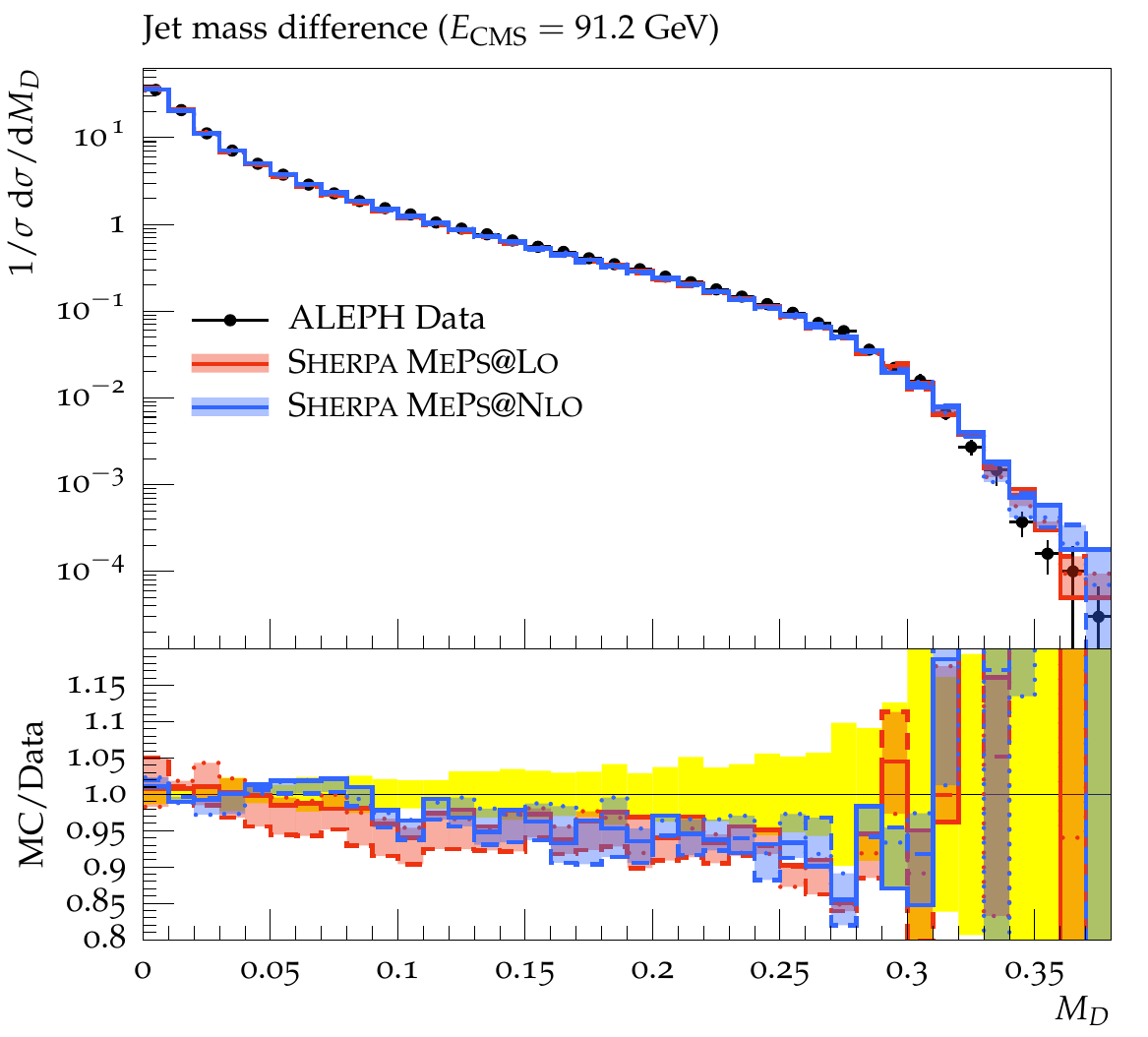}}\\
\end{center}
\caption{Variation of jet mass variables at LEP1 from ALEPH under scale variations.  SCF is varied from $0.5$ (dotted lines) to $2.0$ (dashed lines) for both the LO and NLO curves.}
\label{fig:scal2}
\end{figure}

The first of these is the thrust distribution from ALEPH, shown in Fig. \ref{fig:scal1}.  Here, we see that varying SCF results in changes to the thrust distribution of ${\cal O}(5\%)$ from its central value for the LO SHERPA MC.  The NLO SHERPA MC displays smaller variations of a few percent.  Similar conclusions hold for the jet mass variable distributions, shown in Fig. \ref{fig:scal2}. 

\begin{figure}[h]
\begin{center}
\subfigure[]{\includegraphics[width=3.0in,bb=0 0 330 330]{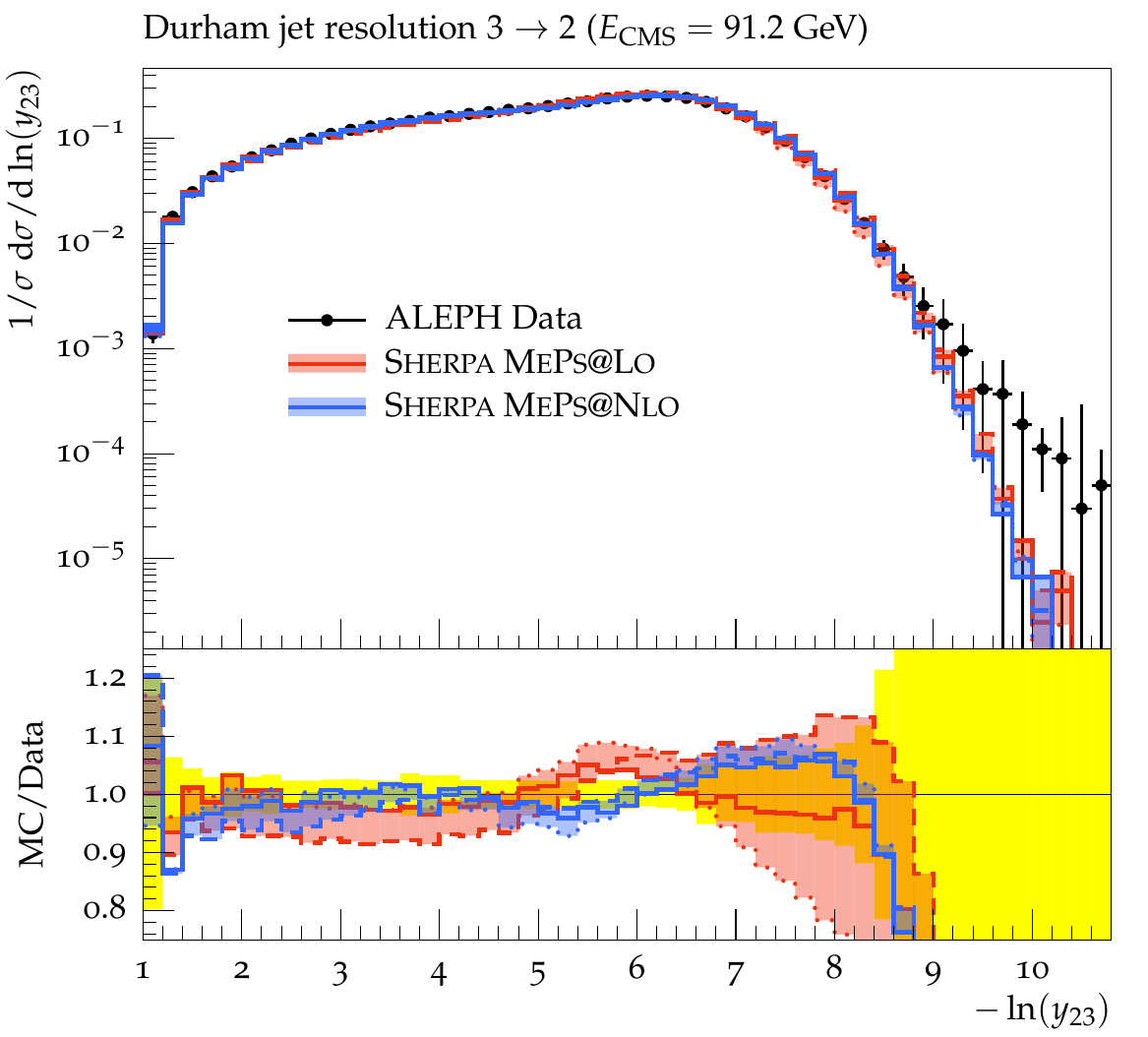}}\hspace{.4in}
\subfigure[]{\includegraphics[width=3.0in,bb=0 0 330 330]{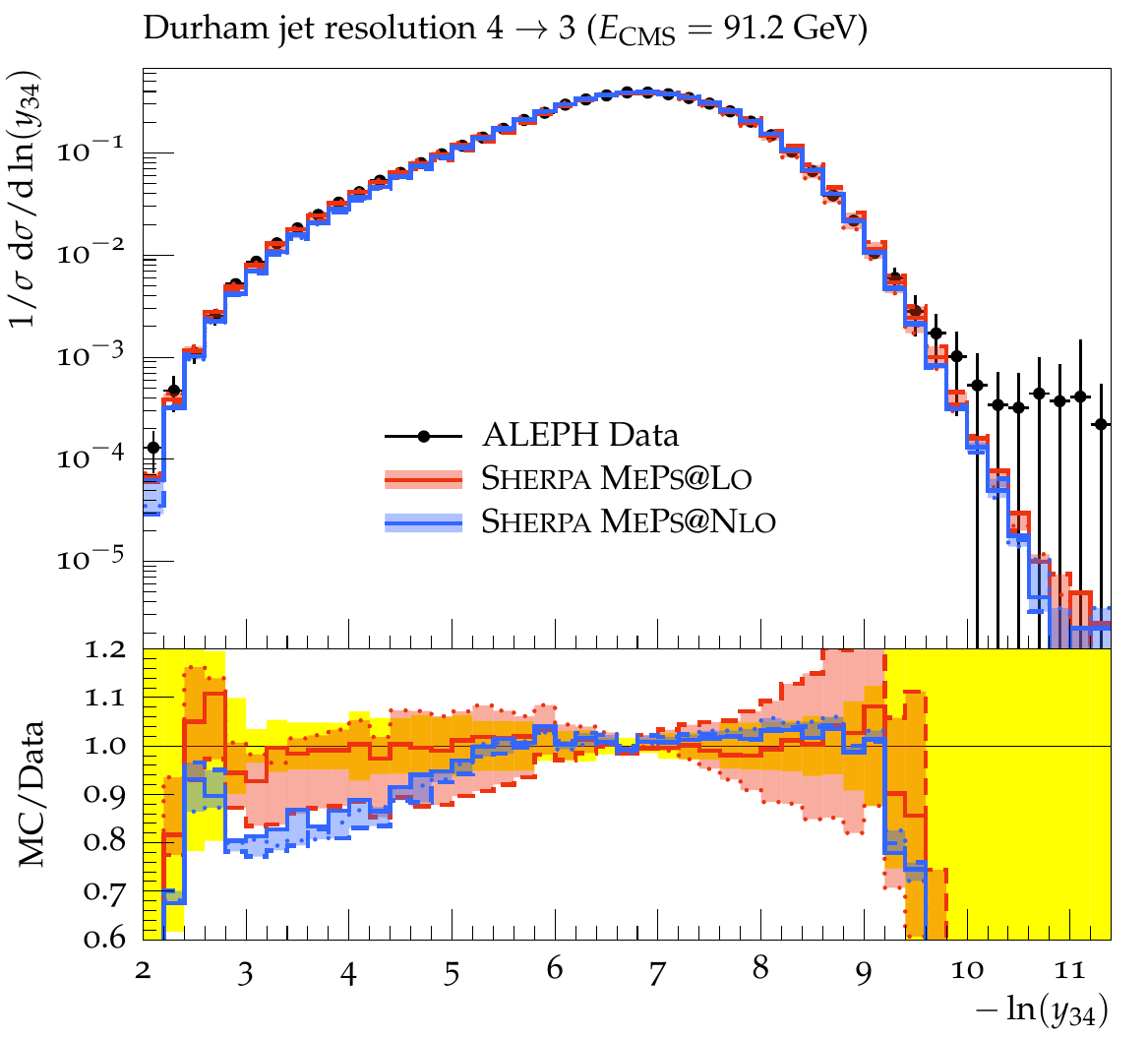}}\\
\subfigure[]{\includegraphics[width=3.0in,bb=0 0 330 330]{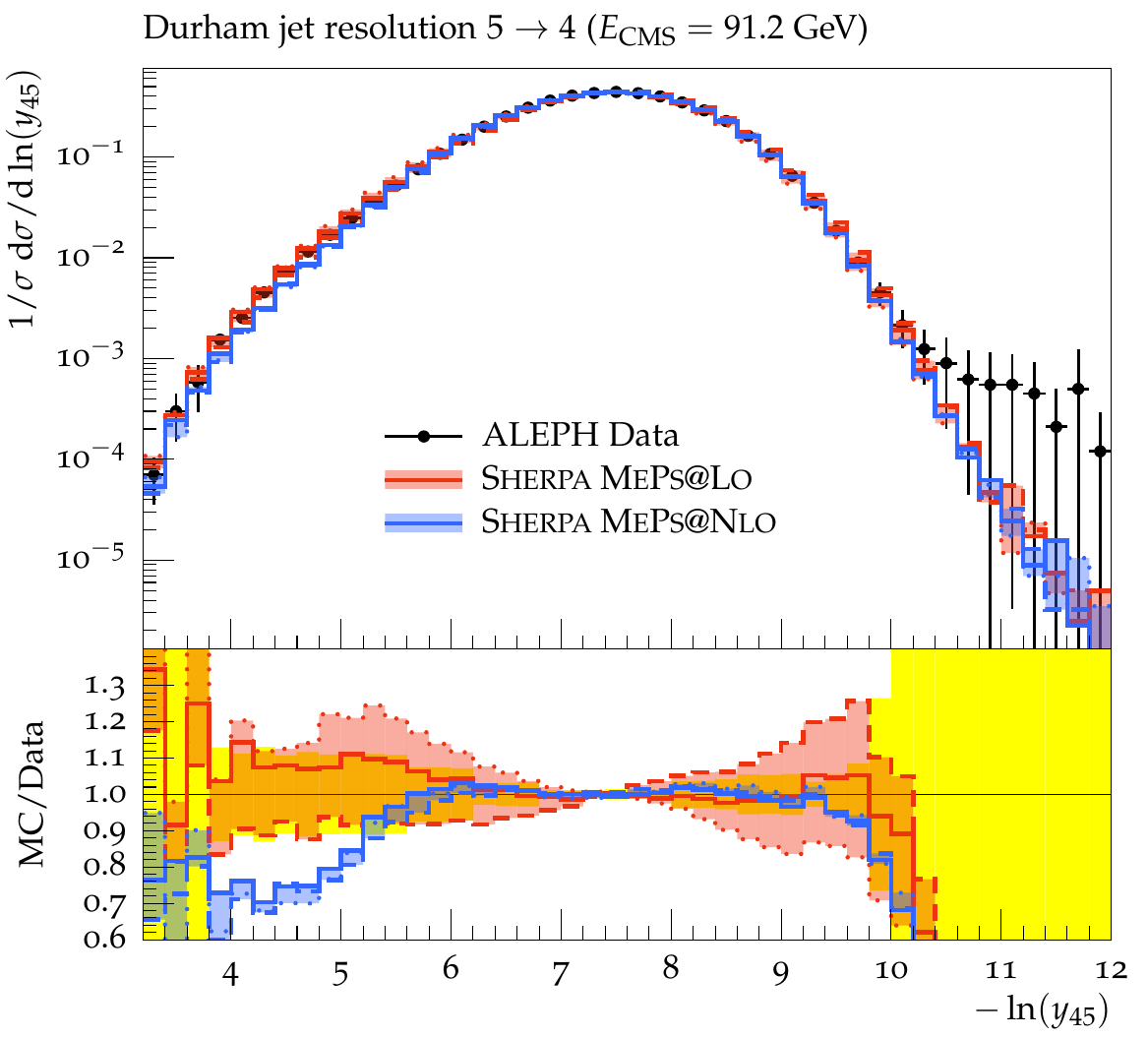}}\hspace{.4in}
\subfigure[]{\includegraphics[width=3.0in,bb=0 0 330 330]{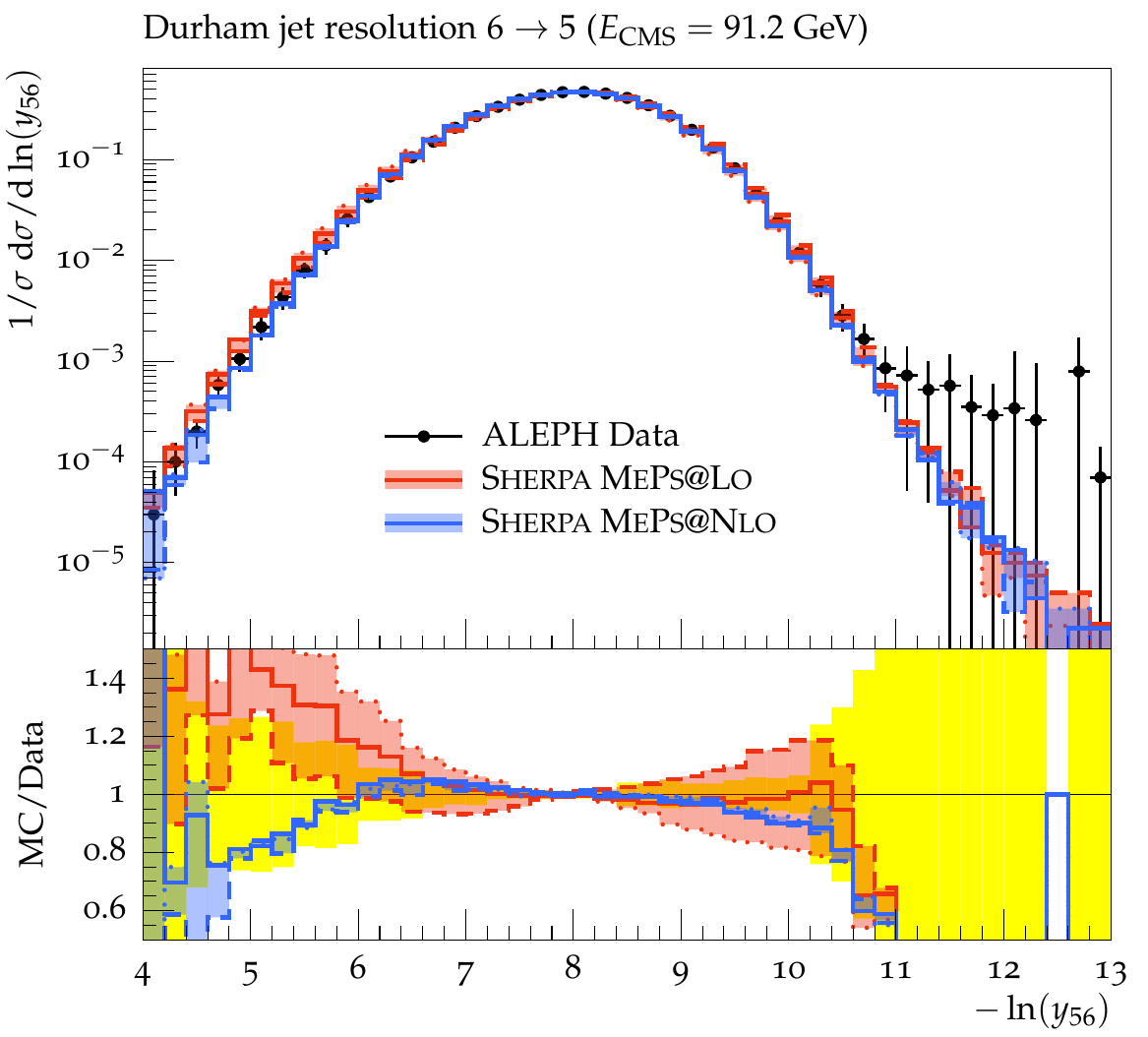}}\\
\end{center}
\caption{Variation of $y_{ij}$ at LEP1 from ALEPH under scale variations.  SCF is varied from $0.5$ (dotted lines) to $2.0$ (dashed lines) for both the LO and NLO curves.}
\label{fig:scale3}
\end{figure}

\begin{figure}[h]
\begin{center}
\subfigure[]{\includegraphics[width=3.0in,bb=0 0 330 330]{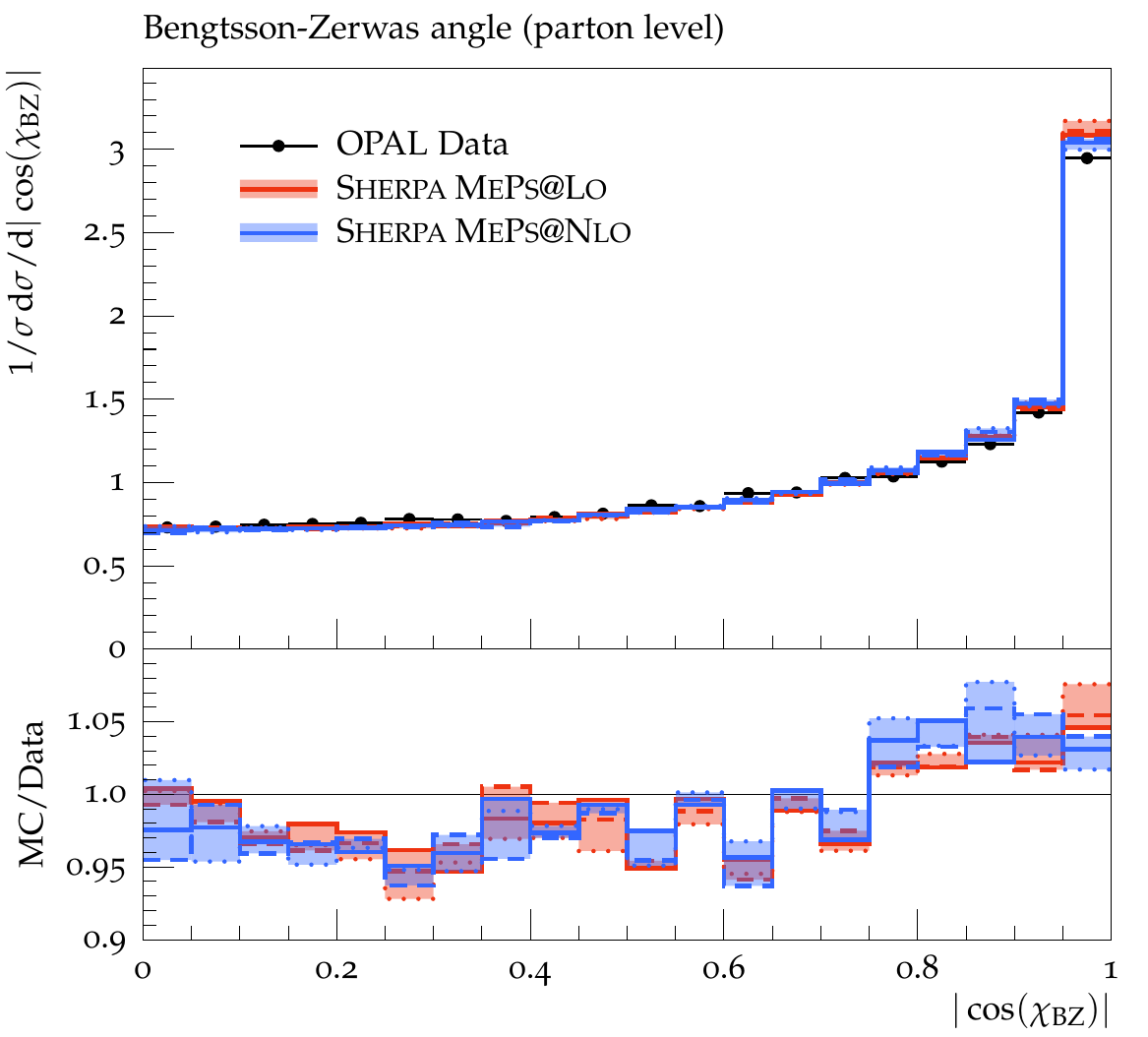}}\hspace{.4in}
\subfigure[]{\includegraphics[width=3.0in,bb=0 0 330 330]{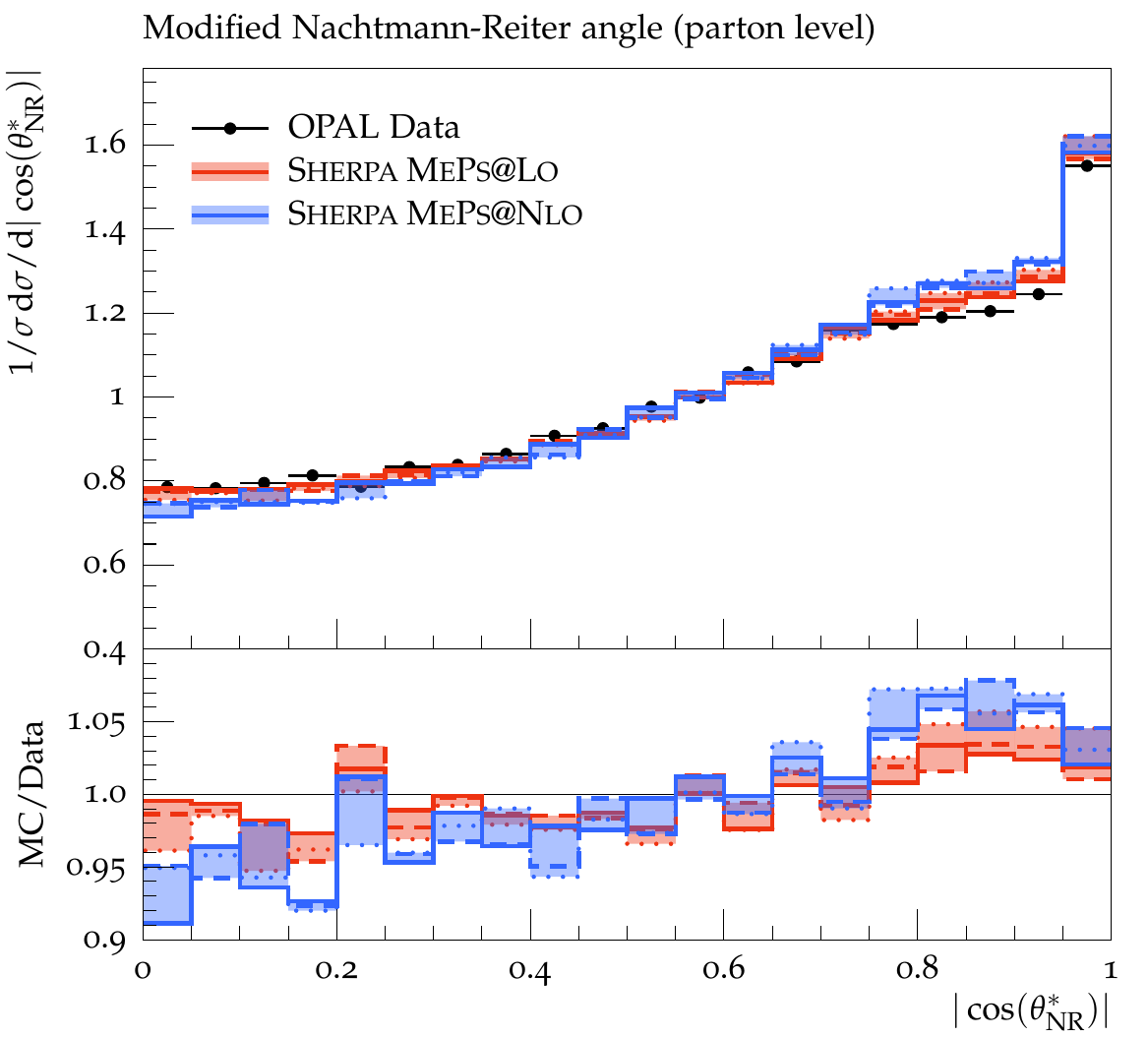}}\\
\subfigure[]{\includegraphics[width=3.0in,bb=0 0 330 330]{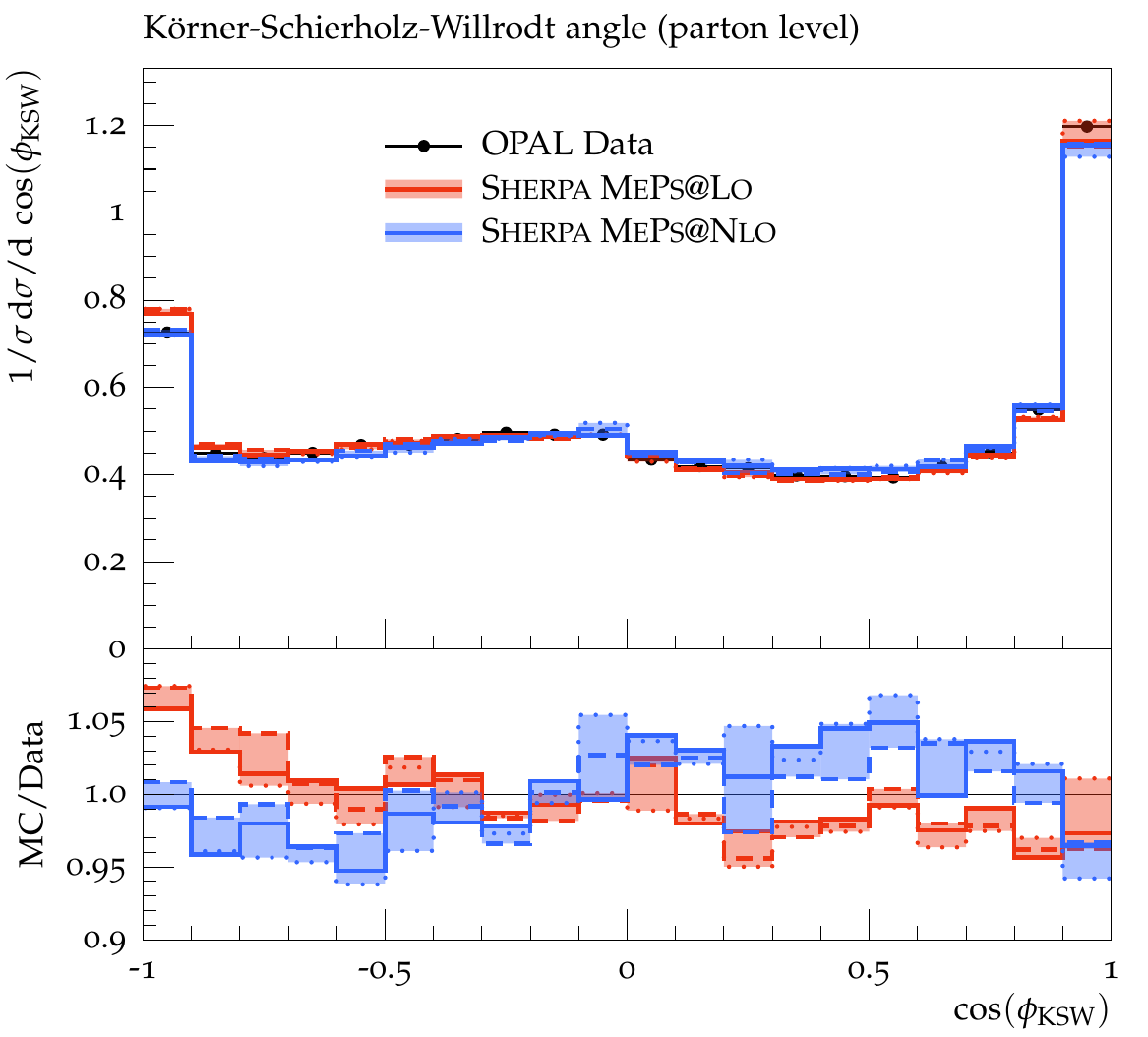}}\hspace{.4in}
\subfigure[]{\includegraphics[width=3.0in,bb=0 0 330 330]{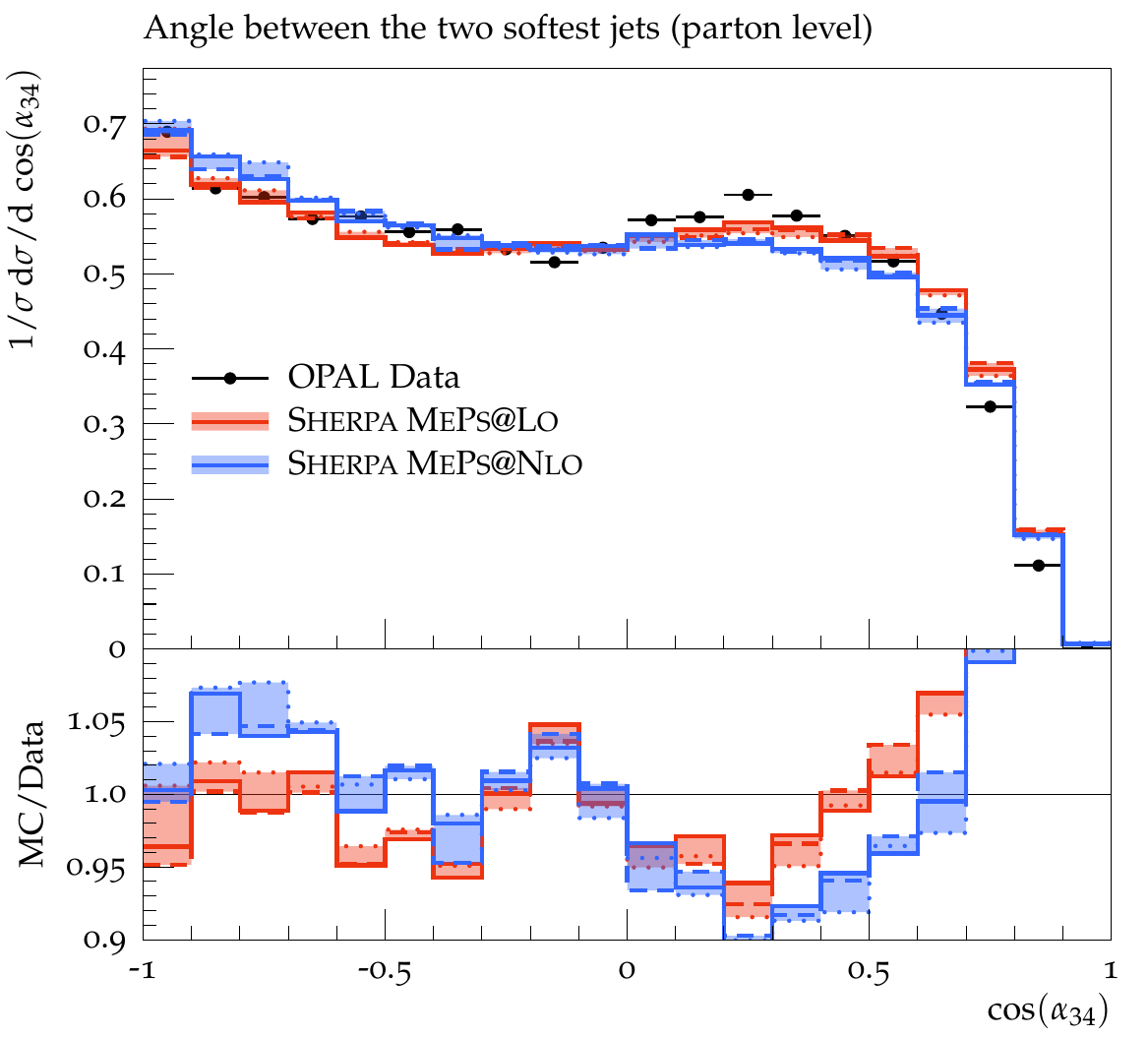}}\\
\end{center}
\caption{Variation of four-jet variables at LEP1 from OPAL under scale variations.  SCF is varied from $0.5$ (dotted lines) to $2.0$ (dashed lines) for both the LO and NLO curves.}
\label{fig:scale4}
\end{figure}

In Fig. \ref{fig:scale3}, we show the changes in the distributions of the Durham jet resolutions $y_{ij}$ with variations in SCF, compared to unfolded data from ALEPH.  The variations in the LO SHERPA distribution with changes in SCF range from several percent up to ${\cal O}(10\%)$.  The NLO distributions, however, are remarkably robust under changes in SCF; the $y_{34}$, $y_{45}$, and $y_{56}$ show variations of ${\cal O}(1-2\%)$.  We also observe good stability under scale variations for the OPAL four-jet variables shown in Fig. \ref{fig:scale4}.  Both the LO and NLO distributions only vary by a few percent as SCF varies from $0.5$ to $2.0$; however, this is of the same order as the size of statistical fluctuations in the bins, so the actual variation with SCF may be smaller.

\section{Conclusions}
\label{conc}

In this paper, we have explored the tuning of the SHERPA generator to produce MC for $e^+e^-\rightarrow\mbox{hadrons}$ applicable to LEP analyses.  Our main aim is to produce an MC sample which suitably describes the QCD four-jet background in order to study the events which show an excess in Ref. \cite{paper3}.  For this reason, the SHERPA MC generator, which contains the four-jet matrix element with proper matching and merging, was tuned using publicly-available LEP analyses in the Rivet package.  We describe our tuning procedure, give the values which we determined for the tuning parameters, and compare the resulting MC with LEP data and with MC from the KK2f generator.

We find that both our LO and NLO SHERPA tunes produce samples which are comparable to those of KK2f for observables which do not depend on clustering events into jets.  For Durham jet resolutions, we typically see improvement in going from KK2f to the SHERPA tunes, and we find that both SHERPA tunes describe four-jet variables significantly better than KK2f does. We also find that the LO SHERPA sample experiences changes of order a few percent to ${\cal O}(10\%)$ under scale variations; the NLO SHERPA sample is more stable, with variations of only a few percent.  However, we find that the overall agreement with data is best for the LO sample, and we will adopt this for our four-jet studies, retaining the NLO sample as well as the KK2f MC for systematic studies.

We thus find the SHERPA MC to be an improvement over the KK2f MC for our purposes.  In an accompanying paper \cite{paper2}, we submit these MC samples to full detector simulation and compare them directly to ALEPH data at LEP1 and LEP2.  There we study a wider range of observables related to the four-jet structure of events and perform more systematic studies.  Our results here suggest that modern MC generators such as SHERPA show great promise for future analyses and particularly for studies of four-jet states.  These are particularly relevant for analyses of LEP data, as well as those at future lepton colliders.

\begin{acknowledgments}
The authors would like to thank Thomas McElmurry for helpful discussions and comments.  The work of JK was supported in part by the Funda\c c\~ao para a Ci\^encia e a Tecnologia (FCT, Portugal), project UID/FIS/00777/2013.
\end{acknowledgments}

\appendix

\section{Professor Weight Files}
\label{weights}

\subsection{Flavor Tune Weight File}
\label{flavtundet}
\footnotesize
\begin{verbatim}
/ALEPH_1996_S3486095/d01-x01-y01  1  # Sphericity, $S$ (charged)
/ALEPH_1996_S3486095/d02-x01-y01  1  # Aplanarity, $A$ (charged)
/ALEPH_1996_S3486095/d03-x01-y01  1  # 1-Thrust, $1-T$ (charged)
/ALEPH_1996_S3486095/d04-x01-y01  1  # Thrust minor, $m$ (charged)
/ALEPH_1996_S3486095/d05-x01-y01  1  # Two-jet resolution variable, $Y_3$ (charged)
/ALEPH_1996_S3486095/d06-x01-y01  1  # Heavy jet mass (charged)
/ALEPH_1996_S3486095/d07-x01-y01  1  # $C$ parameter (charged)
/ALEPH_1996_S3486095/d08-x01-y01  1  # Oblateness, $M - m$ (charged)
/ALEPH_1996_S3486095/d09-x01-y01  1  # Scaled momentum, $x_p = |p|/|p_\text{beam}|$ (charged)
/ALEPH_1996_S3486095/d10-x01-y01  1  # Rapidity w.r.t. thrust axes, $y_T$ (charged)
/ALEPH_1996_S3486095/d11-x01-y01  1  # In-plane $p_T$ in GeV w.r.t. sphericity axes (charged)
/ALEPH_1996_S3486095/d12-x01-y01  1  # Out-of-plane $p_T$ in GeV w.r.t. sphericity axes (charged)
/ALEPH_1996_S3486095/d17-x01-y01  1  # Log of scaled momentum, $\log(1/x_p)$ (charged)
/ALEPH_1996_S3486095/d18-x01-y01  1  # Charged multiplicity distribution
/ALEPH_1996_S3486095/d19-x01-y01  10 # Mean charged multiplicity
/ALEPH_1996_S3486095/d20-x01-y01  5  # Mean charged multiplicity for rapidity $|Y| < 0.5$
/ALEPH_1996_S3486095/d21-x01-y01  5  # Mean charged multiplicity for rapidity $|Y| < 1.0$
/ALEPH_1996_S3486095/d22-x01-y01  5  # Mean charged multiplicity for rapidity $|Y| < 1.5$
/ALEPH_1996_S3486095/d23-x01-y01  5  # Mean charged multiplicity for rapidity $|Y| < 2.0$
/ALEPH_1996_S3486095/d25-x01-y01  1  # $\pi^\pm$ spectrum
/ALEPH_1996_S3486095/d26-x01-y01  1  # $K^\pm$ spectrum
/ALEPH_1996_S3486095/d27-x01-y01  1  # $p$ spectrum
/ALEPH_1996_S3486095/d28-x01-y01  1  # $\gamma$ spectrum
/ALEPH_1996_S3486095/d29-x01-y01  1  # $\pi^0$ spectrum
/ALEPH_1996_S3486095/d30-x01-y01  1  # $\eta$ spectrum
/ALEPH_1996_S3486095/d31-x01-y01  1  # $\eta'$ spectrum
/ALEPH_1996_S3486095/d32-x01-y01  1  # $K^0$ spectrum
/ALEPH_1996_S3486095/d33-x01-y01  1  # $\Lambda^0$ spectrum
/ALEPH_1996_S3486095/d34-x01-y01  1  # $\Xi^-$ spectrum
/ALEPH_1996_S3486095/d35-x01-y01  1  # $\Sigma^\pm(1385)$ spectrum
/ALEPH_1996_S3486095/d36-x01-y01  1  # $\Xi^0(1530)$ spectrum
/ALEPH_1996_S3486095/d37-x01-y01  1  # $\rho$ spectrum
/ALEPH_1996_S3486095/d38-x01-y01  1  # $\omega(782)$ spectrum 
/ALEPH_1996_S3486095/d39-x01-y01  1  # $K^{*0}(892)$ spectrum
/ALEPH_1996_S3486095/d40-x01-y01  1  # $\phi$ spectrum
/ALEPH_1996_S3486095/d43-x01-y01  1  # $K^{*\pm}(892)$ spectrum
/ALEPH_1996_S3486095/d44-x01-y02  10  # Mean $\pi^0$ multiplicity
/ALEPH_1996_S3486095/d44-x01-y03  10  # Mean $\eta$ multiplicity
/ALEPH_1996_S3486095/d44-x01-y04  10  # Mean $\eta'$ multiplicity
/ALEPH_1996_S3486095/d44-x01-y05  10  # Mean $K_S + K_L$ multiplicity
/ALEPH_1996_S3486095/d44-x01-y06  10  # Mean $\rho^0$ multiplicity
/ALEPH_1996_S3486095/d44-x01-y07  10  # Mean $\omega(782)$ multiplicity
/ALEPH_1996_S3486095/d44-x01-y08  10  # Mean $\phi$ multiplicity
/ALEPH_1996_S3486095/d44-x01-y09  10  # Mean $K^{*\pm}$ multiplicity
/ALEPH_1996_S3486095/d44-x01-y10  10  # Mean $K^{*0}$ multiplicity
/ALEPH_1996_S3486095/d44-x01-y11  10  # Mean $\Lambda$ multiplicity
/ALEPH_1996_S3486095/d44-x01-y12  10  # Mean $\Sigma$ multiplicity
/ALEPH_1996_S3486095/d44-x01-y13  10  # Mean $\Xi$ multiplicity
/ALEPH_1996_S3486095/d44-x01-y14  10  # Mean $\Sigma(1385)$ multiplicity
/ALEPH_1996_S3486095/d44-x01-y15  10  # Mean $\Xi(1530)$ multiplicity
/ALEPH_1996_S3486095/d44-x01-y16  10  # Mean $\Omega^\mp$ multiplicity

/ALEPH_2004_S5765862/d157-x01-y01 10  # Durham jet resolution $3 \to 2$ ($E_\mathrm{CMS}=91.2$ GeV)
/ALEPH_2004_S5765862/d165-x01-y01 10  # Durham jet resolution $4 \to 3$ ($E_\mathrm{CMS}=91.2$ GeV)
/ALEPH_2004_S5765862/d173-x01-y01 10  # Durham jet resolution $5 \to 4$ ($E_\mathrm{CMS}=91.2$ GeV)
/ALEPH_2004_S5765862/d180-x01-y01 10  # Durham jet resolution $6 \to 5$ ($E_\mathrm{CMS}=91.2$ GeV)
\end{verbatim}

\subsection{LO Tune Weight File}
\label{lotundet}
\begin{verbatim}
/ALEPH_1996_S3486095/d01-x01-y01  1  # Sphericity, $S$ (charged)
/ALEPH_1996_S3486095/d02-x01-y01  1  # Aplanarity, $A$ (charged)
/ALEPH_1996_S3486095/d03-x01-y01  1  # 1-Thrust, $1-T$ (charged)
/ALEPH_1996_S3486095/d04-x01-y01  1  # Thrust minor, $m$ (charged)
/ALEPH_1996_S3486095/d05-x01-y01  1  # Two-jet resolution variable, $Y_3$ (charged)
/ALEPH_1996_S3486095/d06-x01-y01  1  # Heavy jet mass (charged)
/ALEPH_1996_S3486095/d07-x01-y01  1  # $C$ parameter (charged)
/ALEPH_1996_S3486095/d08-x01-y01  1  # Oblateness, $M - m$ (charged)
/ALEPH_1996_S3486095/d09-x01-y01  1  # Scaled momentum, $x_p = |p|/|p_\text{beam}|$ (charged)
/ALEPH_1996_S3486095/d10-x01-y01  1  # Rapidity w.r.t. thrust axes, $y_T$ (charged)
/ALEPH_1996_S3486095/d11-x01-y01  1  # In-plane $p_T$ in GeV w.r.t. sphericity axes (charged)
/ALEPH_1996_S3486095/d12-x01-y01  1  # Out-of-plane $p_T$ in GeV w.r.t. sphericity axes (charged)
/ALEPH_1996_S3486095/d17-x01-y01  1  # Log of scaled momentum, $\log(1/x_p)$ (charged)
/ALEPH_1996_S3486095/d18-x01-y01 10  # Charged multiplicity distribution
/ALEPH_1996_S3486095/d19-x01-y01 50  # Mean charged multiplicity

/ALEPH_2004_S5765862/d102-x01-y01  1  # Thrust minor ($E_\mathrm{CMS}=91.2$ GeV)
/ALEPH_2004_S5765862/d110-x01-y01  1  # Jet mass difference ($E_\mathrm{CMS}=91.2$ GeV)
/ALEPH_2004_S5765862/d118-x01-y01  1  # Aplanarity ($E_\mathrm{CMS}=91.2$ GeV)
/ALEPH_2004_S5765862/d133-x01-y01  1  # Oblateness ($E_\mathrm{CMS}=91.2$ GeV)
/ALEPH_2004_S5765862/d141-x01-y01  1  # Sphericity ($E_\mathrm{CMS}=91.2$ GeV)
/ALEPH_2004_S5765862/d149-x01-y01  2  # Durham jet resolution $2 \to 1$ ($E_\mathrm{CMS}=91.2$ GeV)
/ALEPH_2004_S5765862/d157-x01-y01  6  # Durham jet resolution $3 \to 2$ ($E_\mathrm{CMS}=91.2$ GeV)
/ALEPH_2004_S5765862/d165-x01-y01  8  # Durham jet resolution $4 \to 3$ ($E_\mathrm{CMS}=91.2$ GeV)
/ALEPH_2004_S5765862/d173-x01-y01 14  # Durham jet resolution $5 \to 4$ ($E_\mathrm{CMS}=91.2$ GeV)
/ALEPH_2004_S5765862/d180-x01-y01 20  # Durham jet resolution $6 \to 5$ ($E_\mathrm{CMS}=91.2$ GeV)
/ALEPH_2004_S5765862/d187-x01-y01  1  # 1-jet fraction ($E_\mathrm{CMS}=91.2$ GeV)
/ALEPH_2004_S5765862/d195-x01-y01  2  # 2-jet fraction ($E_\mathrm{CMS}=91.2$ GeV)
/ALEPH_2004_S5765862/d203-x01-y01  6  # 3-jet fraction ($E_\mathrm{CMS}=91.2$ GeV)
/ALEPH_2004_S5765862/d211-x01-y01  8  # 4-jet fraction ($E_\mathrm{CMS}=91.2$ GeV)
/ALEPH_2004_S5765862/d219-x01-y01 14  # 5-jet fraction ($E_\mathrm{CMS}=91.2$ GeV)
/ALEPH_2004_S5765862/d227-x01-y01 20  # $\geq$6-jet fraction ($E_\mathrm{CMS}=91.2$ GeV)
/ALEPH_2004_S5765862/d54-x01-y01   1  # Thrust ($E_\mathrm{CMS}=91.2$ GeV)
/ALEPH_2004_S5765862/d62-x01-y01   1  # Heavy jet mass ($E_\mathrm{CMS}=91.2$ GeV)
/ALEPH_2004_S5765862/d70-x01-y01   1  # Total jet broadening ($E_\mathrm{CMS}=91.2$ GeV)
/ALEPH_2004_S5765862/d78-x01-y01   1  # Wide jet broadening ($E_\mathrm{CMS}=91.2$ GeV)
/ALEPH_2004_S5765862/d86-x01-y01   1  # C-Parameter ($E_\mathrm{CMS}=91.2$ GeV)
/ALEPH_2004_S5765862/d94-x01-y01   1  # Thrust major ($E_\mathrm{CMS}=91.2$ GeV)
\end{verbatim}

\subsection{NLO Tune Weight File}
\label{nlotundet}
\begin{verbatim}
/ALEPH_1996_S3486095/d01-x01-y01  1  # Sphericity, $S$ (charged)
/ALEPH_1996_S3486095/d02-x01-y01  1  # Aplanarity, $A$ (charged)
/ALEPH_1996_S3486095/d03-x01-y01  1  # 1-Thrust, $1-T$ (charged)
/ALEPH_1996_S3486095/d04-x01-y01  1  # Thrust minor, $m$ (charged)
/ALEPH_1996_S3486095/d05-x01-y01  1  # Two-jet resolution variable, $Y_3$ (charged)
/ALEPH_1996_S3486095/d06-x01-y01  1  # Heavy jet mass (charged)
/ALEPH_1996_S3486095/d07-x01-y01  1  # $C$ parameter (charged)
/ALEPH_1996_S3486095/d09-x01-y01  1  # Scaled momentum, $x_p = |p|/|p_\text{beam}|$ (charged)
/ALEPH_1996_S3486095/d10-x01-y01  1  # Rapidity w.r.t. thrust axes, $y_T$ (charged)
/ALEPH_1996_S3486095/d17-x01-y01  1  # Log of scaled momentum, $\log(1/x_p)$ (charged)
/ALEPH_1996_S3486095/d18-x01-y01  5  # Charged multiplicity distribution
/ALEPH_1996_S3486095/d19-x01-y01  100  # Mean charged multiplicity
/ALEPH_1996_S3486095/d20-x01-y01  2  # Mean charged multiplicity for rapidity $|Y| < 0.5$
/ALEPH_1996_S3486095/d21-x01-y01  2  # Mean charged multiplicity for rapidity $|Y| < 1.0$
/ALEPH_1996_S3486095/d22-x01-y01  2  # Mean charged multiplicity for rapidity $|Y| < 1.5$
/ALEPH_1996_S3486095/d23-x01-y01  2  # Mean charged multiplicity for rapidity $|Y| < 2.0$

/ALEPH_2004_S5765862/d102-x01-y01  2  # Thrust minor ($E_\mathrm{CMS}=91.2$ GeV)
/ALEPH_2004_S5765862/d110-x01-y01  2  # Jet mass difference ($E_\mathrm{CMS}=91.2$ GeV)
/ALEPH_2004_S5765862/d118-x01-y01  2  # Aplanarity ($E_\mathrm{CMS}=91.2$ GeV)
/ALEPH_2004_S5765862/d141-x01-y01  2  # Sphericity ($E_\mathrm{CMS}=91.2$ GeV)
/ALEPH_2004_S5765862/d157-x01-y01  5  # Durham jet resolution $3 \to 2$ ($E_\mathrm{CMS}=91.2$ GeV)
/ALEPH_2004_S5765862/d165-x01-y01  5  # Durham jet resolution $4 \to 3$ ($E_\mathrm{CMS}=91.2$ GeV)
/ALEPH_2004_S5765862/d173-x01-y01  5  # Durham jet resolution $5 \to 4$ ($E_\mathrm{CMS}=91.2$ GeV)
/ALEPH_2004_S5765862/d180-x01-y01  5  # Durham jet resolution $6 \to 5$ ($E_\mathrm{CMS}=91.2$ GeV)
/ALEPH_2004_S5765862/d203-x01-y01  2  # 3-jet fraction ($E_\mathrm{CMS}=91.2$ GeV)
/ALEPH_2004_S5765862/d211-x01-y01  2  # 4-jet fraction ($E_\mathrm{CMS}=91.2$ GeV)
/ALEPH_2004_S5765862/d219-x01-y01  2  # 5-jet fraction ($E_\mathrm{CMS}=91.2$ GeV)
/ALEPH_2004_S5765862/d227-x01-y01  2  # $\geq$6-jet fraction ($E_\mathrm{CMS}=91.2$ GeV)
/ALEPH_2004_S5765862/d54-x01-y01  10  # Thrust ($E_\mathrm{CMS}=91.2$ GeV)
/ALEPH_2004_S5765862/d62-x01-y01  10  # Heavy jet mass ($E_\mathrm{CMS}=91.2$ GeV)
/ALEPH_2004_S5765862/d70-x01-y01   2  # Total jet broadening ($E_\mathrm{CMS}=91.2$ GeV)
/ALEPH_2004_S5765862/d78-x01-y01   2  # Wide jet broadening ($E_\mathrm{CMS}=91.2$ GeV)
/ALEPH_2004_S5765862/d86-x01-y01   2  # C-Parameter ($E_\mathrm{CMS}=91.2$ GeV)
/ALEPH_2004_S5765862/d94-x01-y01   2  # Thrust major ($E_\mathrm{CMS}=91.2$ GeV)
\end{verbatim}

\end{document}